\documentclass[prd,aps,preprint,tightenlines,nofootinbib,superscriptaddress]{revtex4}
\usepackage{mathrsfs}
\usepackage{amsfonts}
\usepackage{amsmath}
\usepackage{array}
\usepackage{verbatim}
\usepackage{epsfig}
\usepackage{graphicx}

\begin{document}
\title{TMD Evolution: Matching SIDIS to Drell-Yan and W/Z Boson Production}

\author{Peng Sun}
\affiliation{Nuclear Science Division, Lawrence Berkeley National
Laboratory, Berkeley, CA 94720, USA}
\affiliation{Center for High Energy Physics, Peking University, Beijing 100871,
China}
\author{Feng Yuan}
\affiliation{Nuclear Science Division, Lawrence Berkeley National
Laboratory, Berkeley, CA 94720, USA}
\affiliation{Center for High Energy Physics, Peking University, Beijing 100871,
China}

\begin{abstract}
We examine the QCD evolution for the transverse momentum dependent observables
in hard processes of semi-inclusive hadron production in deep inelastic scattering and
Drell-Yan lepton pair production in $pp$ collisions, including the spin-average cross sections
and Sivers single transverse spin asymmetries. We show that the evolution equations
derived by a direct integral of the Collins-Soper-Sterman evolution kernel from low to high
$Q$ can describe well the transverse momentum distribution of the unpolarized cross sections
in the $Q^2$ range from 2 to 100 GeV$^2$. In addition, the  matching is established
between our evolution and the Collins-Soper-Sterman resummation with $b_*$-prescription
and Konychev-Nodalsky parameterization of the non-perturbative form factors, which are
formulated to describe the Drell-Yan lepton pair and W/Z boson production in hadronic
collisions. With these
results, we present the predictions for the Sivers single transverse spin asymmetries
in Drell-Yan lepton pair production and $W^\pm$ boson production in polarized
$pp$ and $\pi^- p$ collisions for several proposed experiments. We emphasize that
these experiments will not only provide crucial test of the sign change of the Sivers
asymmetry, but also provide important opportunities to study the QCD evolution effects.
\end{abstract}

 \maketitle

\section{introduction}

Transverse momentum dependent (TMD) parton distributions and fragmentation
functions are formally introduced as an extension to the parton model
description of nucleon structure and an important tool to calculate hadronic
processes. In the last few years, these distribution functions have attracted great attentions
in hadron physics community. In particular, the novel single transverse spin
asymmetries (SSAs) in semi-inclusive hadron production in deep inelastic scattering
processes (SIDIS) observed by the HERMES, COMPASS, and JLab Hall A collaborations,
have stimulated much theoretical developments. The TMD factorization
provides a solid theoretical framework to understand these spin asymmetries.
Moreover, together with the generalized parton distributions (GPDs), the TMDs
unveil the internal structure of nucleon in a three dimension fashion,
the so-called nucleon tomography. These topics are major emphases in the
planed electron-ion collider~\cite{Boer:2011fh}.

An important theoretical aspect of the TMD parton distribution and fragmentation
functions is the energy evolution, which was thoroughly studied in the early
paper by Collins and Soper~\cite{Collins:1981uk}. This evolution is referred as the Collins-Soper
(CS) evolution equation. It has been applied to formulate the perturbative
resummation of large double logarithms in hard scattering processes where
transverse momentum distribution are measured. The associated resummation
is called Collins-Soper-Sterman (CSS)~\cite{Collins:1984kg} resummation, or
transverse momentum resummation. In these hard processes,
because of two separate scales, there exist large double logarithms in
each order of perturbative calculations (originally from a QED calculation
by Sudakov~\cite{Sudakov:1954sw}), and the relevant resummation
has to be taken in the calculation~\cite{Dokshitzer:1978dr,Parisi:1979se,Collins:1984kg}.
For example, in Drell-Yan lepton
pair production in $pp$ collisions, the invariant mass $Q$ is much larger than
the total transverse momentum of the lepton pair $q_\perp$, $Q\gg q_\perp$,
where perturbative corrections will induce large logarithms
$\alpha_s^i\left(\ln Q^2/q_\perp^2\right)^{2i-1}$. The resummation of these
large logarithms are performed by applying the TMD factorization
and the CS evolution. Successful applications have been made to
study the low transverse momentum distribution of Drell-Yan type of
processes in hadronic collisions from fixed target experiments to
highest collider energy experiments, such as the Tevatron at Fermilab
and the large hadron collider (LHC) at CERN, see, for example, the relevant
publications in Refs.~\cite{Landry:2002ix,Konychev:2005iy,Qiu:2000ga,Kulesza:2002rh,Catani:2000vq,
Catani:2003zt,Bozzi:2003jy}.

The resummation for the hard processes are based on the TMD
factorization for these processes~\cite{Collins:1981uk,Collins:1984kg,Ji:2004wu,Collins:2004nx,Collins,Aybat:2011zv,Aybat:2011ge}.
Since the definition of the TMDs contains the light-cone singularity~\cite{Collins:1981uk},
the detailed calculations depend on the scheme to regulate this
singularity. In the original paper of Collins-Soper~\cite{Collins:1981uk},
an axial gauge has been used. This was followed by a gauge invariant
approach in Ji-Ma-Yuan with a slight off-light-cone gauge link in covariant gauge~\cite{Ji:2004wu}
(referred as Ji-Ma-Yuan scheme in the following).
A new definition for the TMD and the associated soft factor has been proposed in Ref.~\cite{Collins}
where a subtraction method was used to regulate the light-cone singularity
(referred as Collins-11 in the following),
and the phenomenological applications were presented in Refs.~\cite{Aybat:2011zv,Aybat:2011ge}.
Although the TMDs depend on the regularization scheme, the resummation
for the physical observables, such as the differential cross sections and the spin
asymmetries, is independent on the scheme.
We will present detailed discussions in the below between the two formalisms of Ref.~\cite{Ji:2004wu} and Ref.~\cite{Collins}.

To understand the energy evolution of the spin-dependent hard processes,
such as the SSA in SIDIS and Drell-Yan lepton pair production, we need
to extend the CS and CSS derivation to the interested observables~\cite{Boer:2001he}.
The CS evolution equations for the TMDs
was extensively discussed in Ref.~\cite{Idilbi:2004vb}, where the evolution
kernel was derived for all the leading order TMD quark distributions.
In particular, for the so-called $k_\perp$-even TMDs, the evolution
is exactly the same as the original CS evolution. For the
$k_\perp$-odd ones, a slightly different form has to be used,
but with the same kernel. These evolution equations
can be cross checked with the finite order perturbative calculations,
which has been shown to yield consistent results~\cite{Kang:2011mr}.

Besides the above developments in the investigation of the TMDs
in full QCD, recently an effective theory approach
based on the soft-collinear-effective-theory has been applied to
the evolution of the TMDs. Several different schemes are
proposed in the literature~\cite{Mantry:2009qz,Becher:2010tm,
GarciaEchevarria:2011rb,Chiu:2012ir}. It has been shown in Ref.~\cite{Collins:2012uy}
that one of the effective theory approach~\cite{GarciaEchevarria:2011rb}
(referred as EIS in the following)
is equivalent to the Collins-11 formalism~\cite{Collins}.

Although there are different ways to formulate the TMD distribution
and fragmentation functions, the energy evolution and resummation
for the physical observables (including the differential cross
sections and spin asymmetries) will always take the same form
as they should be. Therefore, in this paper, we will focus on the
energy evolution for the differential cross section and spin
asymmetries. Of course, to have a solid prediction for the
physical observables, we need to have the TMD factorization proven
for the relevant processes. The SIDIS and Drell-Yan lepton
pair production in $pp$ collisions are two examples that
a rigorous TMD factorization has been proven.

The main goal of this paper is to make predictions for the
Sivers single spin asymmetries in Drell-Yan lepton pair
production in $pp$ collisions from the constraints from the
Sivers asymmetries observed in SIDIS from HERMES/COMPASS
experiments. The TMD factorization and universality has predicted
that the Sivers asymmetries in these two processes differ by
a sign, because of difference between the initial/final state interaction ~\cite{Brodsky:2002cx,Collins:2002kn}.
The Sivers single spin asymmetries in SIDIS have been observed
by HERMES/COMPASS/JLab Hall A collaborations with $Q^2$ at the region from
2 to 4 GeV$^2$~\cite{Airapetian:2009ae,Airapetian:2010ds,Alekseev:2008aa,Adolph:2012sp,Qian:2011py}.
However, the typical Drell-Yan measurements
will be around the region from
$4^2$ to $ 91.19^2 $  GeV$^2$~\cite{compassdy,fermilabdy,futurerhic}. Therefore, the energy evolution
of the associated TMDs is important to carry out a rigorous
test of the sign change prediction. Early calculations are based
on the TMD factorization, however, without the energy evolution effects
in the derivation~\cite{Vogelsang:2005cs,Collins:2005rq,Anselmino:2005ea,Anselmino:2005an,Anselmino:2009st,Kang:2009sm,
Bacchetta:2011gx,Gamberg:2013kla}.
Recently, several studies
have started to take into account the evolution effects~\cite{Boer:2001he,Anselmino:2012aa,Sun:2013dya,Aybat:2011ta}.
In particular, in Ref.~\cite{Aybat:2011ta}, a strong decreasing was found in comparing
the SSA in typical Drell-Yan processes to those observed
by HERMES/COMPASS. In this paper, we will carefully examine
these predictions, and present a consistent calculation
for the energy evolution in both spin-average and
single-spin dependent cross sections. A brief summary
has been published earlier~\cite{Sun:2013dya}.

The starting point of our calculations is to build the correct
evolution framework which can describe the known experimental
data of the unpolarized cross sections in the associated processes.
One has to test the TMD evolution with the unpolarized
cross sections before they can be applied to spin-dependent
cross sections and the spin asymmetries. This is a very important
point, which, unfortunately, is often forgotten in the phenomenological
studies.

We will make use of the successful approach in the CSS resummation.
In these formulations, a non-perturbative form factor has to be included.
We follow the BLNY and KN calculations~\cite{Landry:2002ix,Konychev:2005iy}, where
$b_*$-prescription of CSS resummation is applied: $b\rightarrow b_*=b/\sqrt{1+b^2/b_{max}^2}$
with $b$ the impact parameter. This prescription guarantees
that $1/b_*>1/b_{max}\gg\Lambda_{\rm QCD}$. The non-perturbative form factor
takes a form as $(g_1+g_2\ln Q/2Q_0+g_1g_3\ln(100x_1x_2))b^2$
in the impact parameter space with $x_1$ and $x_2$ the longitudinal
momentum fractions of the incoming nucleons carried by
the initial state quark and antiquark. The parametrization
was fitted to the typical Drell-Yan lepton pair production
with $4\textmd{GeV} < Q < 12\textmd{GeV}$ and $W/Z$ production ($Q\sim 90\textmd{GeV}$).
By applying the universality argument, these parameterizations
should be able to apply in the SIDIS processes for the associated
quark distribution part.
However, if we extrapolate the above parameterization
down to the typical HERMES/COMPASS kinematics
where $Q^2$ is around $ 3\textmd{GeV}^2$, we can not describe the
transverse momentum distribution of hadron production
in these experiments (see the discussion in Sec. III D). The main reason is that the logarithmic
dependence leads to a strong change around low $Q^2$,
which, however, contradicts with the smooth dependence from
the experimental observation. It will be interesting to check
other forms of non-perturative form factors to see if they can
be extrapolated to HERMES/COMPASS
energy region~\cite{Meng:1995yn,Nadolsky:1999kb}.
We will come back to this issue in a future publication.

\begin{figure}[tbp]
\centering
\includegraphics[width=9cm]{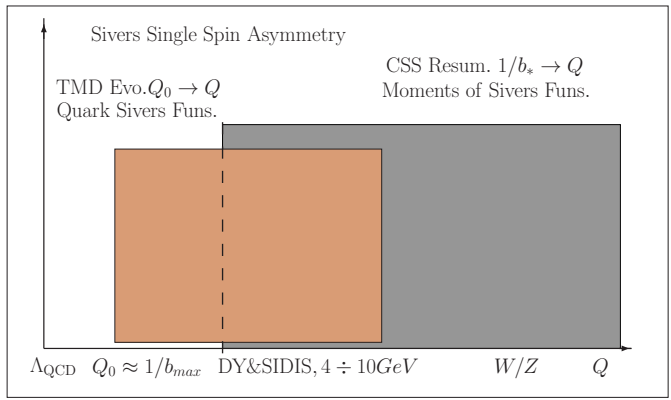}
\caption{Schematic matching for the Sivers single
spin asymmetries in hard processes in the region of $Q$ from 4 to 10 \textmd{GeV}: left, apply
the TMD evolution directly from $Q_0\approx 1/b_{max}$
to $Q$; right, apply CSS resummation with
integral from $b_*$ to $Q$. The connections between
the two evolutions are the TMD Sivers functions
and their transverse-momentum moments. In the overlap
region, both shall yield consistent results for the
asymmetries.}
\end{figure}

Meanwhile, for moderate $Q^2$ variation, there is an alternative
approach to apply, from which we can directly solve the evolution by an integral of
the kernel from low to high $Q$~\cite{Ji:2004wu}.
This is, in particular, useful at relative low $Q$ region, and
can be applied to describe the transverse momentum distributions
in SIDIS from HERMES/COMPASS experiments and fixed target
Drell-Yan lepton pair production experiments~\cite{Sun:2013dya}.
This will also help to build a connection to the ultimate CSS resummation
in Drell-Yan and $W/Z$ production. As illustrated in Fig.~1, in the moderate $Q$ region
(including HERMES/COMPASS kinematics of SIDIS and Drell-Yan process
in fixed target experiments), we apply the evolution by a direct integral of
the kernel from relative low $Q$ to relative high $Q$. In the high $Q$ region which covers
Drell-Yan lepton pair production and $W/Z$ production, we apply the complete
CSS resummation with $b_*$-prescription (following BLNY/KN parameterization
of the non-perturbative form factors). In the overlap region, we shall obtain a
consistent picture for the transverse momentum distribution of the
cross section and the spin asymmetries.

Following this procedure, we will determine the quark Sivers functions
from the HERMES/COMPASS experiments in SIDIS with
$Q^2$-evolution taken into account using direct integral of the kernel.
In particular, we constrain the transverse momentum moments of the
quark Sivers functions, which correspond to the twist-three quark-gluon-quark
correlation functions (so-called Qiu-Sterman matrix elements~\cite{Efremov:1981sh,Qiu:pp}).
These are the bases to evaluate the Sivers single spin asymmetries in the
CSS resummation formalism~\cite{Kang:2011mr}. We then
calculate the Sivers asymmetries in Drell-Yan processes with
the constrained Sivers functions.
The consistent check is carried out by comparing the predictions
between the evolutions done with direct integral of the kernel from
$Q_0$ to $Q$ and that with CSS resummation with integral of the kernel
from $1/b_*$ to $Q$. We notice that in the original BLNY parameterization,
there is a strong $x$-dependence (which is correlated to the $Q^2$-dependence)
in the non-perturbative form factor~\cite{Landry:2002ix}. To avoid this strong dependence, we
follow an updated fit by Konychev and Nadolsky~\cite{Konychev:2005iy}
which describes equally well the Drell-Yan and W/Z boson data with a
mild dependence on $x$. We would like to emphasize that the Sivers
asymmetries observed in HERMES/COMPASS experiments mainly focus
on the moderate $x$-region around 0.1, which is also the typical $x$-range
for the Drell-Yan fixed target experiments~\footnote{Drell-Yan process
at RHIC will be able to probe, for the first time, the wide range of $x$.
This will be important to check the $x$-dependence of the non-perturbative form
factor. Hope this experiment can be carried out soon.}.

The rest of the paper is organized as follows. In Sec.II, we present
a brief review on the theory of low transverse momentum hard processes
as a self-contained introduction. In particular, we will present
the detailed derivations of our previous publication of Ref.~\cite{Kang:2011mr}.
We will also discuss various TMD factorizations.
In Sec.III, we discuss the TMD evolution and resummation
in the context of the transverse momentum dependent differential
cross sections and the Sivers single spin asymmetries. We will
illustrate the incompatibility between the BLNY parameterization
of the CSS resummation and the HERMES/COMPASS measurements
of the $p_\perp$ distribution in SIDIS process. We will also discuss
in detail our approach to calculate the transverse
momentum distribution in this kinematic region, and compare to
the experimental data on multiplicity distribution in SIDIS from HERMES/COMPASS
experiments and Drell-Yan fixed target experiments,
and demonstrate that our approach consistently describe these
data with the TMD evolution taken into account.
In Sec.IV, we extend the evolution effects to the Sivers single
spin asymmetries measured by the HERMES/COMPASS
collaborations, and perform a combined analysis.
In Sec. V, we present the predictions for the Sivers asymmetries in Drell-Yan
lepton pair productions in the planed experiments, and $W^\pm$
production at RHIC. We demonstrate the matching between
two different calculations. With this, we show the results
for the proposed Drell-Yan experiments. We will emphasize
the test of the sign change between SIDIS and Drell-Yan, and
highlight the ability to separate the flavor dependence by combining
Drell-Yan/W measurements with SIDIS results as well. In Sec. VI, we
summarize our paper, and discuss further developments.

\section{Theory Review of Low Transverse Momentum Hard Processes}

In this section, we present a brief review of the theory background for
low transverse momentum hard processes. Under the context of
this paper, the hard processes are hadronic processes with two separate
scales: the invariant mass of virtual photon $Q$ and the transverse momentum
of observed particles $q_\perp$ for lepton pair in Drell-Yan process
or $P_{h\perp}$ for final state hadron in SIDIS.

TMD factorization applies in the kinematic region of low transverse momentum:
$q_\perp\ll Q$. As mentioned in the Introduction, large double logarithms will
arise from perturbative gluon radiation. These large
logs have been demonstrated in the single transverse spin dependent
differential cross sections as well~\cite{Ji:2006ub,Kang:2011mr}. In the following, we will summarize
these calculations, and, in particular, present detailed derivations of
our previous publication~\cite{Kang:2011mr}. We will start with the low transverse momentum
Drell-Yan lepton pair productions for both spin averaged and spin-dependent
cross sections. We then examine the TMD factorization. Finally, we extend the discussions to the SIDIS process.

\subsection{Low Transverse Momentum Drell-Yan}

In the Drell-Yan lepton pair production in $pp$ collisions, we have
\begin{equation}
A (P_A,S_\perp)+B(P_B) \to \gamma^* (q) +X \to \ell^+ + \ell^ -
+X,
\end{equation}
where $P_A$ and $P_B$ represent the momenta of hadrons $A$ and $B$,
and $S_\perp$ for the transverse polarization vector of $A$,
respectively. We further assume hadron $A$ moving in the $+\hat z$
direction. Light-cone momentum $p^\pm$ is defined as $p^\pm=1/\sqrt{2}(p^0\pm p^z)$.
Therefore, $P_A$ is dominated by its plus component, whereas $P_B$ by
its minus component.
The single transverse spin dependent differential cross section can be expressed as
\begin{eqnarray}
\frac{d\Delta\sigma(S_\perp)}{dydQ^2d^2q_\perp}&=&
\sigma_0^{\rm (DY)}\left(W_{UU}(Q;q_\perp)+\epsilon^{\alpha\beta}S_\perp^\alpha W_{UT}^\beta (Q;q_\perp)\right)\ ,
\end{eqnarray}
where $q_\perp$ and $y$ are transverse momentum and rapidity of the lepton pair,
$\sigma_0^{\rm (DY)}=4\pi\alpha_{em}^2/3N_csQ^2$ with $s=(P_A+P_B)^2$,
and $\epsilon^{\alpha\beta}$ is defined
as $\epsilon^{\alpha\beta\mu\nu}P_{A\mu}P_{B\nu}/P_A\cdot P_B$.
When $q_\perp\ll Q$, the structure function $W_{UT}$ can be formulated
in terms of the TMD factorization where the quark Sivers function is
involved~\cite{Brodsky:2002cx,Collins:2002kn},
whereas when $q_\perp\gg \Lambda_{\rm QCD}$
it can be calculated in the collinear factorization approach in terms of the twist-three
quark-gluon-quark correlation functions~\cite{Ji:2006ub,Efremov:1981sh,Qiu:pp}.
It has been shown that the TMD and collinear twist-three approaches
give the consistent results in the intermediate transverse momentum region:
$\Lambda_{\rm QCD}\ll q_\perp\ll Q$~\cite{Ji:2006ub,bbdm}. This consistency allows
us to separate $W_{UT}$ into two terms~\cite{Collins:1984kg},
\begin{eqnarray}
W_{UU}(Q;q_\perp)&=&\int \frac{d^2b}{(2\pi)^{2}} e^{i \vec{q}_\perp\cdot \vec{b}} \widetilde
{W}_{UU}(Q;b) +Y_{UU}^\alpha(Q;q_\perp) \ , \nonumber\\
W_{UT}^\alpha(Q;q_\perp)&=&\int \frac{d^2b}{(2\pi)^{2}} e^{i \vec{q}_\perp\cdot \vec{b}} \widetilde
{W}_{UT}^\alpha(Q;b) +Y_{UT}^\alpha(Q;q_\perp) \ , \nonumber
\end{eqnarray}
where the first term dominates in the $q_\perp\ll Q$ region, while the
second term dominates in the region of $q_\perp\sim Q$ and $q_\perp > Q$. The latter is obtained
by subtracting the the leading term of $q_\perp^2/Q^2$ from
the full perturbative calculation. In this paper, we focus
on the low transverse momentum region, where a TMD factorization is appropriate. We will review how perturbative
corrections modify the differential cross sections, in particular,
from the large logarithms in fixed order calculations.
The results for $\widetilde{W}_{UU,UT}$ up to one-loop
corrections will be shown.

\subsection{Perturbative Contribution in the Small $b_\perp$ Region}

To study the QCD dynamics, in particular, to understand the scale
evolution of the TMDs, it is illustrative to have a perturbative calculation
for the above quantities at small $b_\perp$ limit. It is straightforward to
write down the leading Born diagram contributions to $\widetilde{W}_{UU}$
and $\widetilde{W}_{UT}$,
\begin{eqnarray}
\widetilde{W}_{UU}^{(0)}(Q,b)&=&q(z_1)\bar q(z_2) \ ,\nonumber\\
\widetilde{W}_{UT}^{(0)\alpha}(Q,b)&=&\left(\frac{-ib_\perp^\alpha}{2}\right)T_F(z_1,z_1)\bar q(z_2) \ ,
\end{eqnarray}
where $z_1=Q/\sqrt{s}e^y$, $z_2=Q/\sqrt{s}e^{-y}$, $q(z_1)$ and $\bar q(z_2)$
are the integrated quark and antiquark distribution functions. The single
transverse spin asymmetry comes from the quark Sivers function
$T_F(z_1,z_1)=\int d^2k_\perp k_\perp^2/M f_{1T}^{\perp (DY)}(z_1,k_\perp)$\footnote{Transverse-momentum
moment of the Sivers function defined in~\cite{Bacchetta:2004jz} as $f_{1T}^{\perp(1)}(z_1)$ differs from
$T_F$ by a normalization factor $1/2M$. In this paper, $T_F$ follows the definition of
Ref.~\cite{Ji:2006ub}.}.
The Sivers function $f_{1T}^\perp$ follows the Trento convention~\cite{Bacchetta:2004jz}. Since it
is process dependent, we adopt that in the Drell-Yan process
to calculate the transverse-momentum moment, which is also defined as twist-three
quark-gluon-quark correlation function,
\begin{eqnarray}
T_F(x_1,x_2)& =&
\int\frac{d\xi^-d\eta^-}{4\pi}e^{i(k_{q1}^+\eta^-+k_g^+\xi^-)}
\, \epsilon_\perp^{\beta\alpha}S_{\perp\beta}\nonumber\\
&&\times \left\langle PS|\overline\psi(0){\cal L}(0,\xi^-)\gamma^+
gF_{\alpha}^{\ +}(\xi^-){\cal L}(\xi^-,\eta^-)
\psi(\eta^-)|PS\right\rangle \ ,
\end{eqnarray}
where $x_1=k_{q1}^+/P^+$ and $x_2=k_{q2}^+/P^+$ ,
while $x_g=k_g^+/P^+ = x_2-x_1$,
${\cal L}$ is the light-cone gauge link to make the above definition
gauge invariant.

At one-loop order, the gluon radiation contribution comes from
real and virtual diagrams. The real diagrams have been calculated
in the literature~\cite{Ji:2006ub}, and we can write down the results as~\cite{Ji:2006ub}
\begin{eqnarray}
W_{UU}(Q,q_\perp)|_{q_\perp\ll Q}&=&\frac{\alpha_s}{2\pi^2}C_F\frac{1}{q_\perp^2}\int\frac{dx}{x}\frac{dx'}{x'}
q(x)\bar q(x')\left\{
\left[\left(\frac{1+\xi_1^2}{(1-\xi_1)_+}+\frac{D-2}{2}(1-\xi_2)\right)\delta (1-\xi_1)\right.\right.\nonumber\\
&&\left.\left.+(\xi_1\leftrightarrow \xi_2)\right]
+2\delta(1-\xi_1)\delta(1-\xi_2)\ln\frac{Q^2}{q_\perp^2}\right\}\\
W_{UT}^\alpha(Q,q_\perp)|_{q_\perp\ll Q}&=&\frac{\alpha_s}{2\pi^2}\frac{q_\perp^\alpha}{(q_\perp^2)^2}\int\frac{dx}{x}\frac{dx'}{x'}
\bar q(x')\left\{T_F(x,x)\delta(1-\xi_1)\left(\frac{1+\xi_2^2}{(1-\xi_2)_+}+\frac{D-2}{2}(1-\xi_2)\right)\right.\nonumber\\
&&\left.+2C_F T_F(x,x)\delta(1-\xi_1)\delta(1-\xi_2)\ln\frac{Q^2}{q_\perp^2}+\delta(1-\xi_2)\frac{C_A}{2}
T_F(x,z_1)\frac{1+\xi_1}{(1-\xi_1)_+}\right.\nonumber\\
&&\left.+\delta(1-\xi_2)\frac{1}{2N_C}\left[ \left(x\frac{\partial}{\partial
x}T_F(x,x)\right)(1+\xi_1^2)+T_F(x,x)\frac{(1-\xi_1)^2(2\xi_1+1)-2}{(1-\xi_1)_+}\right]\right.\nonumber\\
&&\left.+\delta(1-\xi_2)\frac{1}{2N_C} T_F(x,x)\frac{D-2}{2}(1-\xi_1)\right\} \ , \label{wutp}
\end{eqnarray}
where $\xi_1=z_1/x$ and $\xi_2=z_2/x'$, and
we have kept the $\epsilon=(2-D)/2$ (with $D$ represents the
transverse dimension in the dimension regulation) term in the above calculations. After
Fourier transformation into the impact parameter space, this will lead to
a finite contribution. In the above results, we only keep the soft
and hard gluon pole contributions in the $q\bar q$ channel
for the single spin asymmetry calculations. All other contributions
can be formulated similarly.

Applying the Fourier transform formulas we listed in the Appendix, we
obtain the following result for the real gluon radiation contribution to
$\widetilde{W}_{UU}(Q;b)$ at one-loop order,
\begin{eqnarray}
\widetilde{W}_{UU}^{(r )}(b)&=& \frac{\alpha_s}{2\pi}C_F\left\{\left[-\frac{1}{\epsilon}+\ln\frac{c_0^2}{b_\perp^2\mu^2}\right]
\left[\frac{1+\xi_1^2}{(1-\xi_1)_+}\delta(1-\xi_2)+\frac{1+\xi_2^2}{(1-\xi_2)_+}\delta(1-\xi_1)\right]\right.\nonumber\\
&&+2\delta(1-\xi_1)\delta(1-\xi_2)\left[\frac{1}{\epsilon^2}-\frac{1}{\epsilon}\ln\frac{Q^2}{\mu^2}
+\frac{1}{2}\left(\ln\frac{Q^2}{\mu^2}\right)^2-\frac{1}{2}\left(\ln\frac{Q^2b_\perp^2}{c_0^2}\right)^2-\frac{\pi^2}{12}\right]\nonumber\\
&&\left.+(1-\xi_1)\delta(1-\xi_2)+(1-\xi_2)\delta(1-\xi_1)\right\} \ ,
\end{eqnarray}
where $c_0=2e^{-\gamma_E}$ and a common integral $\int dxdx'/xx'$ as that in Eq.~(\ref{wutp}) has been omitted
for simplicity. To arrive the above result, we have applied the $\overline{\rm MS}$
subtraction scheme with $\mu^2\rightarrow 4\pi e^{-\gamma_E}\mu^2$. This
is different from the $\overline{\rm MS}$ used in the Collins book~\cite{Collins} (see the discussions below).

The above result contains collinear and soft divergences. The soft divergences shall be
cancelled by the virtual diagrams, whereas the collinear divergences absorbed
into the renormalization of the parton distributions. The virtual diagrams contributes
\begin{eqnarray}
\widetilde{W}_{UU}^{(v)}=\frac{\alpha_s}{2\pi}\left[-\frac{2}{\epsilon^2}-\frac{3}{\epsilon}+\frac{2}{\epsilon}\ln\frac{Q^2}{\mu^2}
+\frac{7}{6}\pi^2+3\ln\frac{Q^2}{\mu^2}-\left(\ln\frac{Q^2}{\mu^2}\right)^2-8\right] \delta(1-\xi_1)\delta(1-\xi_2)\ .\label{wuuv}
\end{eqnarray}
Adding them together, we will have
\begin{eqnarray}
\widetilde{W}_{UU}(Q,b)&=&\frac{\alpha_s}{2\pi}C_F\left\{\left(-\frac{1}{\epsilon}+\ln\frac{c_0^2}{b^2\mu^2}\right)
\left({\cal P}_{q\to q}(\xi_1)\delta(1-\xi_2)+{\cal P}_{q\to q}(\xi_2)\delta(1-\xi_1)\right)\right.\nonumber\\
&&+(1-\xi_1)\delta(1-\xi_2)+(1-\xi_2)\delta(1-\xi_1)\nonumber\\
&&\left.+\delta(1-\xi_1)\delta(1-\xi_2)\left[3\ln\frac{Q^2b^2}{c_0^2}-\left(\ln\frac{Q^2b^2}{c_0^2}\right)^2+\pi^2-8\right]\right\} \ ,\label{euu}
\end{eqnarray}
where ${\cal P}_{q\to q}(\xi)=\left(\frac{1+\xi^2}{1-\xi}\right)_+$ is the
quark splitting kernel.

Similarly, for the single-spin dependent cross section, we have for the real diagram
contributions,
\begin{eqnarray}
\widetilde{W}_{UT}^{\alpha(r )}&=& \frac{\alpha_s}{2\pi}\left(\frac{-ib_\perp^\alpha}{2}\right)\bar q(x')
\left\{\left(-\frac{1}{\epsilon}+\ln\frac{c_0^2}{b_\perp^2\mu^2}\right)
\left[\frac{1+\xi_2^2}{(1-\xi_2)_+}T_F(x,x) \delta(1-\xi_1)+\delta(1-\xi_2)\right.\right.\nonumber\\
&&\left.\times \left(\frac{C_A}{2}T_F(x,z_1)\frac{1+\xi_1}{(1-\xi_1)_+}+\frac{1}{2N_C}T_F(x,x)\left(\frac{-1-\xi_1^2}{(1-\xi_1)_+}
-2\delta(1-\xi_1)\right)\right)\right]\nonumber\\
&&+2C_FT_F(x,x)\delta(1-\xi_1)\delta(1-\xi_2)\left[\frac{1}{\epsilon^2}-\frac{1}{\epsilon}\ln\frac{Q^2}{\mu^2}
+\frac{1}{2}\left(\ln\frac{Q^2}{\mu^2}\right)^2-\frac{1}{2}\left(\ln\frac{Q^2b_\perp^2}{c_0^2}\right)^2-\frac{\pi^2}{12}\right]\nonumber\\
&&-2C_FT_F(x,x)\delta(1-\xi_1)\delta(1-\xi_2)\left(-\frac{1}{\epsilon}+\ln\frac{c_0^2}{b^2\mu^2}\right)\nonumber\\
&&\left.+\left(-\frac{1}{2N_c}\right)T_F(x,x)(1-\xi_1)\delta(1-\xi_2)+C_FT_F(x,x)(1-\xi_2)\delta(1-\xi_1)\right\} \ ,
\end{eqnarray}
where we have simplified the expression by integrating out the partial derivative in Eq.~(\ref{wutp}).
The last second line comes from the second term in the Fourier transform
formula of Eq.~(\ref{ftlog}) in the Appendix~\footnote{This term accounts
for the partial difference between previous calculations of Refs.~\cite{Vogelsang:2009pj,Kang:2008ey,Zhou:2008mz}
and Ref.~\cite{Braun:2009mi} on the splitting kernel for $T_F(x,x)$. In particular, after adding
a similar contribution in the calculation of Ref.~\cite{Vogelsang:2009pj}, it can be shown that
the derivation in Ref.~\cite{Vogelsang:2009pj} agrees with that in Ref.~\cite{Braun:2009mi}.}.
The virtual contribution has the similar form as Eq.~(\ref{wuuv}). After adding them together,
we find that the total contribution at one-loop order,
\begin{eqnarray}
W_{UT}&=&\frac{\alpha_s}{2\pi}\left(\frac{-ib_\perp^\alpha}{2}\right)\left\{\left(-\frac{1}{\epsilon}+\ln\frac{c_0^2}{b^2\mu^2}\right)
\left({\cal P}_{qg\to qg}^T\otimes T_F(z_1)\delta(1-\xi_2)+C_F{\cal P}_{q\to q}(\xi_2)T_F(z_1,z_1)\delta(1-\xi_1)\right)\right.\nonumber\\
&&+T_F(x,x)\left[\left(-\frac{1}{2N_c}\right)(1-\xi_1)\delta(1-\xi_2)+C_F(1-\xi_2)\delta(1-\xi_1)\right]\nonumber\\
&&\left.+\delta(1-\xi_1)\delta(1-\xi_2)T_F(z_1,z_1)C_F\left[3\ln\frac{Q^2b^2}{c_0^2}-\left(\ln\frac{Q^2b^2}{c_0^2}\right)^2+\pi^2-8\right]\right\} \ ,\label{eut}
\end{eqnarray}
where ${\cal P}_{q\to q}(\xi)$ is the same as above, and the splitting kernel for the
Sivers function can be written as
\begin{eqnarray}
{\cal P}_{qg\to qg}^T\otimes T_F(z)&=&\int\frac{dx}{x} \left\{T_F(x,x)\left[C_F\left(\frac{1+\xi^2}{1-\xi}\right)_+-C_A\delta(1-\xi)\right]\right.\nonumber\\
&&\left.+\frac{C_A}{2}\left(T_F(x,z)\frac{1+\xi}{1-\xi}-T_F(x,x)\frac{1+\xi^2}{1-\xi}\right)\right\}  \ ,
\end{eqnarray}
which agrees with recent calculations for the splitting kernel for the part involved in the
above calculations~\cite{Braun:2009mi,Schafer:2012ra,Kang:2012em,Ma:2012xn}.

In the above results, the one-loop corrections Eqs.~(\ref{euu},\ref{eut}) clearly demonstrate
large logarithms. To resum these large logs, we need to apply
the TMD factorization, and solve the relevant evolution equation. Although the different TMD
schemes have been used in the literature, the final evolution for the structure functions
$\widetilde{W}_{UU,UT}$ remain the same.
First, we examine the TMD factorization for the perturbative calculations at one-loop order.

\subsection{Sivers Quark Distributions and TMD factorization}

To demonstrate the factorization, we calculate the TMD
quark and antiquark distributions, and show that the collinear part of the structure
functions calculated in the last subsection can be absorbed into these TMD
distributions. It has been known that,
however, there is scheme-dependence in the TMD definitions. Therefore, the hard
factors will also depend on which scheme you choose to calculate the TMDs.
The scheme dependence comes from the fact that the TMD distributions have
light-cone singularity, and different ways to regulate this singularity define different
schemes of the TMD distributions.

\subsubsection{Ji-Ma-Yuan Scheme}

In the Ji-Ma-Yuan scheme, the light-cone gauge link in the TMD definition
is chosen to be slightly off-light-cone, $n=(1^-,0^+,0_\perp)\to v=(v^-,v^+,0_\perp)$
with $v^-\gg v^+$. Similarly, for the TMD for the antiquark distribution, $\tilde v$
was introduced, $\tilde v=(\tilde v^-,\tilde v^+,0_\perp)$ with $\tilde v^+\gg \tilde v^-$.
Because of the additional $v$ and $\tilde v$, there are additional invariants:
$\zeta_1^2=(2v\cdot P_A)^2/v^2$, $\zeta_2^2=(2\tilde{v}\cdot P_B)^2/\tilde{v}^2$,
and $\rho^2=(2v\cdot \tilde v)^2/v^2\tilde{v}^2$.
The TMD quark distributions of a polarized proton is defined
through the following matrix:
\begin{eqnarray}
    {\cal M}^{\alpha\beta} &=&   P^+\int
        \frac{d\xi^-}{2\pi}e^{-ix\xi^-P^+}\int
        \frac{d^2b_\perp}{(2\pi)^2} e^{i\vec{b}_\perp\cdot
        \vec{k}_\perp} \nonumber\\
        &&\times \left\langle
PS\left|\overline\psi^\beta(\xi^-,0,\vec{b}_\perp){\cal L}_{v}^\dagger(-\infty;\xi){\cal L}_{v}(-\infty;0)
        \psi^\alpha(0)\right|PS\right\rangle\ ,
\end{eqnarray}
with the gauge link
\begin{equation}
 {\cal L}_{v}(-\infty;\xi) \equiv \exp\left(-ig\int^{-\infty}_0 d\lambda
\, v\cdot A(\lambda v +\xi)\right) \ .\label{glink}
\end{equation}
This gauge link goes to $-\infty$, indicating that we adopt the
definition for the TMD quark distributions for the Drell-Yan
process. Keeping only the unpolarized
quark distribution and the Sivers function, we have the following
expansion for the matrix ${\cal M}$:
\begin{equation}
{\cal M} = \frac{1}{2}
\left[q(x,k_\perp)\gamma_\mu P^\mu
    + \frac{1}{M}\, f_{1T}^\perp(x,k_\perp)\,
\epsilon_{\mu\nu\alpha\beta}\gamma^\mu P^\nu k^\alpha S^\beta +
\ldots\right] \label{matrixexp}
\end{equation}
where $q(x,k_\perp)$ is the TMD distribution in an unpolarized
proton, $f_{1T}^\perp(x, k_\perp)$ is the Sivers function, and $M$ is a
hadron mass, used to normalize $q(x,k_\perp)$ and $f_{1T}^\perp(x,k_\perp)$ to
the same mass dimension.

First, the soft factor has been calculated,
\begin{equation}
S(b_\perp)=\frac{\alpha_s}{2\pi}C_F\ln\frac{b^2\mu^2}{c_0^2}\left(2-\ln\rho^2\right) \ .
\end{equation}
The calculations of the TMDs in the Ji-Ma-Yuan is straightforward, and we find that the
quark distribution can be written as,
\begin{equation}
q(z,k_\perp)|_{real}=\frac{\alpha_s}{2\pi^2}\frac{1}{k_\perp^2}C_F
\int \frac{dx}{x}q(x)\left\{\frac{1+\xi^2}{(1-\xi)_+}+\frac{D-2}{2}(1-\xi)+\delta(1-\xi)\left(\ln\frac{z^2\zeta^2}{k_\perp^2}-1\right)\right\} \ ,
\end{equation}
where $\xi=z/x$ and in the impact parameter space,
\begin{eqnarray}
q(z,b_\perp)|_{real}&=&\frac{\alpha_s}{2\pi}C_F\left\{\left(-\frac{1}{\epsilon}
+\ln\frac{c_0^2}{b^2\mu^2}\right)\left[\frac{1+\xi^2}{(1-\xi_1)_+}-\delta(1-\xi)\right]+(1-\xi)\right.\nonumber\\
&&\left.+\delta(1-\xi)\left[\frac{1}{\epsilon^2}-\frac{1}{\epsilon}\ln\frac{z^2\zeta^2}{\mu^2}
+\frac{1}{2}\left(\ln\frac{z^2\zeta^2}{\mu^2}\right)^2-\frac{1}{2}\left(\ln\frac{z^2\zeta^2b_\perp^2}{c_0^2}\right)^2-\frac{\pi^2}{12}\right]\right\} \ .
\end{eqnarray}
The virtual diagram contributes,
\begin{eqnarray}
q(z,b_\perp)|_{vir}&=&\frac{\alpha_s}{2\pi}\delta(1-\xi)
\left[-\frac{1}{\epsilon^2}-\frac{5}{2\epsilon}+\frac{1}{\epsilon}\ln\frac{\zeta^2}{\mu^2}
+\ln\frac{\zeta^2}{\mu^2}-\frac{1}{2}\left(\ln\frac{z^2\zeta^2}{\mu^2}\right)^2-\frac{5}{12}\pi^2-2\right]  \ .\label{tmd-v}
\end{eqnarray}
Adding them together, we have
\begin{eqnarray}
q(z,b_\perp)&=&\frac{\alpha_s}{2\pi}C_F\left\{\left(-\frac{1}{\epsilon}+\ln\frac{c_0^2}{b^2\bar\mu^2}\right){\cal P}_{q\to q}(\xi)
-\delta(1-\xi)\ln\frac{c_0^2}{b^2\mu^2}+(1-\xi)\right.\nonumber\\
&&\left.+\delta(1-\xi)\left[\frac{3}{2}\ln\frac{b^2\mu^2}{c_0^2}+\ln\frac{z^2\zeta^2}{\mu^2}
-\frac{1}{2}\left(\ln\frac{z^2\zeta^2b_\perp^2}{c_0^2}\right)^2-2-\frac{\pi^2}{2}\right]\right\} \ .
\end{eqnarray}
Similar expression can be written for the antiquark distribution.
According to the TMD factorization, we can subtract the quark distribution, antiquark distribution
and the soft factor, and obtain the hard factor,
\begin{eqnarray}
W_{UU}(Q;b)=q(z_1,b,\zeta_1)\bar q(z_2,b,\zeta_2)H_{UU}(Q)\left(S(b,\rho)\right)^{-1} \ .
\end{eqnarray}
Applying these results, we have the following result for the hard factor,
\begin{equation}
H_{UU}(Q)=\frac{\alpha_s}{2\pi}C_F\left[\ln\frac{Q^2}{\mu^2}+\ln\rho^2\ln\frac{Q^2}{\mu^2}-\ln\rho^2+\ln^2\rho+2\pi^2-4\right] \ ,
\end{equation}
where $z_1^2z_2^2\zeta_1^2\zeta_2^2=\rho^2Q^4$ has been used to simplify the hard factor.

We can follow the same procedure to calculate that for the Sivers asymmetry. The TMD
quark Sivers function can be written as,
\begin{eqnarray}
f_{1T}^\perp(z,k_\perp)&=&\frac{\alpha_s}{2\pi^2}\frac{M}{(k_\perp^2)^2}
\int\frac{dx}{x}\left\{\frac{C_A}{2}T_F(x,z)\frac{1+\xi}{(1-\xi)_+}+T_F(x,x)\frac{-1}{2N_c}\frac{D-2}{2}(1-\xi)\right.\nonumber\\
&&+\frac{1}{2N_C}\left[ \left(x\frac{\partial}{\partial
x}T_F(x,x)\right)(1+\xi^2)+T_F(x,x)\frac{(1-\xi)^2(2\xi+1)-2}{(1-\xi)_+}\right]\nonumber\\
&&\left.
+T_F(x,x)\delta(1-\xi)C_F\left(\ln\frac{x^2\zeta^2}{k_\perp^2}-2\right)\right\} \ .
\end{eqnarray}
We note a factor of (-2) in the last term, which is different from that in Ref.~\cite{Ji:2006ub}. This
comes from a sub-leading expansion contribution from the soft-pole and hard-pole diagrams,
which was omitted in Ref.~\cite{Ji:2006ub}. This term will contribute to the collinear singularity
when Fourier transforming into the impact parameter space.
Adding the virtual diagram contributions, we will have total result in the impact parameter space,
\begin{eqnarray}
\tilde{f}_{1T}^\alpha(z,b)&=&\frac{\alpha_s}{2\pi}\left(\frac{-ib_\perp^\alpha}{2}\right)\left\{
\left(-\frac{1}{\epsilon}+\ln\frac{c_0^2}{b^2\mu^2}\right){\cal P}_{qg\to qg}^T\otimes T_F(z)\right.\nonumber\\
&&-\delta(1-\xi)T_F(x,x)C_F\ln\frac{c_0^2}{b^2\mu^2}-\frac{1}{2N_c}T_F(x,x)(1-\xi)\nonumber\\
&&\left.+\delta(1-\xi)T_F(x,x)C_F\left[\frac{3}{2}\ln\frac{b^2\mu^2}{c_0^2}+\ln\frac{z^2\zeta^2}{\mu^2}
-\frac{1}{2}\left(\ln\frac{z^2\zeta^2b_\perp^2}{c_0^2}\right)^2-2-\frac{\pi^2}{2}\right]\right\} \  ,
\end{eqnarray}
at one-loop order.
By subtraction, we obtain the hard factor for the Sivers single spin asymmetry in Drell-Yan
process,
\begin{equation}
H_{UT}(Q)=H_{UU}(Q)=\frac{\alpha_s}{2\pi}C_F\left[\ln\frac{Q^2}{\mu^2}+\ln\rho^2\ln\frac{Q^2}{\mu^2}-\ln\rho^2+\ln^2\rho+2\pi^2-4\right] \ .
\end{equation}
This is an important result, as it shows that the hard factor is spin-independent.

\subsubsection{Collins-11}

In 2011, Collins introduces a new definition for the TMDs, where the soft
gluon and light-cone singularities are subtracted in the TMDs from the beginning.
As a result, there is no soft factor in the factorization formula, which is absorbed into the definition of PDF.

From its definition, we find that the real diagram contribution can be written as~\cite{Collins}
\begin{equation}
q_{\rm JCC}(z,k_\perp)|_{real}=\frac{\alpha_s}{2\pi^2}\frac{1}{k_\perp^2}C_F
\int \frac{dx}{x}q(x)\left\{\frac{1+\xi^2}{(1-\xi)_+}+\frac{D-2}{2}(1-\xi)+\delta(1-\xi)\left(\ln\frac{\zeta_c^2}{k_\perp^2}\right)\right\} \ ,
\end{equation}
where $\zeta_c$ is defined as $\zeta_c^2=(z_1P_A^+)^2e^{-2y_n}$ with $y_n$
the rapidity cutoff to regulate the light-cone singularity.
The virtual diagram for $q_{\rm JCC}$ only contributes to the counter terms,
\begin{eqnarray}
q_{\rm JCC}(z,b_\perp)|_{vir}&=&\frac{\alpha_s}{2\pi}\delta(1-\xi)
\left[-\frac{1}{\epsilon^2}-\frac{3}{2\epsilon}+\frac{1}{\epsilon}\ln\frac{\zeta_c^2}{\mu^2}
\right]  \ ,
\end{eqnarray}
where, to be consistent, we have followed the $S_\epsilon=(4\pi)^\epsilon/\Gamma(1-\epsilon)$ prescription of $\overline{\rm MS}$ subtraction
of Ref.~\cite{Collins}.

Therefore, the total quark distribution can be written as,
\begin{eqnarray}
q_{\rm JCC}(z,b_\perp)&=&\frac{\alpha_s}{2\pi}C_F\left\{\left(-\frac{1}{\epsilon}+\ln\frac{c_0^2}{b^2\bar\mu^2}\right){\cal P}_{q\to q}(\xi)
+(1-\xi)\right.\nonumber\\
&&\left.+\delta(1-\xi)\left[\frac{3}{2}\ln\frac{b^2\mu^2}{c_0^2}
+\frac{1}{2}\left(\ln\frac{\zeta_c^2}{\mu^2}\right)^2-\frac{1}{2}\left(\ln\frac{\zeta_c^2b_\perp^2}{c_0^2}\right)^2
\right]\right\} \ .
\end{eqnarray}
We notice that an additional term of $(-\frac{\pi^2}{12})$ shall be added to the above
equation if we use the $\overline{\rm MS}$ subtraction method of the last subsection,
see, also the detailed discussions in Ref.~\cite{Collins:2012uy}.
To calculate the hard factor in this scheme, we apply the factorization
\begin{eqnarray}
W_{UU}(Q;b)=q_{\rm JCC}(z_1,b,\zeta_c)\bar q_{\rm JCC}(z_2,b,\zeta_{\bar{c}})H_{UU}^{(\rm JCC)}(Q) \ .
\end{eqnarray}
The hard factor can be calculated~\cite{Collins},
\begin{equation}
H_{UU}^{(\rm JCC)}(Q)=\frac{\alpha_s}{2\pi}C_F\left[3\ln\frac{Q^2}{\mu^2}+\ln^2\frac{Q^2}{\mu^2}+\frac{1}{2}\pi^2-8\right] \ ,
\end{equation}
We notice that the different $\overline{\rm MS}$
subtraction method will lead to different hard factors in Collins-11 scheme for the TMD
definition~\cite{Collins:2012uy}. In particular, the $\overline{\rm MS}$ subtraction used
in the last sub-section will add additional term of $\pi^2/6$ in the above hard factor.
This is because the Collins-11 definition of the TMD distribution, there is a double pole
$1/\epsilon^2$ in the UV divergence in the dimensional regulation for the virtual diagram~\cite{Collins:2012uy}.

For the Sivers function, the calculations can follow similarly.
The real diagram contribution for the quark Sivers function can be
calculated in the Collins-11 definition,
\begin{eqnarray}
f_{1T}^{({\rm JCC} )\perp}(z,k_\perp)&=&\frac{\alpha_s}{2\pi^2}\frac{M}{(k_\perp^2)^2}
\int\frac{dx}{x}\left\{\frac{C_A}{2}T_F(x,z)\frac{1+\xi}{(1-\xi)_+}+T_F(x,x)\frac{-1}{2N_c}\frac{D-2}{2}(1-\xi)\right.\nonumber\\
&&+\frac{1}{2N_C}\left[ \left(x\frac{\partial}{\partial
x}T_F(x,x)\right)(1+\xi^2)+T_F(x,x)\frac{(1-\xi)^2(2\xi+1)-2}{(1-\xi)_+}\right]\nonumber\\
&&\left.
+T_F(x,x)\delta(1-\xi)C_F\left(\ln\frac{\zeta_c^2}{k_\perp^2}-1\right)\right\} \ .
\end{eqnarray}
Virtual diagram is the same as the unpolarized case, and the total quark
Sivers function in $b$-space,
\begin{eqnarray}
\tilde{f}_{1T}^{({\rm JCC} )\alpha}(z,b)&=&\frac{\alpha_s}{2\pi}\left(\frac{-b_\perp^\alpha}{2}\right)\left\{
\left(-\frac{1}{\epsilon}+\ln\frac{c_0^2}{b^2\mu^2}\right){\cal P}_{qg\to qg}^T\otimes T_F(z)-\frac{1}{2N_c}T_F(x,x)(1-\xi)\right.\nonumber\\
&&\left.+\delta(1-\xi)C_F\left[\frac{3}{2}\ln\frac{b^2\mu^2}{c_0^2}
+\frac{1}{2}\left(\ln\frac{\zeta_c^2}{\mu^2}\right)^2-\frac{1}{2}\left(\ln\frac{\zeta_c^2b_\perp^2}{c_0^2}\right)^2
\right]\right\} \ ,
\end{eqnarray}
where, again, we have followed the $S_\epsilon$ prescription for $\overline{\rm MS}$ subtraction in
Collins-11 definition of the TMDs.
Again, we find that the hard factor can be calculated
\begin{equation}
H_{UT}^{(c )}(Q)=H_{UU}^{(c )}(Q)=\frac{\alpha_s}{2\pi}C_F\left[3\ln\frac{Q^2}{\mu^2}+\ln^2\frac{Q^2}{\mu^2}+\frac{1}{2}\pi^2-8\right] \ .
\end{equation}
Again, if we choose the $\overline{\rm MS}$ subtraction method of the last
sub-section, we would add additional term of $\pi^2/6$ to the
above hard factor.

\subsubsection{Echevarria-Idilbi-Scimemi (EIS)}

In a recent publication by Collins and Rogers~\cite{Collins:2012uy}, it has been shown that
EIS version~\cite{GarciaEchevarria:2011rb} of the soft-collinear-effective-theory
approach for the TMD quark distribution is equivalent to that of the Collins-11 approach.
Therefore, the calculations in the previous subsection can be carried out similarly for EIS TMD
quark distributions. We omit the details of this calculation.

\subsection{Semi-inclusive DIS}

In this subsection, we briefly review the calculations for the semi-inclusive
hadron production in deep inelastic scattering. Much of the results presented
above can be followed. For SIDIS, we have,
\begin{equation}
e (\ell)+p(P)\to e(\ell') + h
(P_h) + X\ ,
\end{equation}
which proceeds through exchange of a virtual photon
with momentum $q_\mu=\ell_\mu-\ell'_\mu$ and invariant mass
$Q^2=-q^2$.  When $P_{h\perp}\ll Q$, the TMD
factorization applies, according which the
differential  SIDIS cross section may be written as
\begin{eqnarray}
    \frac{d\sigma(S_\perp)}{dx_Bdydz_hd^2\vec {P}_{h\perp}}
      &=& \sigma_0^{\rm (DIS)}\times\left[F_{UU}+\epsilon^{\alpha\beta}S_\perp^\alpha
 F_{\rm sivers}^{\beta}
      \right] \ ,
\end{eqnarray}
where $\sigma_0^{\rm (DIS)}=4\pi\alpha^2_{\rm em}S_{ep}/{Q^4}\times (1-y+y^2/2)
x_B$ with usual DIS kinematic variables $y$, $x_B$ and $Q^2$, $z_h=P_h\cdot P/q\cdot P$,
$P_{h\perp}$ the transverse momentum of the final state hadron respect to the
lepton plane,
and where $\phi_{S}$ and $\phi_{h}$ are the azimuthal angles
of the proton's transverse polarization vector and the transverse
momentum vector of the final-state hadron, respectively. We
only keep the terms we are interested in: $F_{UU}$ corresponds to
the unpolarized cross section, and $F_{\rm sivers}$ to the Sivers
function contribution to the single-transverse-spin asymmetry.
$F_{UU}$ and $F_{\rm sivers}$ depend on the kinematical variables,
$x_B$, $z_h$, $Q^2$, $y$, and $P_{h\perp}$. Similar to
that in the Drell-Yan process, at low transverse momentum ($P_{h\perp}\ll Q$)
the structure functions can be formulated
in terms of the TMD factorization, and they can be
written into two terms,
\begin{eqnarray}
F_{UU}(Q;q_\perp)&=&\int \frac{d^2b}{(2\pi)^{2}} e^{i \vec{q}_\perp\cdot \vec{b}} \widetilde
{F}_{UU}(Q;b) +Y_{UU}(Q;q_\perp) \ , \\
F_{UT}^\alpha(Q;q_\perp)&=&\int \frac{d^2b}{(2\pi)^{2}} e^{i \vec{q}_\perp\cdot \vec{b}} \widetilde
{F}_{UT}^\alpha(Q;b) +Y_{UT}^\alpha(Q;q_\perp) \ ,
\end{eqnarray}
where the first term dominates in $P_{h\perp}\ll Q$ region, and the
second term dominates in the region of $P_{h\perp}\sim Q$ and $P_{h\perp}> Q$. Again, the latter is obtained
by subtracting the leading term of $P_{h\perp}^2/Q^2$ from
the full perturbative calculation.

One perturbative gluon radiation contributes to finite $k_\perp$
for the differential cross section,
\begin{eqnarray}
   F_{UU}|_{P_{h\perp}\ll Q}&=& \frac{\alpha_s}{2\pi^2}\frac{1}{\vec{P}_{h\perp}^2}
C_F\int \frac{dxdz}{xz}q(x)D
      (z)\left\{\left(\frac{1+\xi^2}{(1-\xi)_+}+\frac{D-2}{2}(1-\xi)\right)\delta(\hat\xi-1)\nonumber\right.\\
      &&\left.+\left(\frac{1+\hat\xi^2}{(1-\hat\xi)_+}+\frac{D-2}{2}(1-\hat\xi)\right)\delta(\xi-1)+2
      \delta(\xi-1)\delta(\hat\xi-1)\ln\frac{z_h^2Q^2}{\vec{P}_{h\perp}^2}
\right\}\
      , \label{unx}
\end{eqnarray}
where  $\xi=x_B/x$ and $\hat\xi=z_h/z$,  $q(x)$ represents
the integrated quark distribution, $D(z)$ the fragmentation
function. Similarly, for the single-transverse-spin dependent cross section,
we have
\begin{eqnarray}
  F_{UT}^\beta |_{P_{h\perp}\ll Q}
      &=& -
      \frac{z_hP_{h\perp}^\beta}{(\vec{P}_{h\perp}^2)^2}
      \frac{\alpha_s}{2\pi^2}\int \frac{dxdz}{xz}D
      (z) \left\{C_F T_F(x,x) \delta(\xi-1)\left(\frac{1+\hat\xi^2}{(1-\hat\xi)_+}
      +\frac{D-2}{2}(1-\hat\xi)\right)\right.\nonumber\\
&&+  \delta(\hat \xi-1)
      \frac{1}{2N_C} \left[ \left(x\frac{\partial}{\partial x}T_F(x,x)\right)(1+\xi^2)
      +T_F(x,x)\frac{(1-\xi)^2(2\xi+1)-2}{(1-\xi)_+}\right]\nonumber\\
 &&+ \delta(\hat \xi-1)\left[\frac{C_A}{2} T_F(x,x-\widehat{x}_g)\frac{1+\xi}{(1-\xi)_+}
 +\left(-\frac{1}{2N_c}\right)T_F(x,x)\frac{D-2}{2}(1-\xi)\right]\nonumber\\
&&\left.+2\delta(\hat\xi-1)\delta(\xi-1)C_FT_F(x,x)\ln\frac{z_h^2Q^2}{\vec{P}_{h\perp}^2}
\right\} \ ,
\end{eqnarray}
where $\hat x_g=(1-\xi)x=x-x_B$.

By applying the Fourier transform (some of the useful integrals are
listed in the Appendix), we obtain the following result
for $\widetilde{F}_{UU}(Q,b)$ and $\widetilde{F}_{UT}^\beta(Q,b)$,
\begin{eqnarray}
\widetilde{F}_{UU}|_{real}&=&  \frac{\alpha_s}{2\pi}C_F\frac{1}{\hat\xi}\left\{\left[-\frac{1}{\epsilon}+\ln\frac{c_0^2}{b_\perp^2\mu^2}\right]
\left[\frac{1+\xi^2}{(1-\xi)_+}\delta(1-\hat\xi)+\frac{1+\hat\xi^2}{(1-\hat\xi)_+}\delta(1-\xi)\right]\right.\nonumber\\
&&+2\delta(1-\xi)\delta(1-\hat\xi)\left[\frac{1}{\epsilon^2}-\frac{1}{\epsilon}\ln\frac{Q^2}{\mu^2}
+\frac{1}{2}\left(\ln\frac{Q^2}{\mu^2}\right)^2-\frac{1}{2}\left(\ln\frac{Q^2b_\perp^2}{c_0^2}\right)^2-\frac{\pi^2}{12}\right]\nonumber\\
&&\left.+(1-\xi)\delta(1-\hat\xi)+(1-\hat\xi)\delta(1-\xi)\right\} \ ,\nonumber\\
\widetilde{F}_{UT}^\beta|_{real}&=& \frac{\alpha_s}{2\pi}\frac{1}{\hat\xi}\left(\frac{ib_\perp^\alpha}{2}\right)D(z)
\left\{\left(-\frac{1}{\epsilon}+\ln\frac{c_0^2}{b_\perp^2\mu^2}\right)
\left[\frac{1+\hat\xi^2}{(1-\hat\xi)_+}T_F(x,x) \delta(1-\xi)+\delta(1-\hat\xi)\right.\right.\nonumber\\
&&\left.\times \left(\frac{C_A}{2}T_F(x,z_1)\frac{1+\xi}{(1-\xi)_+}+\frac{1}{2N_C}T_F(x,x)\left(\frac{-1-\xi^2}{(1-\xi)_+}
-2\delta(1-\xi)\right)\right)\right]\nonumber\\
&&+2C_FT_F(x,x)\delta(1-\xi)\delta(1-\hat\xi)\left[\frac{1}{\epsilon^2}-\frac{1}{\epsilon}\ln\frac{Q^2}{\mu^2}
+\frac{1}{2}\left(\ln\frac{Q^2}{\mu^2}\right)^2-\frac{1}{2}\left(\ln\frac{Q^2b_\perp^2}{c_0^2}\right)^2-\frac{\pi^2}{12}\right]\nonumber\\
&&-2C_FT_F(x,x)\delta(1-\xi)\delta(1-\hat\xi)\left(-\frac{1}{\epsilon}+\ln\frac{c_0^2}{b^2\mu^2}\right)\nonumber\\
&&\left.+\left(-\frac{1}{2N_c}\right)T_F(x,x)(1-\xi)\delta(1-\hat\xi)+C_FT_F(x,x)(1-\hat\xi)\delta(1-\xi)\right\} \ ,
\end{eqnarray}
Clearly, the real diagrams contributions contain soft divergence,
which will be cancelled by the virtual diagrams contributions.
The virtual diagram contributes to a factor,
\begin{equation}
\frac{\alpha_s}{2\pi}\left[-\frac{2}{\epsilon^2}-\frac{3}{\epsilon}+\frac{2}{\epsilon}\ln\frac{Q^2}{\mu^2}
+\frac{1}{6}\pi^2+3\ln\frac{Q^2}{\mu^2}-\left(\ln\frac{Q^2}{\mu^2}\right)^2-8\right]  \ ,
\end{equation}
which differs from that for Drell-Yan process by a term of $\pi^2$.
After canceling out these divergences, we have the total
contribution at one-loop order,
\begin{eqnarray}
\widetilde{F}_{UU}&=&\frac{\alpha_s}{2\pi}C_F\left\{\left(-\frac{1}{\epsilon}+\ln\frac{c_0^2}{b^2\mu^2}\right)
\left({\cal P}_{q\to q}(\xi)\delta(1-\hat\xi)+\frac{1}{\hat\xi}{\cal P}_{q\to q}(\hat \xi)\delta(1-\xi)\right)\right.\nonumber\\
&&+(1-\xi)\delta(1-\hat\xi)+\frac{1}{\hat\xi}(1-\hat\xi)\delta(1-\xi)\nonumber\\
&&\left.+\delta(1-\xi)\delta(1-\hat\xi)\left[3\ln\frac{Q^2b^2}{c_0^2}-\left(\ln\frac{Q^2b^2}{c_0^2}\right)^2-8\right]\right\} \ , \nonumber\\
\widetilde{F}_{UT}^\beta&=&\frac{\alpha_s}{2\pi}\left(\frac{ib_\perp^\alpha}{2}\right)\left\{\left(-\frac{1}{\epsilon}+\ln\frac{c_0^2}{b^2\mu^2}\right)
\left({\cal P}_{qg\to qg}^T(\xi)\delta(1-\hat\xi)+\frac{1}{\hat\xi}{\cal P}_{q\to q}(\hat\xi)C_F\delta(1-\xi)\right)\right.\nonumber\\
&&+\left(-\frac{1}{2N_c}\right)(1-\xi)\delta(1-\hat\xi)+C_F\frac{1}{\hat\xi}(1-\hat\xi)\delta(1-\xi)\nonumber\\
&&\left.+\delta(1-\xi)\delta(1-\hat\xi)C_F\left[3\ln\frac{Q^2b^2}{c_0^2}-\left(\ln\frac{Q^2b^2}{c_0^2}\right)^2-8\right]\right\} \ .
\end{eqnarray}
The sign change between the Sivers single spin asymmetries in DIS and Drell-Yan
lepton pair production in $pp$ collisions can be seen by comparing the above equation with Eq.~(\ref{eut}).
Applying the TMD factorization, we will obtain the hard factors
in the Ji-Ma-Yuan scheme,
\begin{equation}
H_{UT}^{({\rm DIS})}(Q)=H_{UU}^{({\rm DIS})}(Q)=
\frac{\alpha_s}{2\pi}C_F\left[\ln\frac{Q^2}{\mu^2}+\ln\rho^2\ln\frac{Q^2}{\mu^2}-\ln\rho^2+\ln^2\rho+\pi^2-4\right] \ .
\end{equation}
There is difference of $\pi^2$ in the hard factors as compared to
those in the Drell-Yan processes.

Similar calculations can be performed for the Collions-11 TMDs.
We omit these details.


\section{TMD evolution and resummation}

From the above calculations, we find that the fixed order calculations
contain large logarithms. In order to resum these large logarithms, we
have to apply the TMD evolution. The resummation can be performed by
solving various evolution equations and renormalization group equations.
In particular, the TMDs obey the so-called Collins-Soper evolution
equation, whose solution shall resum all the double logarithms. Additional
single logarithms can be resummed by the renormalization group
equation. Although there are different ways to define the TMDs, the
final results for the resummed cross sections take the unique forms,
in particular, in terms of the collinear parton distributions and
correlation functions (in case of azimuthal angular asymmetries
in the hard processes such as the Sivers effects). All the scheme
dependence in the TMD definition cancels out in the final resummation
form. In the following, we will review a straightforward derivation following
Collins-Soper-Sterman 1985. The derivation is carried out for the
differential cross sections, such as the structure functions discussed
in the last section: $\widetilde{W}_{UU,UT}$, $\widetilde{F}_{UU,{\rm sivers}}$.

\subsection{TMD Evolution}

The TMD evolution was first derived in the context for the spin
average cross section. The extension to the $k_\perp$-odd observables
was discussed in Ref.~\cite{Idilbi:2004vb}, which showed that the evolution
kernel is the same as that for the unpolarized case. Following this
derivation, we will find out that the single spin dependent
structure function, e.g., $\widetilde{F}_{\rm sivers}^\alpha$ obey the following
evolution equation,
\begin{eqnarray}
\frac{\partial}{\partial \ln Q^2}{\widetilde F}_{\rm sivers}^\alpha(Q;b)
=\left(K(b,\mu)+G(Q,\mu)\right) {\widetilde F}_{\rm sivers}^\alpha(Q;b) \ ,
\end{eqnarray}
where $K$ and $G$ are the associated soft and hard part
in the evolution kernel. The above evolution can be derived from
the relevant Collins-Soper evolution equation for the TMD quark
distribution and fragmentation functions. The coefficients can also be
obtained by comparing to the one-loop calculation we have showed in the
last section. In particular, at one-loop order, we find that
\begin{equation}
K+G=-\frac{\alpha_sC_F}{\pi}\left(\ln\frac{Q^2b^2}{c_0^2}-\frac{3}{2}\right)\ ,
\end{equation}
which is the same for all the structure functions we discussed in the last section.
The soft part $K(b,\mu)$ can be derived from the evolution of the TMD
parton distribution, and it is known at one-loop order,
\begin{equation}
K(b,\mu)=-\frac{\alpha_sC_F}{\pi}\ln\frac{b^2\mu^2}{c_0^2} \ ,
\end{equation}
which again is the same for all the structure functions. Therefore, at one-loop
order, $G$ can be written as
\begin{equation}
G(Q,\mu)=-\frac{\alpha_sC_F}{\pi}\left(\ln\frac{Q^2}{\mu^2}-\frac{3}{2}\right)\ .
\end{equation}
To solve the evolution equation, we apply the renormalization
equation for  $K$ and $G$ ,
\begin{equation}
\frac{\partial}{\partial \ln \mu}K(b,\mu)=-\gamma_K=-\frac{\partial}{\partial\ln\mu}G(Q,\mu) \ ,
\end{equation}
where $\gamma_K$ is the well-known cusp anomalous dimension. At one-loop
order, $K=-2\alpha_sC_F/\pi$.
By solving the above renormalization equation, we find that,
\begin{equation}
K(b,\mu)+G(Q,\mu)=K(b,\mu_L)+G(Q,C_2/Q)-\int_{\mu_L}^{C_2/Q}\frac{d\mu}{\mu}\gamma_K \ ,
\end{equation}
where we have chosen the upper limit of the integral around scale $Q$, i.e., $C_2$ is order 1. Substituting
the above result into the evolution equation, and taking into account the
running effects in $K$, we will obtain,
\begin{equation}
\widetilde {F}_{\rm sivers}^\alpha(Q;b)=\widetilde
{F}_{\rm sivers}^\alpha(Q_0/C_2;b)e^{-{\cal S}(Q,Q_0,b,C_2)} \  . \label{evo}
\end{equation}
The Sudakov form factor reads as,
\begin{equation}
S(Q,Q_0,b,C_2)=\int_{Q_0}^{C_2Q}\frac{d\bar\mu}{\bar\mu}
\left[\ln\left(\frac{C_2Q^2}{\bar\mu^2}\right)A(bQ_0,\bar\mu)
+B(C_2,bQ_0,\bar\mu)\right] \ ,
\end{equation}
where $A$ and $B$ are defined as
\begin{eqnarray}
A(bQ_0,\bar\mu)&=&\gamma_K(\bar\mu)+\beta\frac{\partial}{\partial g}K(b,Q_0,g(\bar\mu)) \ ,\nonumber\\
B(C_2,bQ_0,\bar\mu)&=&-2K(b,Q_0,g(\bar\mu))-2G(Q,Q/C_2,g(\bar\mu))\ .\label{abji}
\end{eqnarray}
The $A$, $B$ coefficients can be calculated order by order in
perturbation theory.

\subsection{CSS Resummation and $b_*$-prescription}

In the CSS resummation, $Q_0$ has to be set around $1/b$ to further absorb
logarithms in the form factor,
\begin{equation}
\widetilde {F}_{\rm sivers}^\alpha(Q;b)|_{\rm css}=\widetilde
{F}_{\rm sivers}^\alpha(C_1/C_2/b;b)e^{-{\cal S}(Q,C_1/b,b,C_2)} \  . \label{css0}
\end{equation}
With this choice, $A$ and $B$ coefficients can be expanded as
perturbative series $A=\sum_i(\alpha_s/\pi)^iA^{(i)}$.
Furthermore, in the CSS resummation
$\widetilde {F}_{\rm sivers}^\alpha(C_1/C_2/b;b)$ is calculated in the
collinear factorization in terms of collinear parton distributions
and correlation functions,
\begin{equation}
\widetilde {F}_{\rm sivers}^\alpha(C_1/C_2/b;b)=\int \Delta C^T(C_1/C_2,b_\perp\mu)\otimes T_F(z_1,z_2;\mu)
C(C_1/C_2,b_\perp\mu)\otimes D(z;\mu) \ ,
\end{equation}
where $T_F(z_1,z_2;\mu)$ represents the moment of the quark
Sivers function and $D(z,\mu)$ the integrated fragmentation function. From the results in the last section,
we can immediately obtain the associated coefficients.
In practice, a canonical choice is normally made for $C_1$ and
$C_2$: $C_1=c_0$ and $C_2=1$, which we will follow in
our calculations,
\begin{equation}
\widetilde {F}_{\rm sivers}^\alpha(Q;b)|_{\rm css}=\widetilde
{F}_{\rm sivers}^\alpha(c_0/b;b)e^{-{\cal S}_{pert}(Q,b)} \  , \label{css}
\end{equation}
where $S_{pert}$ is written as
\begin{equation}
S_{pert}(Q,b)=\int_{c_0/b}^Q\frac{d\bar\mu}{\bar\mu}\left[A\ln\frac{Q^2}{\bar\mu^2}+B\right] \ .\label{spert}
\end{equation}
We would like to emphasize again, the above resummation formula
do not depend on the scheme to define the transverse momentum dependent
parton distributions. The $A$, $B$,  $C$ coefficients can be calculated
from perturbative diagrams once the factorization been established.
In particular, for unpolarized Drell-Yan process, $A$ coefficient has been calculated
up to $A^{(3)}$, while for $B$ up to $B^{(2)}$. Most recently, $C$
coefficients have been calculated up to $C^{(2)}$ as well.
For single-spin dependent cross section, from the results in the last section,
we shall be able to obtain $A^{(1)}$, $B^{(1)}$, and $C^{(1)}$.
Since $A^{(2,3)}$ are spin-independent, they shall be the same as the unpolarized
cross sections. In the following numeric calculations, we only keep $A^{(1)}$ and $B^{(1)}$
as example to demonstrate the evolution effects.

In the above equations, the Fourier transformation to obtain the transverse momentum
distribution involves the large $b$ region, where the integral will encounter the
so-called Landau pole singularity. In order to avoid the Landau pole
singularity, it was suggested the $b_*$ prescription~\cite{Collins:1984kg}~\footnote{Besides
the $b_*$-prescription, there are other approaches in the literature, see, for example,
Refs.~\cite{Qiu:2000ga,Kulesza:2002rh,Catani:2000vq,
Catani:2003zt,Bozzi:2003jy,Mantry:2009qz,Becher:2010tm,Chiu:2012ir}.},
\begin{equation}
b\Rightarrow b_*=b/\sqrt{1+b^2/b_{max}^2}  \ ,~~b_{max}<1/\Lambda_{QCD}\ ,
\end{equation}
where $b_{max}$ is a parameter. From the above definition, $b_*$ is always in
the perturbative region where $b_{max}$ is normally chosen to be around $1GeV^{-1}$.
Because of the introduction of $b_*$ in the Sudakov form factor, the difference
from the original form factor requires additional non-perturbative form factor,
and a generic form as suggested,
\begin{equation}
S_{NP}=g_2(b)\ln Q/Q_0 +g_1(b) \ .
\end{equation}
Therefore, the final Sudakov form factor can be written as
\begin{equation}
{\cal S}_{sud}\Rightarrow {\cal S}_{pert}(Q;b_*)+S_{NP}(Q;b) \ .
\end{equation}
With the non-perturbative form factor, we can write down final results for
the structure functions as,
\begin{eqnarray}
\widetilde {W}_{UU}(Q;b)&=&e^{-{\cal S}_{pert}(Q^2,b_*)-S_{NP}(Q,b)}
\Sigma_{i,j} C_{qi}^{(DY)}\otimes f_{i/A}(z_1) C_{\bar qj}^{(DY)}\otimes f_{j/B}(z_2') \ ,\\
\widetilde {W}_{UT}^\alpha(Q;b)&=&\left(\frac{-i b_\perp^\alpha}{2}\right)e^{-{\cal S}_{pert}(Q^2,b_*)-S_{NP}^T(Q,b)}
\Sigma_{i,j} \Delta C_{qi}^{T(DY)}\otimes f_{i/A}^{(3)}(z_1',z_1^{\prime\prime}) C_{\bar qj}^{(DY)}\otimes f_{j/B}(z_2') ,\\
\widetilde {F}_{UU}(Q;b)&=&e^{-{\cal S}_{pert}(Q^2,b_*)-S_{NP}(Q,b)}
\Sigma_{i,j} C_{qi}^{(DIS)}\otimes f_{i/A}(z_1) \hat{C}_{qj}^{(DIS)}\otimes D_{j/B}(z_2') \ ,\\
\widetilde {F}_{\rm sivers}^\alpha(Q;b)&=&\left(\frac{-i b_\perp^\alpha}{2}\right)e^{-{\cal S}_{pert}(Q^2,b_*)-S_{NP}^T(Q,b)}
\Sigma_{i,j} \Delta C_{qi}^{T(DIS)}\otimes f_{i/A}^{(3)}(z_1',z_1^{\prime\prime}) \hat{C}_{qj}^{(DIS)}\otimes D_{j/B}(z_2') \ .
\end{eqnarray}
Because the $Q^2$ evolution for the Sivers term is the same as the unpolarized case,
there shall be no difference in the perturbative part of the Sudakov form factor
${\cal S}_{pert}(Q^2,b_*)$. The same argument can be made for the $\ln Q$ term in the non-perturbative form factor.
But they do differ for the constant term. Therefore, generically, we can write down,
\begin{eqnarray}
S_{NP}(Q,b)&=& g_2(b)\ln Q+g_1(b;z_1,z_2)\nonumber\\
S_{NP}^T(Q,b)&=&g_2(b)\ln Q+g_1^T(b;z_1,z_2) \ ,
\end{eqnarray}
where we have included a general dependence on $z_1$ and $z_2$ as well.
Here, $z_1$ and $z_2$ represent the momentum fractions in
the collinear parton distributions or fragmentation functions.

Our calculations in the last subsections lead to the following results of
the $C$ coefficients,
\begin{eqnarray}
C_{qq}^{(DY)}&=&\delta(1-z)+\frac{\alpha_s}{\pi}\left(\frac{C_F}{2}\left(1-z \right) +\frac{C_F}{4}\left(\pi^2 -8\right)\delta(1-z)\right)\ , \\
C_{qq}^{(DIS)}&=&\delta(1-z)+\frac{\alpha_s}{\pi}\left(\frac{C_F}{2}\left(1-z \right) -2C_F\delta(1-z)\right)\ , \\
\hat{C}_{qq}^{(DIS)}&=&\delta(1-z)+\frac{\alpha_s}{\pi}\left(\frac{C_F}{2}\left(1-z \right)+{\cal P}_{q\to q}\ln z -2C_F\delta(1-z)\right)\ ,\\
\Delta C_{qq}^{T(DY)}&=&\delta(1-z)+\frac{\alpha_s}{\pi}\left(-\frac{1}{4N_c}\left(1-z \right) +\frac{C_F}{4}\left(\pi^2 -8\right)\delta(1-z)\right)\ ,\\
\Delta C_{qq}^{T(DIS)}&=&\delta(1-z)+\frac{\alpha_s}{\pi}\left(-\frac{1}{4N_c}\left(1-z \right) -2C_F\delta(1-z)\right) \ ,
\end{eqnarray}
other coefficients can be found the literature~\cite{Balazs:1997xd}.
In the following numeric calculations, however, we only keep the leading term
$C^{(0)}$ in the above equations as a first step estimate. To have a complete
calculations, we have to solve the DGLAP evolution for both integrated quark
distribution and transverse momentum moment of the Sivers function. In Sec.~IV,
we will make an attempt to estimate the partial effects coming from the evolution
of the above distributions.

\subsection{BLNY/KN Parameterizations}

The CSS resummation with $b_*$-prescription has been extensively applied
to describe low transverse momentum Drell-Yan and W/Z boson production
in hadronic collisions, in particular, in a series publications by C.P. Yuan and
P. Nadolsky and their collaborators.
These studies have demonstrated the prediction power
of the CSS resummation formalism. Recent experimental measurements
at the LHC have confirmed the predictions from this resummation calculation.

In the BLNY fit, the following functional form has been chosen,
\begin{equation}
S_{NP}=g_1b^2+g_2b^2\ln\left({Q}/{3.2}\right)+g_1g_3b^2\ln(100 x_1x_2)  \ ,\label{blny0}
\end{equation}
for Drell-Yan type of processes in $pp$ collisions, where $g_{1,2,3}$ are fitting parameters~\cite{Landry:2002ix},
\begin{equation}
g_1=0.21,~~g_2=0.68,~~g_1g_3=-0.2,~~{\rm with}~~b_{max}=0.5\textmd{GeV}^{-1} \ .
\end{equation}
We would like to point out a couple of points on BLNY parameterization.
First, BNLY fit is only applied to the Drell-Yan type of processes (lepton pair production
via virtual photon or W/Z boson) with relative high $Q^2$ ($>20\textmd{GeV}^2$). An attempt
to understand the SIDIS from HERA data in the small-$x$ region has also been tried
with different $x$-dependence in the form factor~\cite{Nadolsky:1999kb}.
Second, the $x$ and $Q^2$ dependence are strongly correlated. This
is simply because $x_1x_2=Q^2/s$. Therefore, $g_2$ coefficient is not
completely reflecting $Q^2$-dependence in the non-perturbative form factor.
Third, the form factor also strongly depends on $b_{max}$. In a later publication~\cite{Konychev:2005iy},
Konychev-Nadolsky (KN) have addressed this issue in great details.
In that paper, they found that the following parameters,
\begin{equation}
g_1=0.20,~~g_2=0.184,~~g_1g_3=-0.026,~~{\rm with}~~b_{max}=1.5\textmd{GeV}^{-1} \ ,\label{kn}
\end{equation}
instead of the original BLNY parameterization. Since this parameterization
has mild $x$-dependence, we will use this form
of the non-perturbative form factor to the Drell-Yan process in the following numeric
calculations.

The CSS resummation formalism has also been applied to study semi-inclusive
hadron production in DIS from HERA experiments as mentioned above. However, these studies focus
on the small-$x$ region. It is interested to notice that, in order to describe the HERA
data in the small-$x$ region, a different non-perturbative form factor was used in these
studies. In this paper, since we focus on the Sivers single spin asymmetries from HERMES
and COMPASS experiments which are mainly in the moderate $x$ range (around 0.1)~\footnote{COMPASS
data also cover a relative small-$x$ region. However, the sizable Sivers asymmetry only exists
around 0.1. To have a complete picture in the small-$x$, we have to take into account the
small-$x$ dependence in the TMD evolution, which is an interesting topic but beyond the
scope of current paper.} , we will not
compare to the HERA data where small-$x$ resummation might be
as important as the transverse momentum resummation.

\subsection{Incompatibility between BLNY/KN and SIDIS Data from HERMES/COMPASS}

In the BLNY (KN) fit to the Drell-Yan lepton pair production, the kinematics cover mostly
the moderate $x$ range which overlaps with the SIDIS data from HERMES and COMPASS,
in particular, where the large Sivers single spin asymmetries were observed around $x\sim 0.1$.
Therefore, from the factorization and universality arguments,
the non-perturbative form factors determined in these fits shall be
used to understand the quark distribution contribution to the SIDIS data
from HERMES and COMPASS.
However, a careful examination has shown that
either BLNY or KN parameterization can not be used to describe the SIDIS data
from HERMES/COMPASS.

\begin{figure}[tbp]
\centering
\includegraphics[width=8cm]{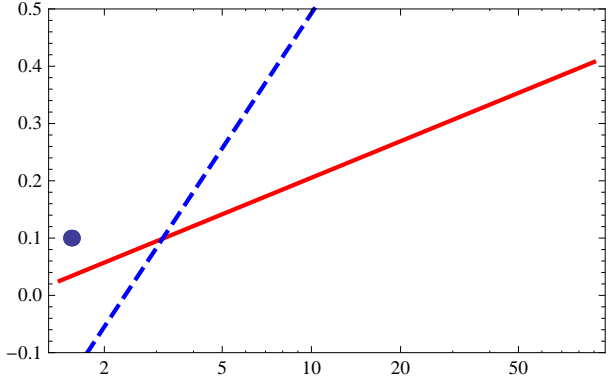}
\caption{Coefficient $a(Q)$ in the non-perturbative form factor $e^{-S_{NP}}=e^{-a(Q)b^2}$
for the TMD quark distribution as function of $Q$: the dot represents the value needed for the SIDIS~\cite{Schweitzer:2010tt}
as compared to the BLNY (dashed line) and KN (solid line) parameterizations for $x=0.1$.}
\label{fig:a2q}
\end{figure}

To illustrate this issue more clearly, in Fig.~1, we plot the non-perturbative form factor
derived from these parameterizations, one from BLNY, and one from KN paper.
If we extrapolate these parameterizations down to $Q^2\approx 3\textmd{GeV}^2$ for SIDIS
at HERMES and COMPASS range, we find that $\left[\ln\left(e^{-S_{NP}}\right)\right]=-a(Q)b^2$
for typical value of $x\approx 0.1$ is too small to describe the data.
For BLNY parameterization, even a negative value for $a(Q)$ will be found around $Q^2\sim 3\textmd{GeV}^2$,
and the whole framework will break down.

The main reason of the above incompatibility is that the relative low
$Q^2$ in current SIDIS experiments from HERMES:
$Q^2$ is in the range of $2\sim 3\textmd{GeV}^2$.
However, in the $b_*$-prescription, $1/b_*\sim 1/b_{max}\approx 1.5\textmd{GeV}$ is also in the
similar range. The consequence is that the $Q^2$-dependence is mainly
coming from the logarithmic dependence in the non-perturbative form
factor, rather than that from the evolution itself. This has to be corrected in order
to describe the SIDIS data from the CSS evolution.

On the other hand, for moderate $Q^2$ variations, we shall be able to
understand the $Q^2$-dependence by directly solving the evolution equation.
For example, in the Sudakov
resummation formula, Eq.~(\ref{evo}), we can, in principle, to study the $Q^2$
dependence by taking the structure functions at lower scale $Q_0$
as input, and calculate the structure function at higher $Q$ using the
direct integral of the kernel from $Q_0$ to $Q$. That is the approach we are
going to take in comparing SIDIS from HERMES/COMPASS to
Drell-Yan lepton pair production. As we briefly shown in Ref.~\cite{Sun:2013dya}, this
approach works well for $Q^2$ range from 2 to 100 GeV$^2$
and covers SIDIS from HERMES and COMPASS and most of the Drell-Yan
processes from the fixed target experiments. Of course, for extreme high
$Q$ such as $W/Z$ boson production, we have to take into account higher
order corrections and back to the complete
CSS resummation.

In the following, we will show that this evolution approach
can describe the transverse momentum distribution in SIDIS and
Drell-Yan processes up to $Q\sim 10\textmd{GeV}$. Since Drell-Yan
data can also be understood from the CSS resummation with BLNY (KN)
parameterization for the non-perturbative form factors,
this provides a nature match between SIDIS and Drell-Yan experiments,
and help us understand the TMD evolution in this particular
energy range.  Once we understand how this works for the unpolarized
cross sections, we will extend to the Sivers single spin asymmetries
in these processes.

\subsection{Sun-Yuan Approach}

In our calculations of the SIDIS from HERMES/COMPASS, we
evolve the cross sections directly from lower to higher scale,
\begin{eqnarray}
\widetilde {W}_{UU}(Q;b)&=&e^{-{\cal S}_{sud}(Q,Q_0,b)}\widetilde {W}_{UU}(Q_0;b)\ ,\label{wuu-sy}\\
\widetilde {W}_{UT}^\alpha(Q;b)&=&e^{-{\cal S}_{sud}(Q,Q_0,b)}\widetilde {W}_{UT}^\alpha(Q_0;b)\ ,\label{wut-sy}\\
\widetilde {F}_{UU}(Q;b)&=&e^{-{\cal S}_{sud}(Q,Q_0,b)}\widetilde {F}_{UU}(Q_0;b)\ ,\\
\widetilde {F}_{\rm sivers}^\alpha(Q;b)&=&e^{-{\cal S}_{sud}(Q,Q_0,b)}\widetilde {F}_{\rm sivers}^\alpha(Q_0;b) \ ,\label{sunyuan}
\end{eqnarray}
where the Sudakov form factor follows the above equation,
\begin{eqnarray}
{\cal
S}_{Sud}={2C_F}\int_{Q_0}^{Q}\frac{d\bar\mu}{\bar\mu}
\frac{\alpha_s(\bar\mu)}{\pi}\left[\ln \left(\frac{Q^2}{\bar\mu^2}\right)
+\ln\frac{Q_0^2b^2}{c_0^2}-\frac{3}{2}\right]\ . \label{sud}
\end{eqnarray}
The above Sudakov form factor comes from the one-loop calculations
of the $A$ and $B$ coefficients of Eq.~(\ref{abji}) in previous subsections. It has been used
by Boer in previous analysis as well~\cite{Boer:2001he}. In the above equation, the second
terms contains $b_\perp$ dependence which will lead to a $p_\perp$ broadening
effects at higher $Q^2$ as compared to lower $Q^2$, whereas the first
and third terms only change the normalization of the cross sections.
We would like to emphasize that the Sudakov form factor is the same for
the spin-average and single-spin dependent cross sections, because the
associated evolution kernel is spin-independent. Moreover, both Drell-Yan and
SIDIS obey the same evolution equations. The difference between the hard
factors in the TMD factorization discussed in the last sections does not affect
the evolution as function of $Q^2$.

It has been well understood that the SIDIS data from HERMES/COMPASS
can be described by a Gaussian assumption for the TMDs Ref.~\cite{Schweitzer:2010tt}. We follow these
suggestions to parameterize the lower $Q_0$ structure functions as,
\begin{eqnarray}
\widetilde{W}_{UU}(Q_0,b)&=& \sum_q\;e_q^2  \; f_q(x,\mu=Q_0)\; f_{\bar q}(x',\mu=Q_0)
e^{-{g_0b^2}-{g_0b^2}}   \ ,\label{wuu0-sy}\\
\widetilde{W}_{UT}^\alpha(Q_0,b)&=&  \frac{-ib_\perp^\alpha M }{2}
\sum_q\;e_q^2  \; \Delta f_q^{\rm sivers}(x)\;  f_{\bar q}(x',\mu=Q_0)
e^{-(g_0-g_s)b^2-{g_0b^2}}\ ,\label{wut0-sy}\\
\widetilde{F}_{UU}(Q_0,b)&=& \sum_q\;e_q^2  \; f_q(x_B,\mu=Q_0)\; D_q(z_h,\mu=Q_0)
e^{-{g_0b^2}-{g_hb^2}/{z_h^2}}   \ ,\label{fuu0}\\
\widetilde{F}_{\rm sivers}^\alpha(Q_0,b)&=&  \frac{ib_\perp^\alpha M }{2}
\sum_q\;e_q^2  \; \Delta f_q^{\rm sivers}(x)\;  D_q(z,\mu=Q_0)
e^{-(g_0-g_s)b^2-{g_hb^2}/{z_h^2}}\ ,
\label{siversq0}
\end{eqnarray}
with $Q_0^2=2.4GeV^2$ chosen around HERMES kinematics,
where $f$ and $D$ represent the integrated quark distribution and
fragmentation functions and they are parameterized at the scale of $Q_0$
and we follow CT10~\cite{Lai:2010vv} and DSS~\cite{deFlorian:2007aj} sets,
respectively.
In the above equations, $g_0$ and $g_h$ are chosen to be $g_0=0.097$
and $g_h=0.045$~\footnote{These
two parameters are not fit to the data but chosen according to
the phenomenology study in Ref.~\cite{Schweitzer:2010tt}.}.
We have also simply assumed that all the quark flavors have the
same parameters of $g_0$ and $g_h$, which shall be improved later on
considering the sea quark distributions ought be different from the
valence ones as demonstrated in a recent calculation~\cite{Schweitzer:2012hh}.
The above Gaussian assumptions are simple parameterizations to
describe low transverse momentum distributions. We can improve
the prediction power of this simple assumption by adding a perturbative
behavior at small $b_\perp$. However, since most of experimental data
are in the low transverse momentum region, the Gaussian approximation
shall be adequate to describe the majority of the data. For relative
moderate transverse momentum, in particular, in high $Q^2$ processes,
we shall improve that. The strategy of our calculations is to build match
between low $Q$ SIDIS and moderate $Q$ Drell-Yan processes. Once
the consistency is shown between the above evolution and the CSS
resummation with BLNY (KN) parameterization of the non-perturbative
form factors, it will be safe to extend to the predictions of the Sivers
single spin asymmetries in Drell-Yan processes.

From the point of view of the TMD factorization, the above parameterizations
correspond to the following choice for the TMD quark distribution and fragmentation
functions~\footnote{Additional hard factors can be included as well. In this paper,
we focus on the single spin asymmetry where the hard factors are the same
for the spin-average and single spin dependent cross sections. We simplify the
expressions without taking into account the hard factors contributions.},
\begin{eqnarray}
q(x,b_\perp)&=&f_q(x,Q_0)e^{-g_0b^2}\ ,\\
D_q(z,b_\perp)&=&D_q(z,Q_0)e^{-g_hb^2/z^2}\ , \\
\tilde{f}_{1T}^{\perp(DY)}(x,b_\perp)&=&\frac{-ib_\perp M}{2}\Delta f^{\rm sivers}_q(x) \ ,\\
\tilde{f}_{1T}^{\perp(DIS)}(x,b_\perp)&=&\frac{ib_\perp M}{2}\Delta f^{\rm sivers}_q(x) \ ,
\end{eqnarray}
at the initial scale $Q_0^2=2.4GeV^2$. In terms of the Ji-Ma-Yuan scheme,
the above expressions contain the soft factor contributions as well.
In the Collins-11 scheme, because the soft factor has already been absorbed
into the TMD quark distribution and fragmentation functions, the above expressions
are just for the quark distribution and fragmentation themselves.
We use the integrated quark distribution and fragmentation functions to
parameterize the TMDs. The scales for the integrated distribution and
fragmentation functions are set around the same scale $\mu=Q_0$.
This is a reasonable choice, though it will introduce additional theoretical
uncertainties. An aspect is that these expressions can phenomenologically
reproduce the integrated distribution of the SIDIS data to a good
approximation~\footnote{Additional
$Y$ terms contributions shall be taken into account for the integrated
distributions.}.

The opposite sign of the Sivers asymmetries between SIDIS and Drell-Yan
processes is reflected by the opposite sign between $\widetilde{W}_{UT}(Q_0)$
and $\widetilde{F}_{\rm sivers}(Q_0)$ in the above equations. This comes
from the opposite sign of the quark Sivers functions in these two
processes.
Comparing the above equations to those in previous sections, we find
that the $\Delta f_q^{\rm sivers}$ parameterize the transverse-momentum
moments of the quark Sivers function, $M\Delta f_q^{\rm sivers}=T_F(z,z;\mu=Q_0)$.
Again, the scale setting is similar to the above argument for the unpolarized
quark distribution.

\begin{figure}[tbp]
\centering
\includegraphics[width=6cm]{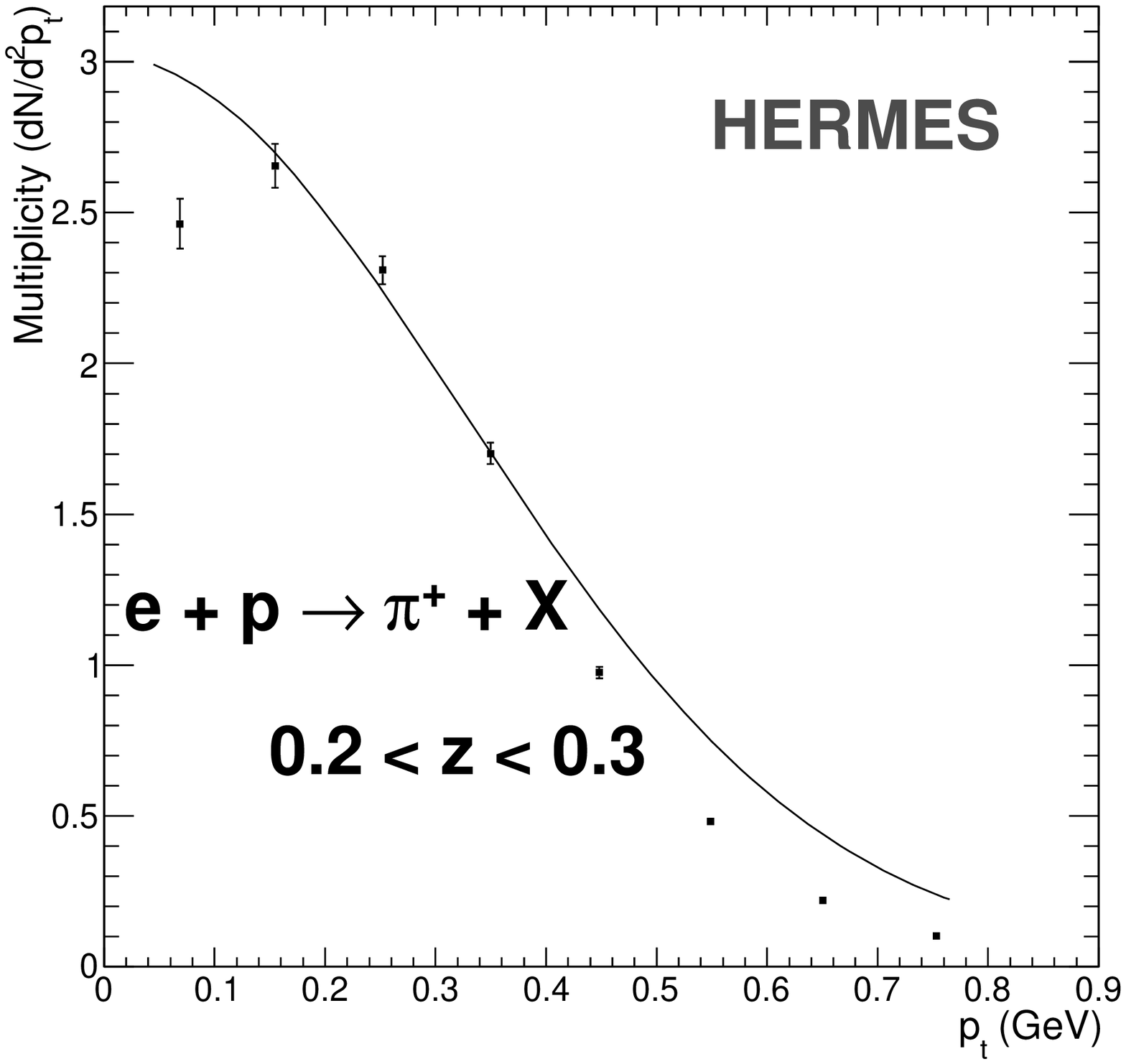}
\includegraphics[width=6cm]{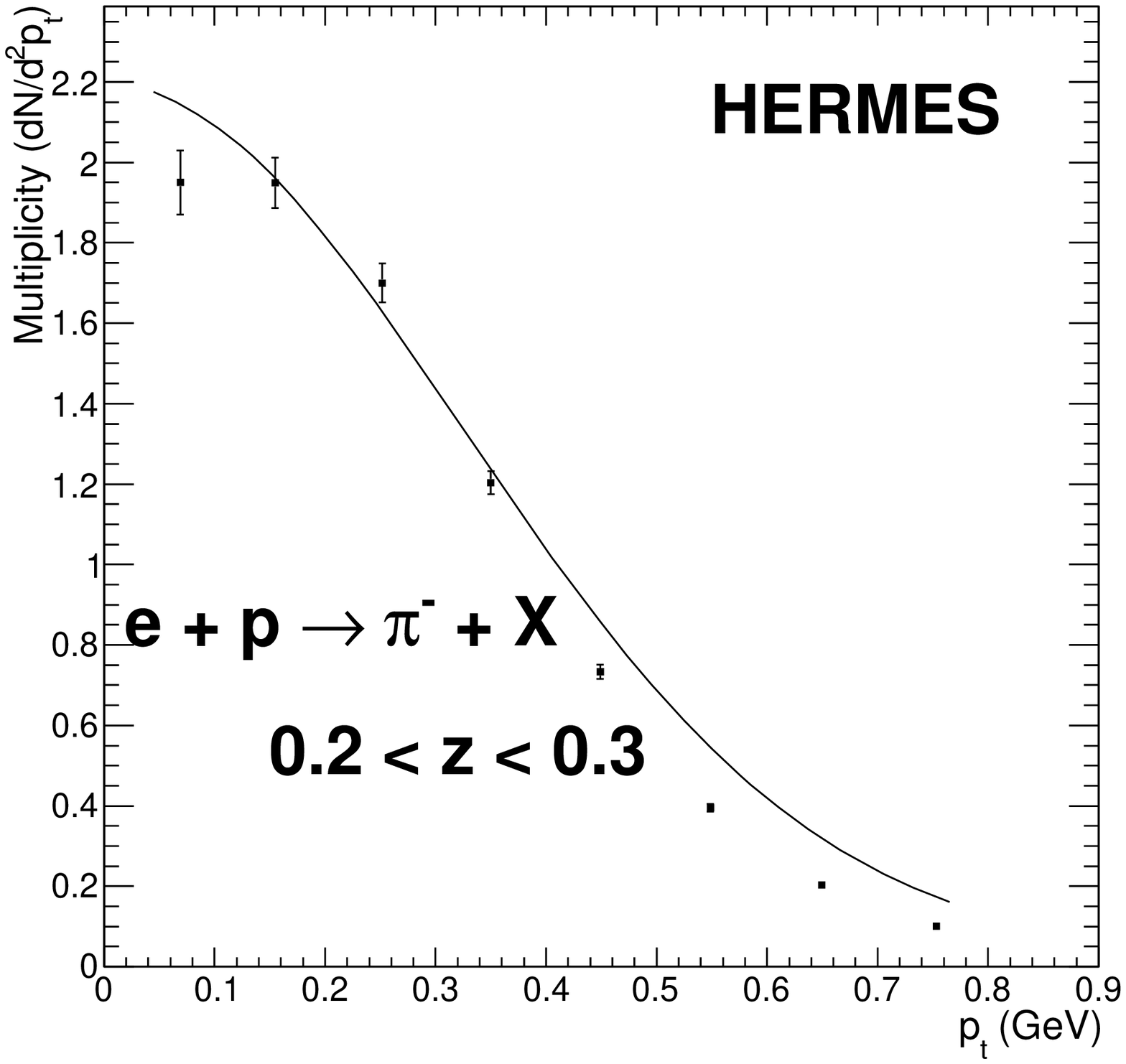}
\includegraphics[width=6cm]{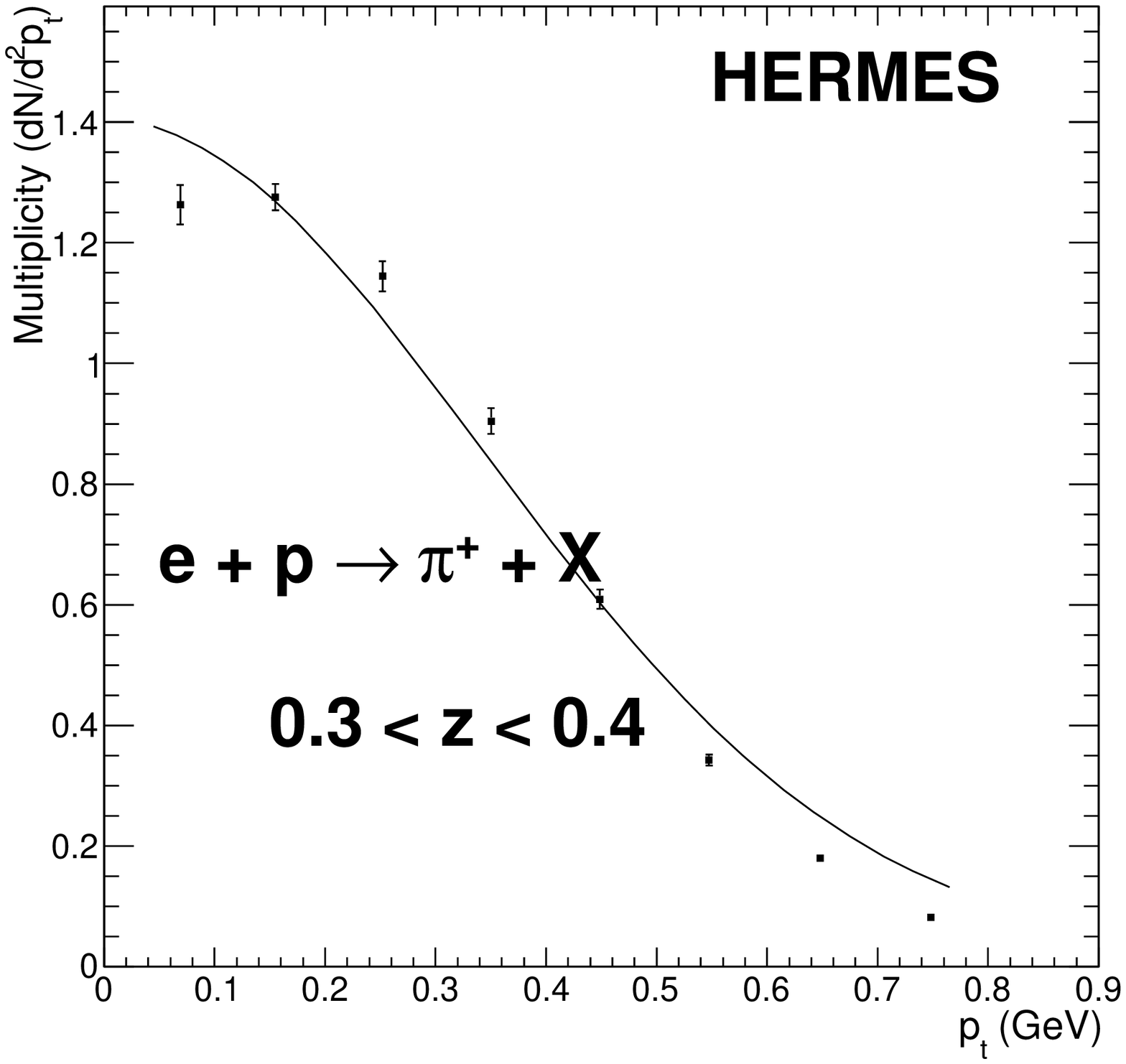}
\includegraphics[width=6cm]{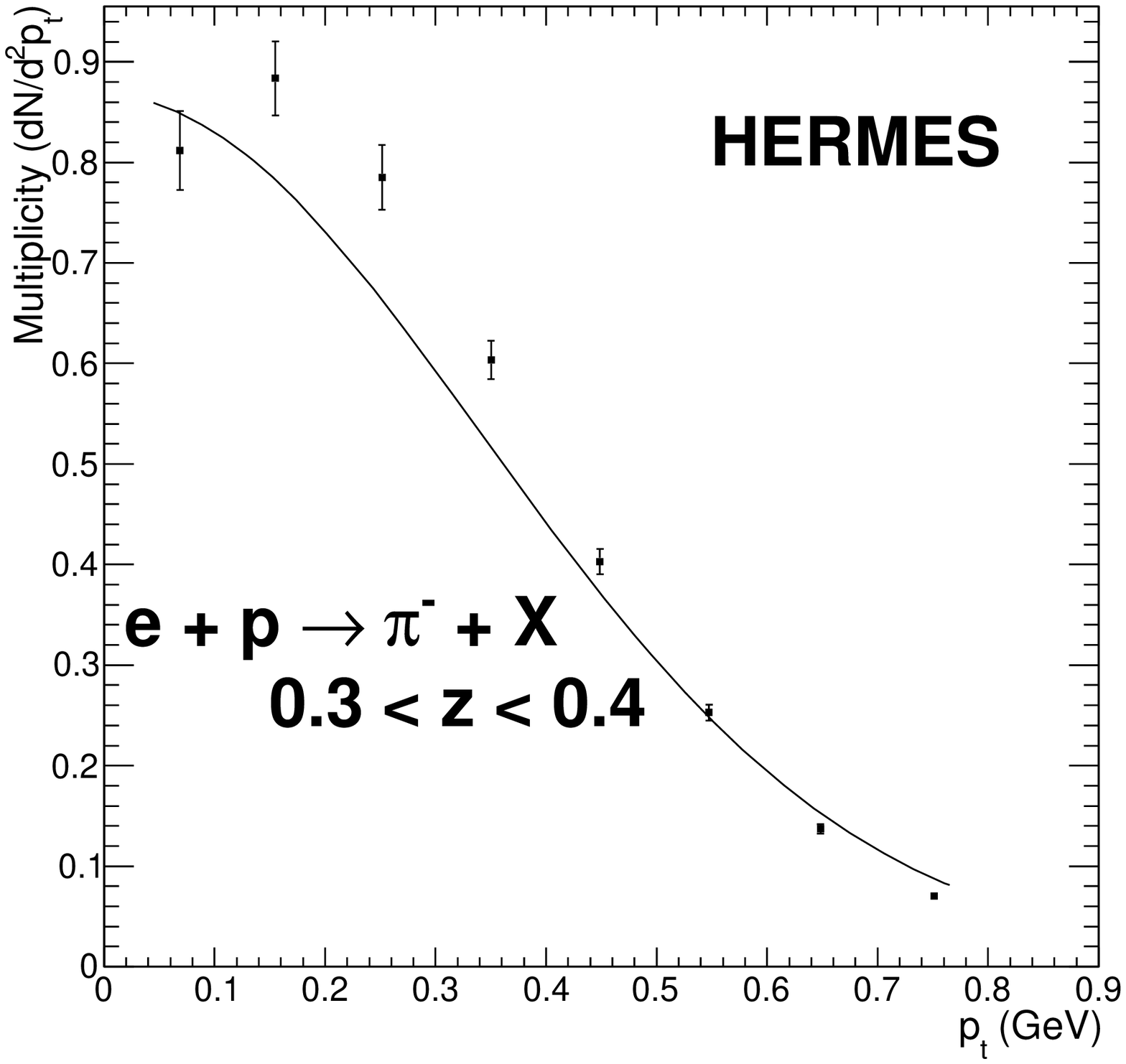}
\includegraphics[width=6cm]{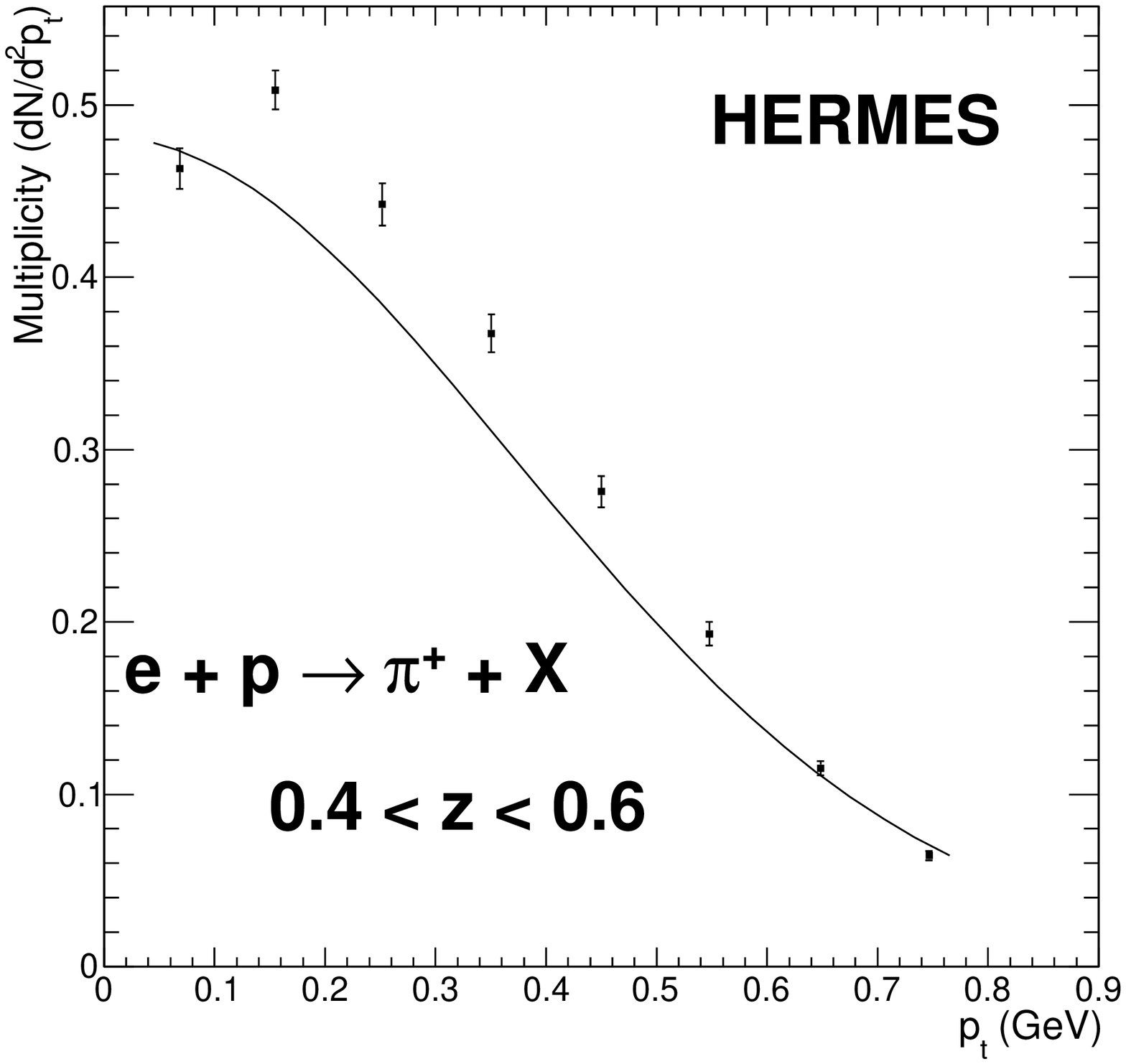}
\includegraphics[width=6cm]{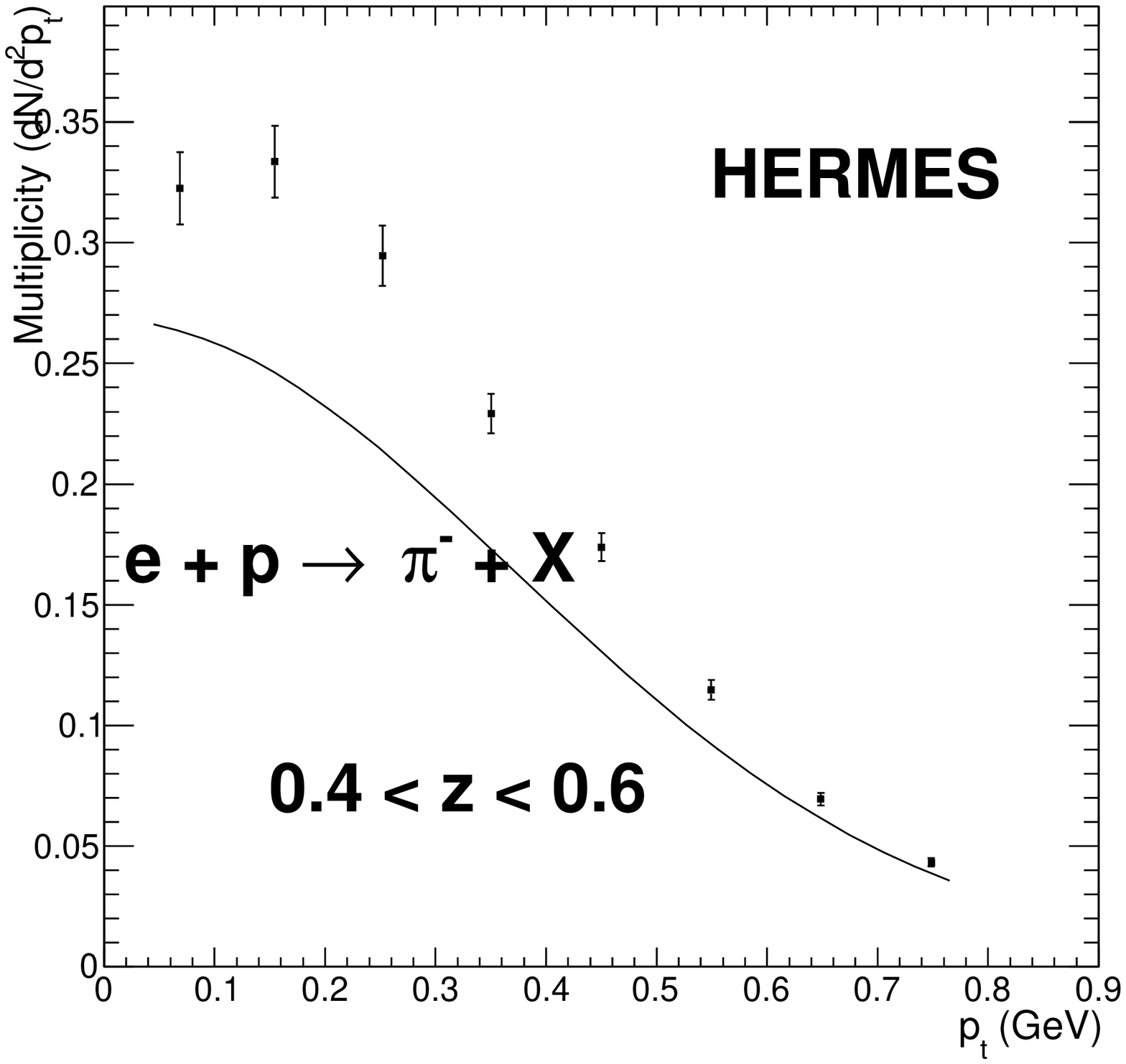}
\caption{Multiplicity distribution as function of transverse
momentum in semi-inclusive hadron production in deep inelastic
scattering compared to the experimental data from HERMES
collaboration at $Q^2=3.14$GeV$^2$ Ref.~\cite{Airapetian:2012ki}.
These data are consistent with a Gaussian assumption
in low energy scale Eq.~(\ref{fuu0}).}
\label{fig:average}
\end{figure}

\begin{figure}[tbp]
\centering
\includegraphics[width=6cm]{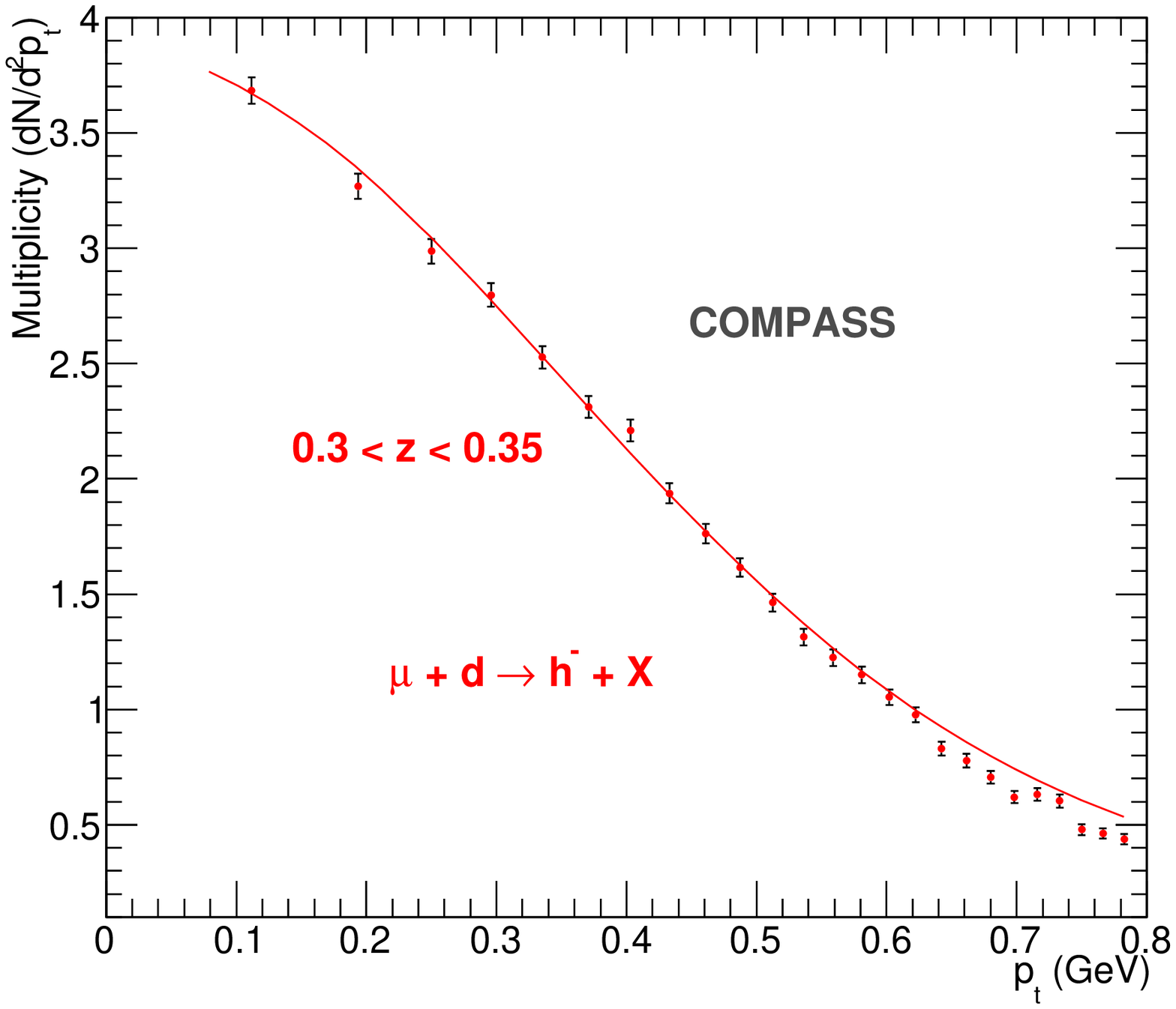}
\includegraphics[width=6cm]{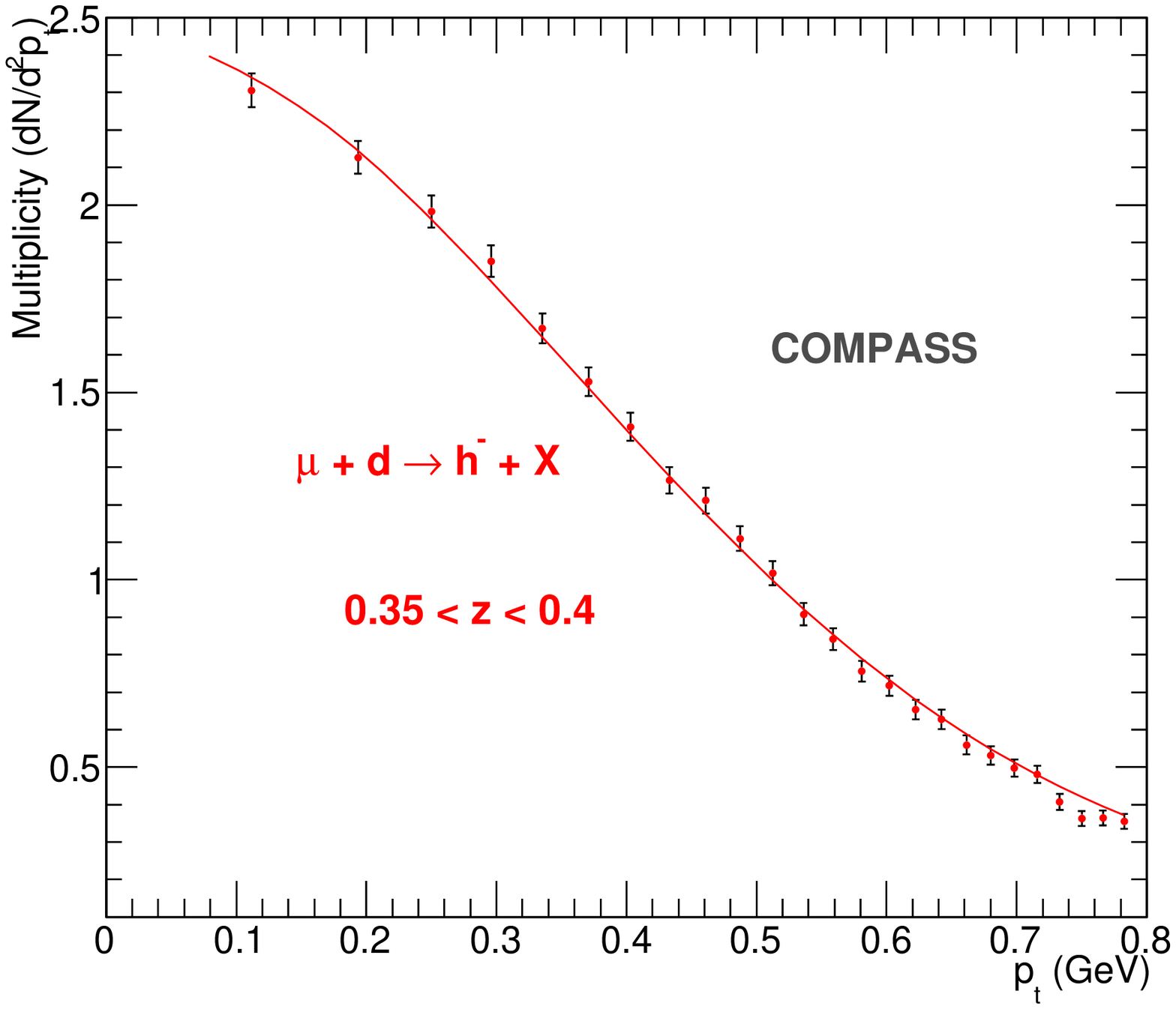}
\includegraphics[width=6cm]{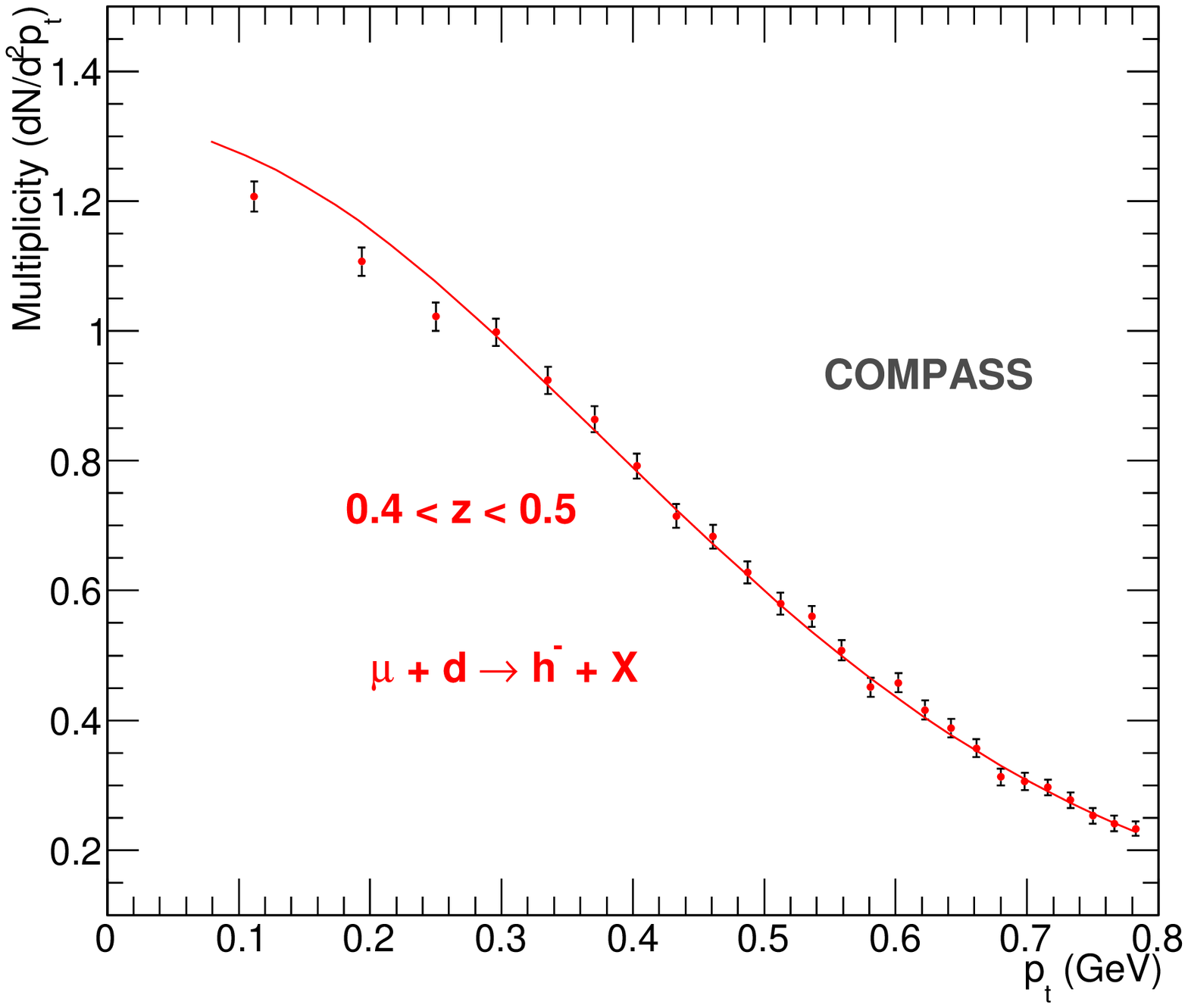}
\includegraphics[width=6cm]{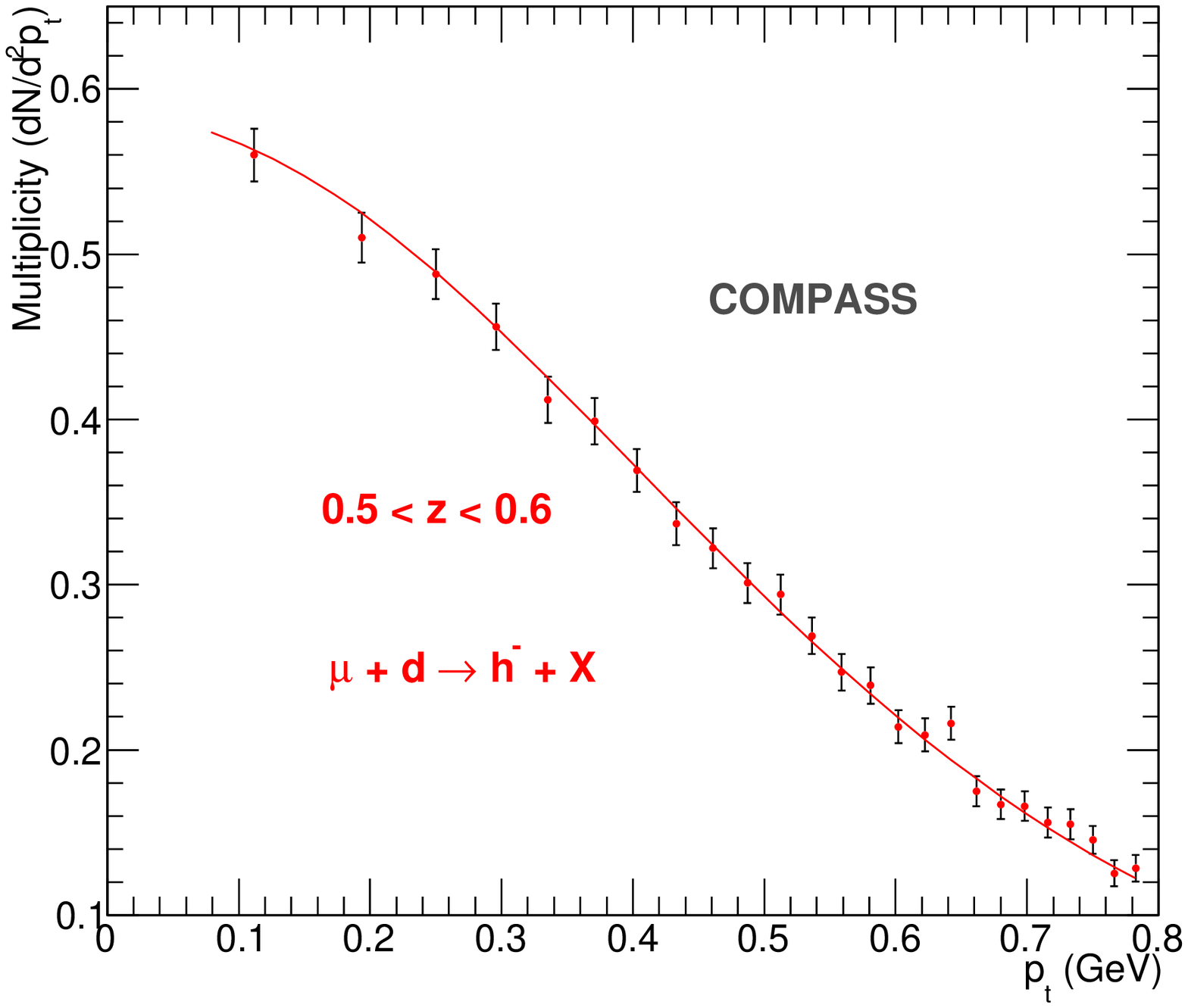}
\includegraphics[width=6cm]{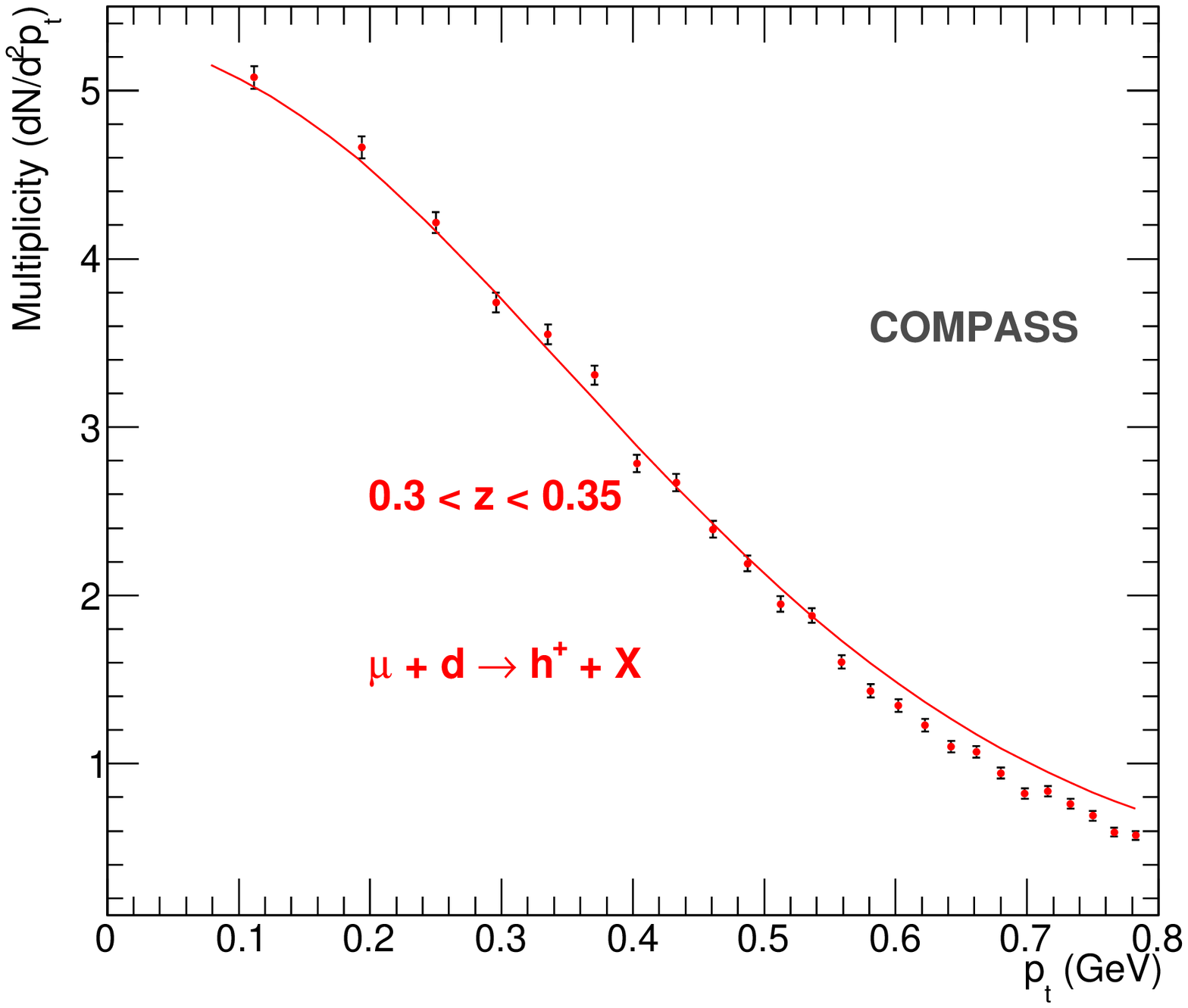}
\includegraphics[width=6cm]{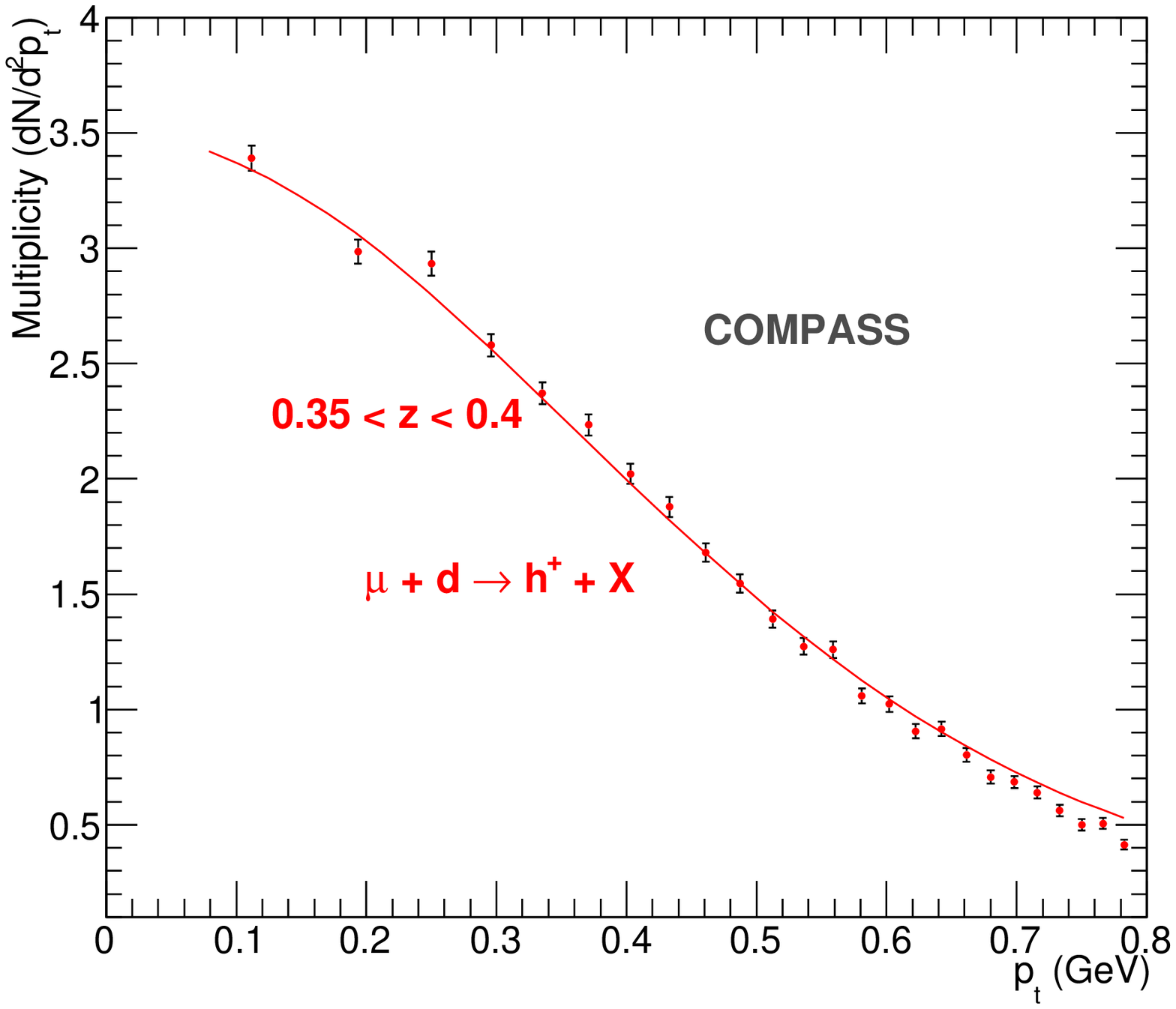}
\includegraphics[width=6cm]{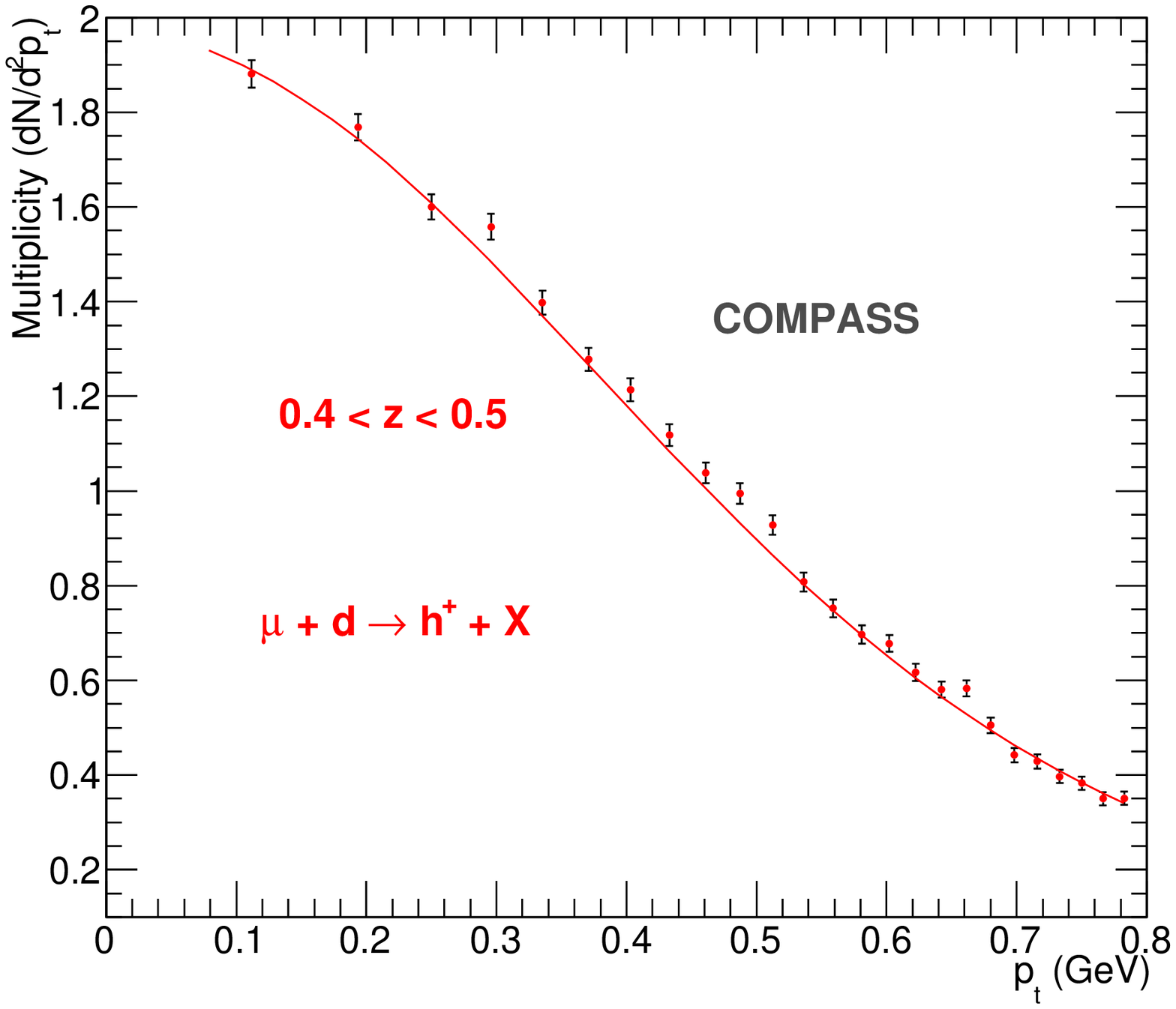}
\includegraphics[width=6cm]{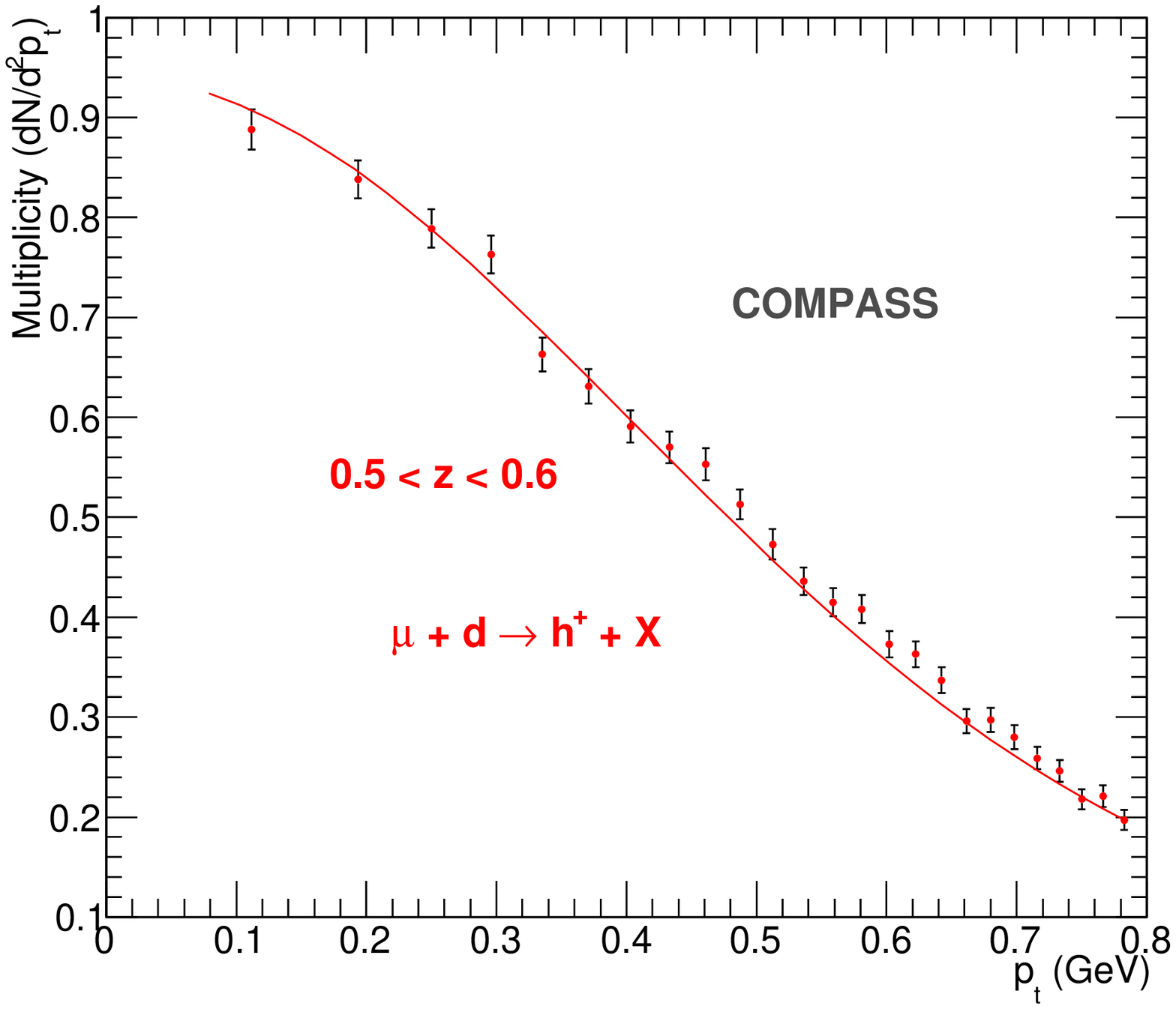}
\caption{Multiplicity distribution as function of transverse
momentum in semi-inclusive hadron production in deep inelastic
scattering compared to the experimental data from COMPASS
collaboration at $Q^2=7.56$GeV$^2$ with moderate $x=0.1$
range of Ref.~\cite{Adolph:2013stb} on deuteron target.
The COMPASS data, in particular, for the $p_\perp$
distributions, are consistent with the Sun-Yuan approach for
the TMD evolution with a Gaussian assumption
in low energy scale Eq.~(\ref{fuu0}).}
\label{compass-m}
\end{figure}

Let us first examine if the above evolution equations can
describe the unpolarized cross sections in SIDIS from HERMES
and COMPASS and the existing Drell-Yan lepton pair
production in $pp$ collisions. Because the low $Q_0$ structure
functions are parameterized according to HERMES
data, we expect they are consistent with the experimental data from
HERMES. Indeed, we show the comparisons between
our calculations and the experimental data from HERMES
collaboration on the charged hadron multiplicity distribution
as function of the transverse momentum of final state hadron
for different $z$ regions. To obtain the multiplicity distribution,
we divide the differential cross section by the leading order
total cross section, which contains an overall normalization
uncertainty. We hope in the future that the differential cross sections
for charged particles can be measured, and directly compared
to the theory calculations. From these plots, we can see that the
$P_{h\perp}$ distributions of the charged particle productions agree with the
simple Gaussian parameterization. Since $Q^2=3.14GeV^2$ is not so
different from the lower scale $Q_0^2=2.4GeV^2$, the evolution effects is
not evident from the above comparison. We notice that the
comparison at higher $z$ bin is not as good as moderate $z$ bins.
This difference has also been noted in Ref.~\cite{Airapetian:2012ki},
where the integrated multiplicity was compared to the quark
fragmentation function parameterization~\cite{deFlorian:2007aj}.
In particular, for $\pi^-$, the data seems larger than the calculation
based on the DSS fragmentation function in the large $z$ region.
Since we have followed the DSS fragmentation functions, our
predictions underestimated the experimental data at large $z$.
We hope future experiments can provide more data in this
region that we can constrain the theory more precisely.

The COMPASS experiment~\cite{Adolph:2013stb} covers a wider range
of $Q^2$. In particular, in the similar $x$-region, the overall $Q^2$ is about a factor
of $2$ larger than that for HERMES experiment. Therefore, there shall be
some $Q^2$ evolution effects in the $P_{h\perp}$ distribution for charged
hadron production. In Fig.~\ref{compass-m}, we compare our predictions
to the COMPASS data.
Again, the multiplicity is obtained by dividing the total cross section.
An additional normalization factor of 1.3 is included in these plots, which shall
account for difference in the luminosity measurements in
these two experiments and the possible higher order corrections.
From these plots, we find an overall agreement between the theory and
experiments. As we emphasized before, in this paper, we focus on
the kinematic region of moderate $x$ range: $x\sim 0.1$. We do
not compare our calculations with relative small-$x$ region of the
COMPASS data. In the future, we hope to come back to this region,
where small-$x$ effects in the transverse momentum distribution
have to be taken into account.

\begin{figure}[tbp]
\centering
\includegraphics[width=7cm]{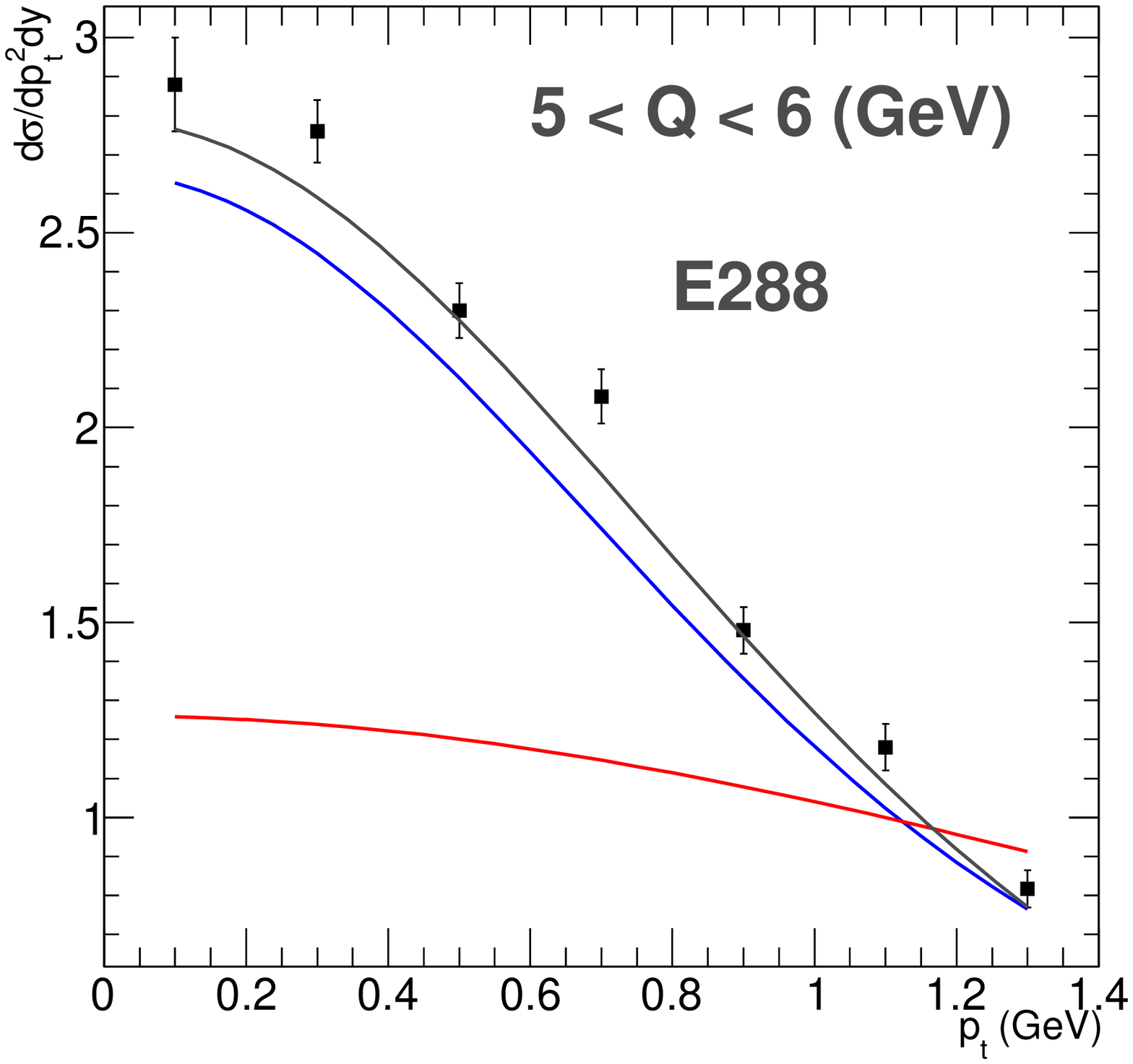}
\includegraphics[width=7cm]{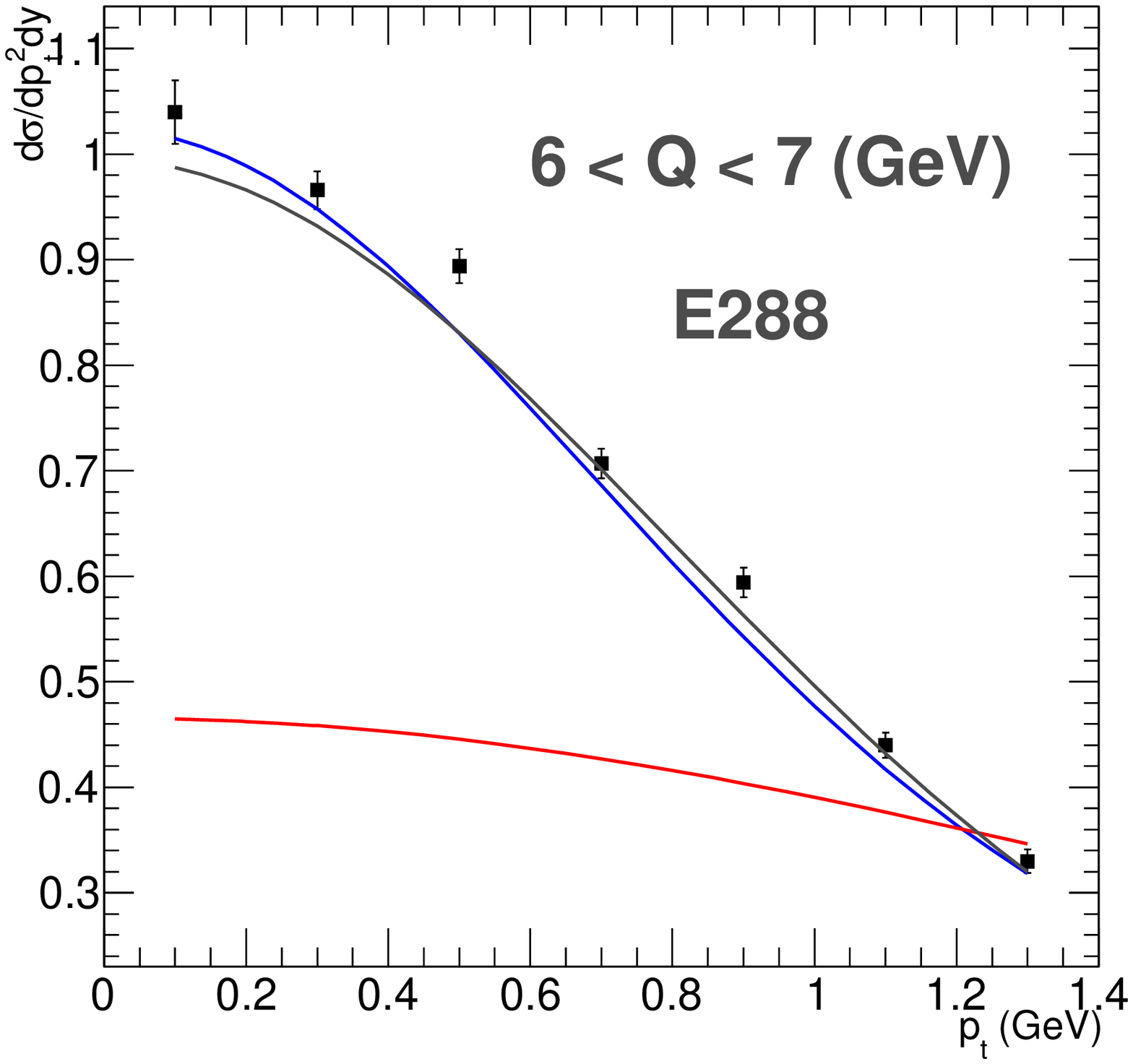}
\includegraphics[width=7cm]{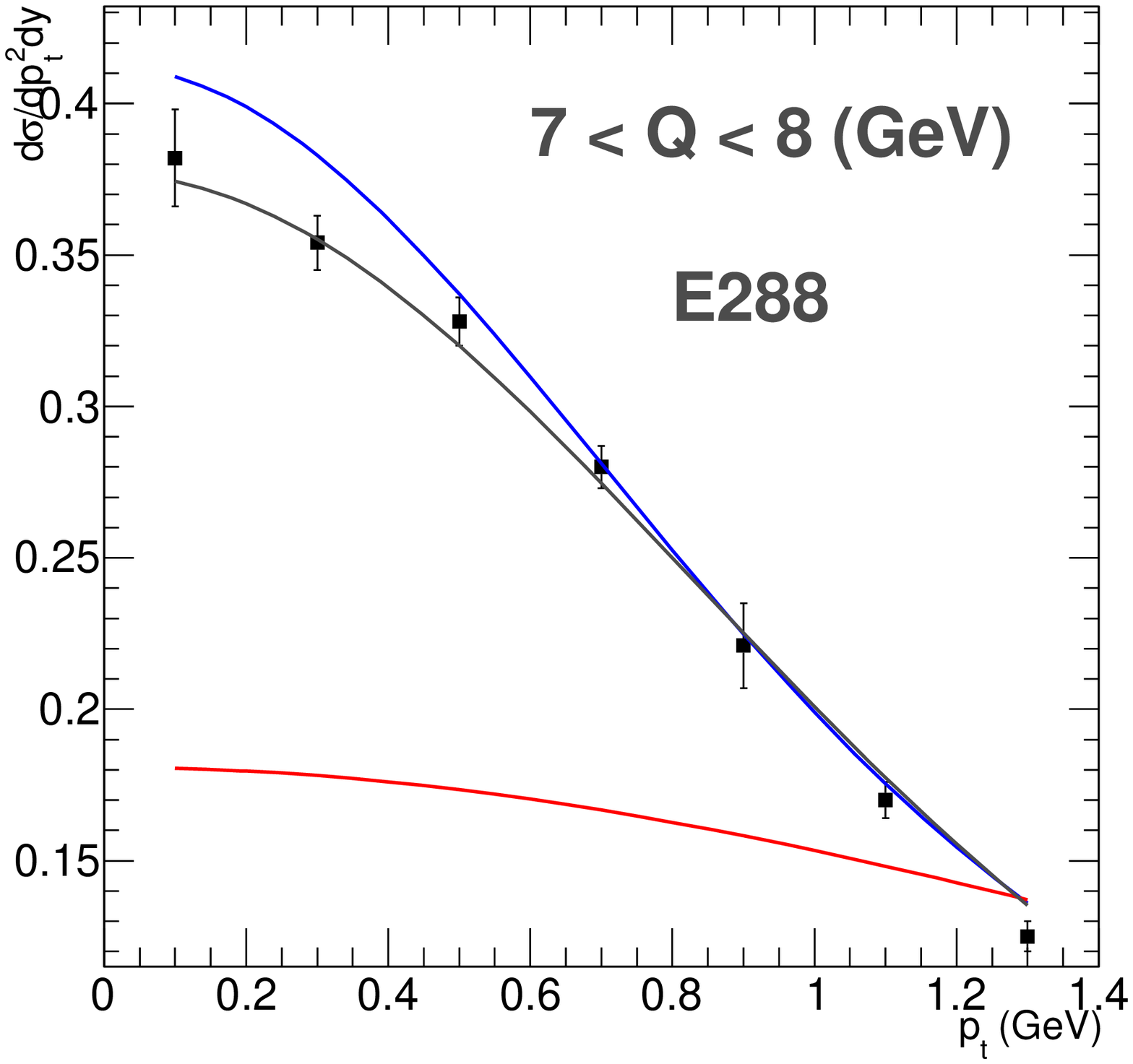}
\includegraphics[width=7cm]{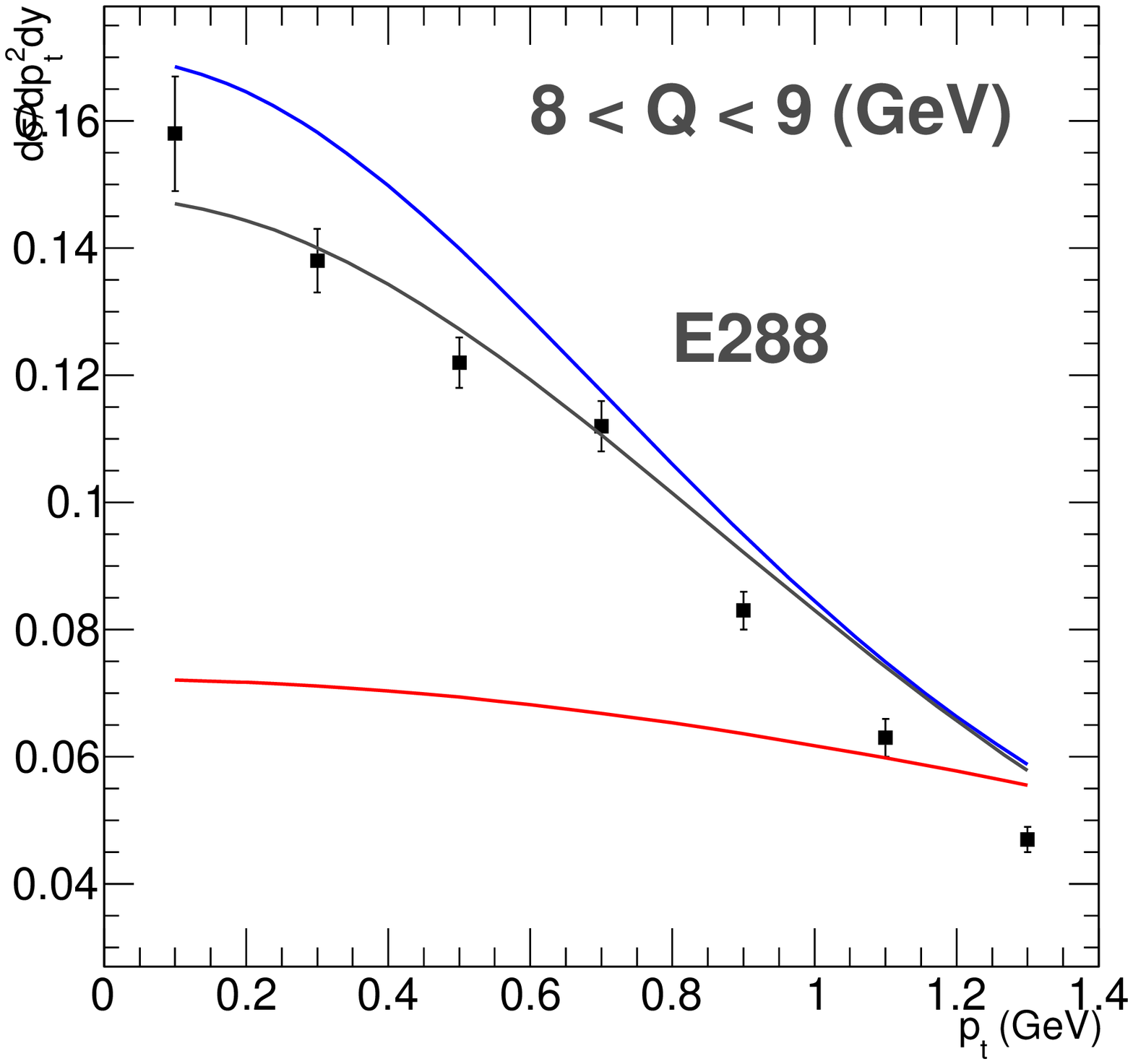}
\caption{Differential cross section for Drell-Yan lepton pair production
in hadronic collisions from E288 collaboration~\cite{Ito:1980ev} compared to the theory predictions with TMD evolution
from low energy scale $Q_0^2=2.4\textmd{GeV}^2$, Eqs.~(\ref{wuu-sy},\ref{sud},\ref{wuu0-sy}).
The predictions calculated from the TMDs from Rogers et al are also shown
as red curves. As a comparison, we also show predictions from the CSS
resummation with the integrated quark distribution set at the scale $\mu=Q_0$,
which gives similar results as distribution set at the scale $\mu=1/b_*$.}
\label{fig:e288}
\end{figure}

\begin{figure}[tbp]
\centering
\includegraphics[width=7cm]{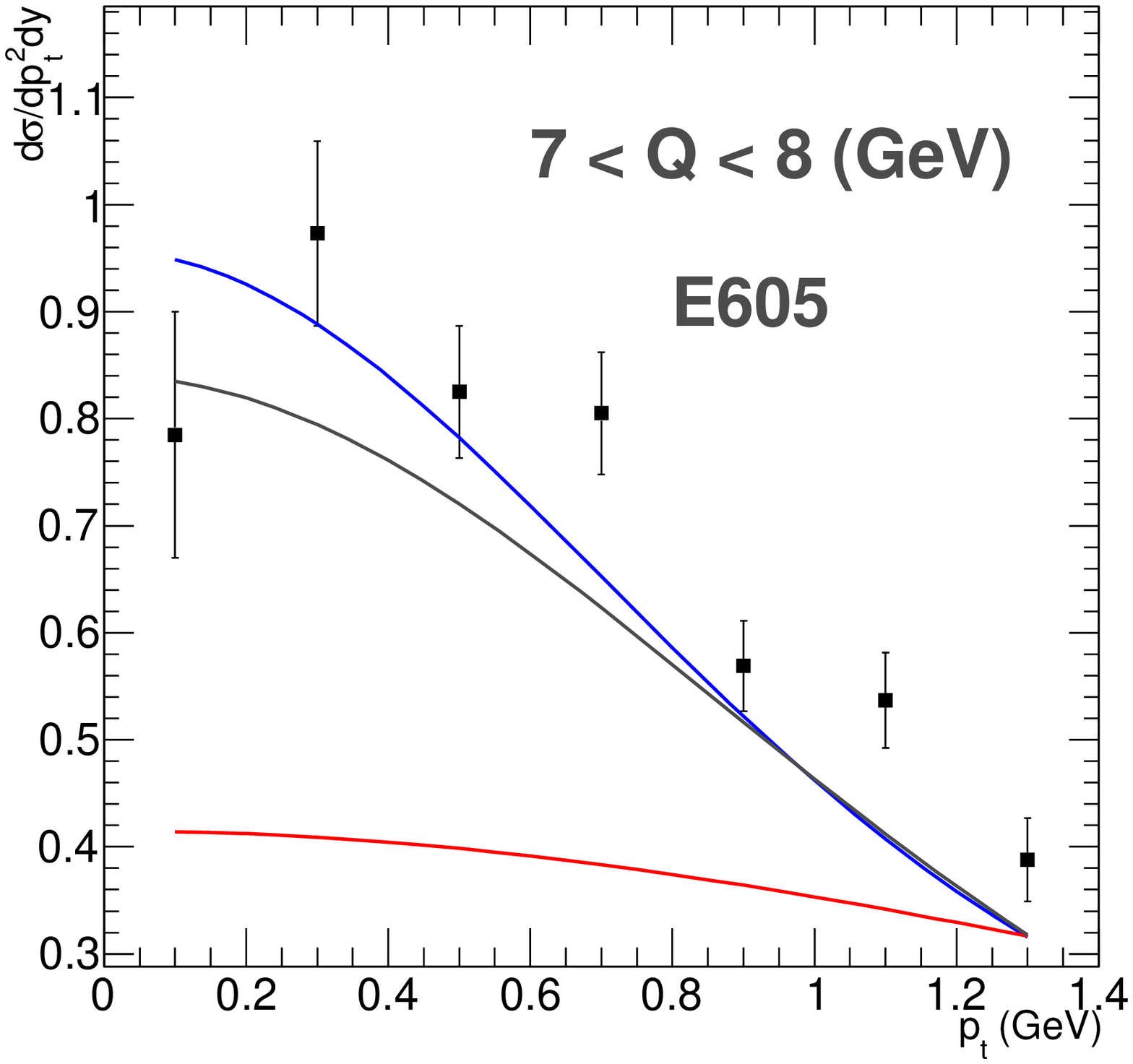}
\includegraphics[width=7cm]{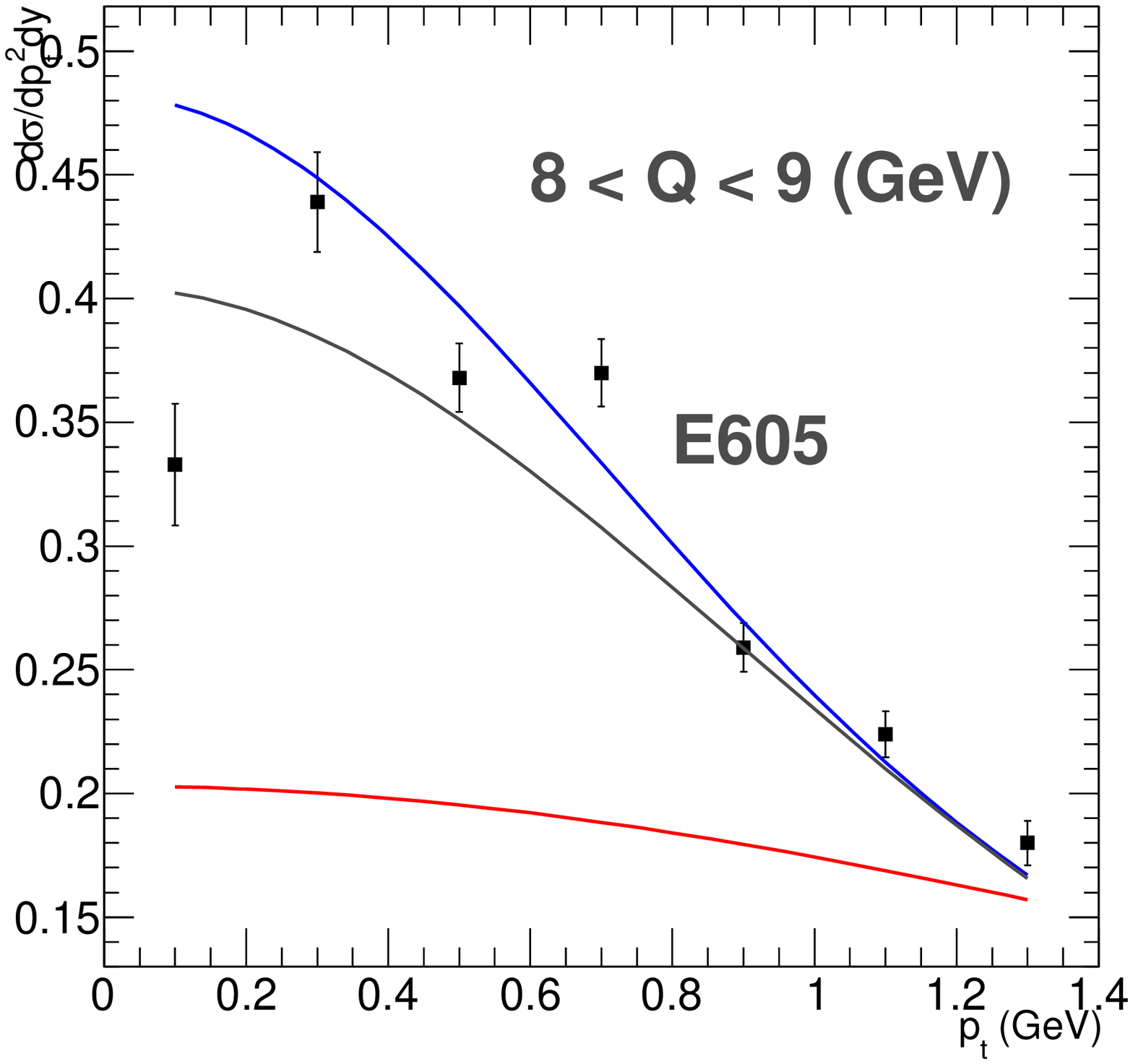}
\caption{Same as Fig.~\ref{fig:e288} for Drell-Yan data from E605 Collaboration~\cite{Moreno:1990sf}.}
\label{fig:e605}
\end{figure}

Now, we turn to the Drell-Yan experimental data, which spans even
higher $Q^2$ region.
To calculate the transverse momentum spectrum for this process,
we apply the universality of the TMD quark distributions, and the evolution
equation from $Q_0$ scale to higher $Q$.
In Fig.~\ref{fig:e288}, we plot the comparisons between the theory calculations with the experimental
data from E288 collaboration, where we have included an overall normalization to
account for the uncertainty in the luminosity in the experiment and higher order
corrections.The broadening effects for the Drell-Yan
processes are well reproduced by the evolution effects of Eqs.~(\ref{evo},\ref{sud}).
More comparisons between the theory calculations with the Drell-Yan data are
plotted in Fig.~\ref{fig:e605}. From these comparisons, we can clearly see that the evolution
effects calculated from the above equations can well describe the experimental
data on the unpolarized cross sections. In particular, in our calculations, the TMDs
are parameterized in a Gaussian form at low scale $Q_0$ with a single parameter
$g_0$, and the $Q^2$ dependence is calculated from the direct integral of the evolution
kernel. It is almost a parameter free prediction.

As we mentioned in the Introduction, the Drell-Yan data of Figs.~\ref{fig:e288} and
\ref{fig:e605} are also extensively studied in the CSS resummation formalism.
These data are actually used to constrain the associated non-perturbative
form factors. To demonstrate the matching between the Sun-Yuan approach
and the CSS resummation with $b_*$-prescription as we outlined
in the Introduction, in these two plots,
we also compare our calculations to the predictions from the CSS resummation
with KN parameterization for the non-perturbative form factor.
In these comparisons, we particularly focus on the normalized $p_\perp$ distribution
which is crucial to estimate the single spin asymmetries in the latter calculations (Sec. V).
Therefore, in the calculation, we neglect the scale dependence in the collinear
parton distributions in the CSS resummation~\footnote{We have also checked these results with the complete
implementation of the CSS resummation, and found they are in a reasonable
agreement, see also, e.g., the detailed calculations of Ref.~\cite{Konychev:2005iy}.},
where we have the following formula for the Drell-Yan lepton pair
production,
\begin{equation}
\widetilde {W}_{UU}(Q;b)=e^{-{\cal S}_{pert}(Q^2,b_*)-S_{NP}(Q,b)}
\Sigma_{q} f_q(z_1,Q_0) f_{\bar q}(z_2,Q_0) \ ,\label{wuu-css0}
\end{equation}
where ${\cal S}_{pert}$ with $A^{(1)}$ and $B^{(1)}$
and $S_{NP}(Q,b)$ take the forms as
in Eqs.~(\ref{spert}) and (\ref{blny0},\ref{kn}), respectively.
The scale of the integrated
quark distributions has been fixed at $\mu^2=Q_0^2=2.4GeV^2$.
These comparisons
aim at a consistent check between our approach and the CSS resummation.
For precise description of the experimental data, we need to implement the complete
CSS resummation with the integrated quark distribution setting
at the scale of $\mu=1/b_*$.
From these comparisons, we also see that the non-perturbative form factor
is the crucial part in the CSS resummation calculations for the $p_\perp$
spectrum for Drell-Yan processes.

Figs.~\ref{fig:e288} and \ref{fig:e605} provide an important evidence that we can match
the different evolution formulas: at relative lower region of $Q^2$ we can apply the
evolution equations of Eqs.~(\ref{sunyuan},\ref{wuu0-sy}); at higher region of $Q^2$ we apply the
CSS resummation with the KN non-perturbative form factors; in the
overlap region, both can be applied and present a consistent description
of the experimental data. From these figures, we also observe that the CSS
resummation with KN form factors describes better the
experimental data than Sun-Yuan approach of the direct integral of the TMD
evolution kernel. This indicates that we shall switch to the CSS resummation
at high $Q^2$ Drell-Yan process, in particular, for $W/Z$ boson production.

The support of the matching from the above analysis encourage
us to extend the above method to the Sivers single spin asymmetries
in the SIDIS and Drell-Yan processes. However, the Sivers asymmetries
are only observed in the SIDIS processes from HERMES and COMPASS
experiments. Therefore, the measurements in the planed Drell-Yan processes
will not only provide crucial test of the sign change between these
two processes, but also provide unique opportunities to study the
energy evolution for the spin asymmetries.
This QCD dynamics shall be extensively investigated
in the planed electron-ion colliders where wide coverage of $Q^2$ will
ultimately help us understand the physics to great precision~\cite{Boer:2011fh}.

\subsection{Rogers et al. Approach}

Before we turn to the Sivers single spin asymmetry study in our
calculation, in this subsection, we comment on the approach used
in Rogers et al.
By following Collins' new definition for the TMD quark distributions,
Rogers {\it et al.} derived an evolution equation the TMDs.
However, in terms of cross section calculations, there is a simple
way to understand the evolution derived by Rogers et al.
For example, we can write down two equations by employing the
CSS resummation with $b_*$-prescription,
\begin{eqnarray}
\widetilde {F}_{\rm sivers}^\alpha(Q;b)&=& e^{-{\cal S}_{pert}(Q^2,b_*)-S_{NP}(Q,b)}\widetilde{F}_{\rm sivers}^\alpha(C_1/b,b)\nonumber\\
\widetilde {F}_{\rm sivers}^\alpha(Q_L;b)&=& e^{-{\cal S}_{pert}(Q_L^2,b_*)-S_{NP}(Q_L,b)}\widetilde{F}_{\rm sivers}^\alpha(C_1/b,b)  \ .
\end{eqnarray}
By combining the above two equations, we find that $F(Q)$ can be written in terms
of $F(Q_L)$,
\begin{eqnarray}
\widetilde {F}_{\rm sivers}^\alpha(Q;b)&=&e^{-\left({\cal S}_{pert}(Q,b_*)-{\cal S}_{pert}(Q_L,b_*)\right)}\nonumber\\
&&\times e^{-\left({ S}_{NP}(Q,b)-S_{NP}(Q_L,b)\right)}\nonumber\\
&&\times \widetilde {F}_{\rm sivers}^\alpha(Q_L;b)\ .
\end{eqnarray}
The second exponential factor can be easily calculated $e^{-g_2b^2\ln\frac{Q}{Q_L}}$.
It is this factor that leads to strong $Q$ dependence in the SSAs calculated in this
approach in the relative low $Q$ region. However,
this behavior over-predicts broadening effects in the Drell-Yan lepton
pair production as compared to the experimental data. In other words, the adoption
used by Rogers et al is not supported by the experimental data. In particular,
the flat distribution of the transverse momentum as shown in Figs.~\ref{fig:e288} and
\ref{fig:e605} from Rogers et al. will lead to almost vanishing Sivers single spin
asymmetries in the Drell-Yan processes in this transverse momentum region.
Therefore, all the previous studies following this approach have to be re-examined,
including, most importantly, the energy evolution for the Sivers single spin asymmetries.

\section{Quark Sivers Functions from Combined Analysis of HERMES and COMPASS Data}

To predict the Sivers single spin asymmetries in Drell-Yan processes, we need to constrain the
quark Sivers functions from the current experimental measurements in SIDIS processes.
In this section, we will perform a combined analysis of these measurements, and obtain
constraints on the quark Sivers functions.
A couple of comments are in order before we perform the combined fit. Our analysis of
quark Sivers functions depends on the following
assumptions: First, we assume that the systematics of the HERMES and COMPASS
experiments are well under control. Both experiments have observed
sizable Sivers asymmetries, in particular for positive charged hadrons ($\pi^+$ for HERMES).
Second, our analysis relies on the applicability of the TMD
factorization in these kinematics. Third, the
experimental data are used to fit the quark Sivers functions, where we assume that
it is the only contribution to the observed azimuthal asymmetries. All these important
issues will be thoroughly addressed in the future SIDIS experiments, including
the 12 GeV upgrade of JLab and the planed electron-ion collider.
In addition, the relevant spin asymmetries in the Drell-Yan lepton pair
production in $pp$ collisions in the proposed experiments
shall also provide important information on the quark Sivers functions. We will
discuss this in more details in Sec. V.

As we have showed in the last section, the evolution equations
we derived for the moderate $Q^2$ range, can well describe the unpolarized
differential cross sections in SIDIS and Drell-Yan processes which cover $
2.5 < Q^2 < 100\textmd{GeV}^2$. This demonstrates that the dominant evolution effects are
taken into account in our derivations. In the following, we extend this approach
to the Sivers single spin asymmetries in SIDIS process, and perform a combined
fit to the HERMES/COMPASS data with the TMD evolution effects. Since the
evolution kernel and the form of the solution is spin-independent,
the structure functions at higher scale $Q$ are calculated from those from
lower scale $Q_0$ with the Sudakov form factor,
where $M=0.94\textmd{GeV}$ is a normalization scale, and
we have chosen an additional parameter $g_s$ for transverse momentum dependence
and the fragmentation part remains the same.
In the following, we will fit Sivers functions by the forms:
\begin{eqnarray}
\Delta f_u^{\rm sivers}&=&N_u x^{\alpha_u} (1-x)^{\beta} \frac{(\alpha_u+\beta)^{\alpha_u+\beta}}
{\alpha_u^{\alpha_u} \; \beta^{\beta}}f_u(x,\mu=Q_0)\ , \nonumber\\
\Delta f_d^{\rm sivers}&=&N_d x^{\alpha_d} (1-x)^{\beta} \frac{(\alpha_d+\beta)^{\alpha_d+\beta}}
{\alpha_d^{\alpha_d} \; \beta^{\beta}}f_d(x,\mu=Q_0) \ ,\nonumber\\
\Delta f_{(\bar u,\bar d,s)}^{\rm sivers}&=&N_{(\bar u,\bar d,s)} x^{\alpha_s} (1-x)^{\beta} \frac{(\alpha_s+\beta)^{\alpha_s+\beta}}
{\alpha_s^{\alpha_s} \; \beta^{\beta}}f_{(\bar u,\bar d,s)}(x,\mu=Q_0) \ .
\end{eqnarray}
where $f_u$, $f_d$ and $f_{(\bar u,\bar d,s)}$ are integrated quark distributions at
initial scale $\mu=Q_0$. As we have showed in the last section, that $\widetilde{F}_{UU}$
from the above equations can describe well the transverse momentum distributions
in SIDIS experiments in HERMES/COMPASS kinematics. Therefore, the observed
Sivers asymmetries can be used to constrain the quark Sivers functions.

\begin{figure}[tbp]
\centering
\includegraphics[width=7.cm]{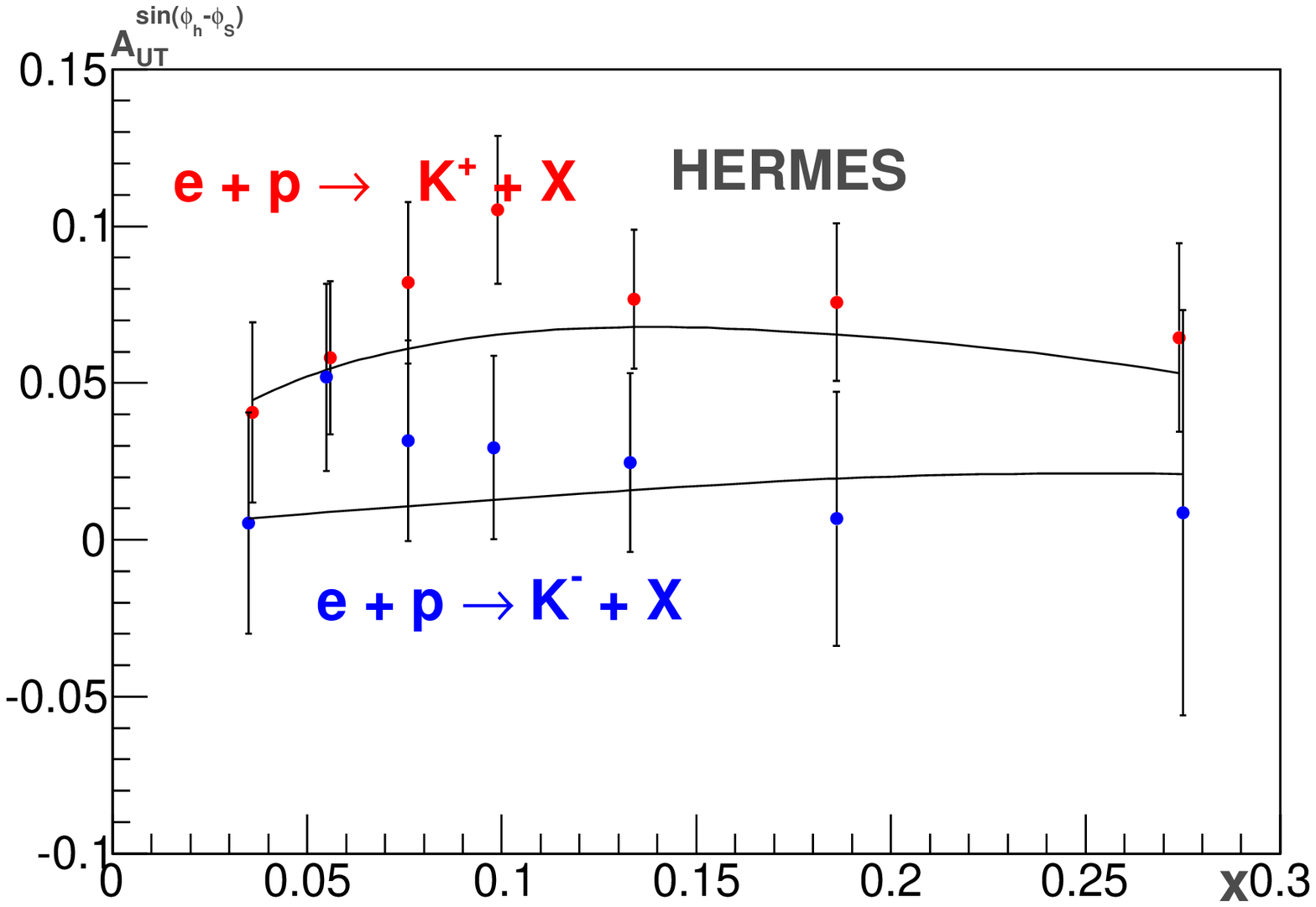}
\includegraphics[width=7.cm]{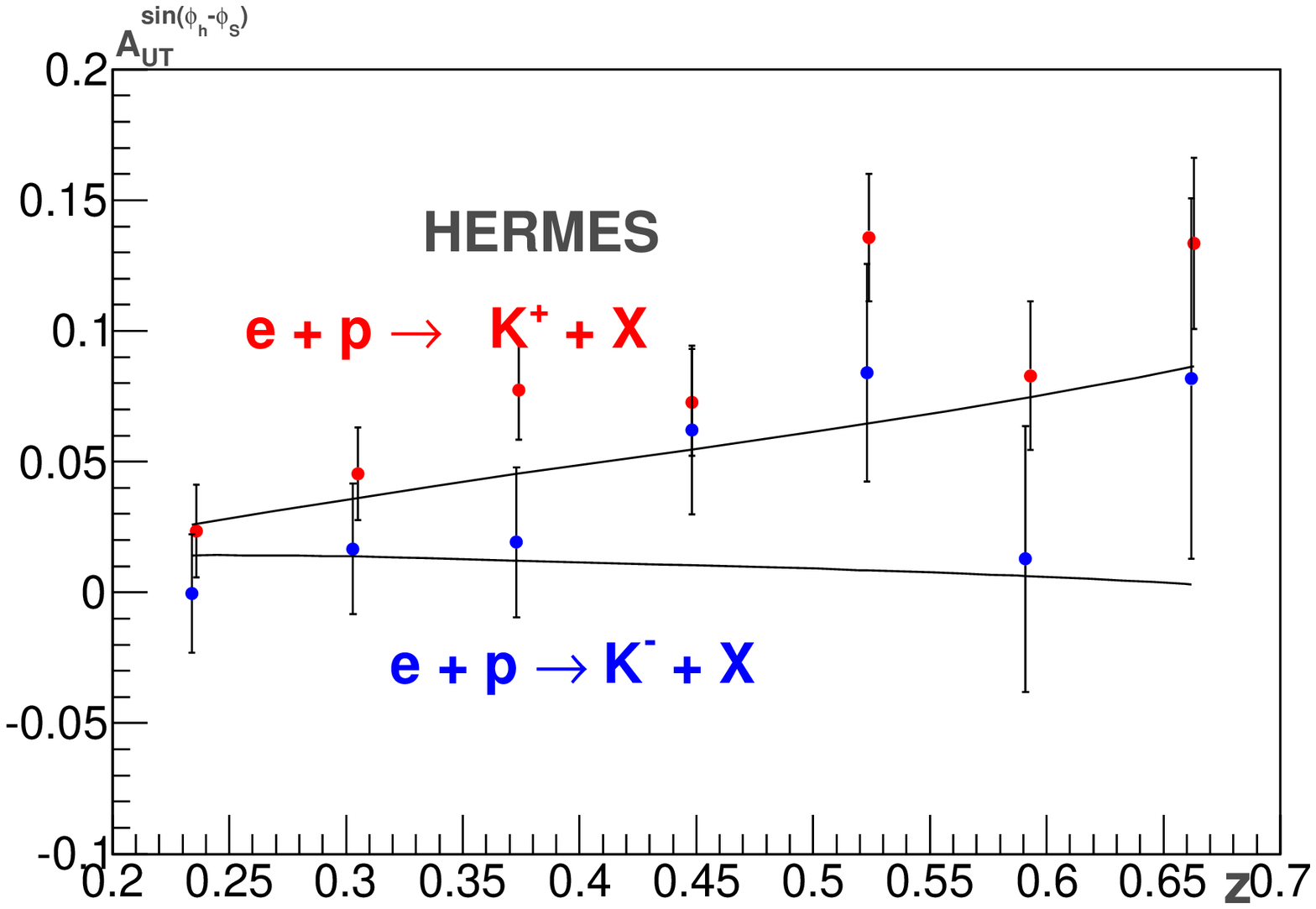}
\includegraphics[width=7.cm]{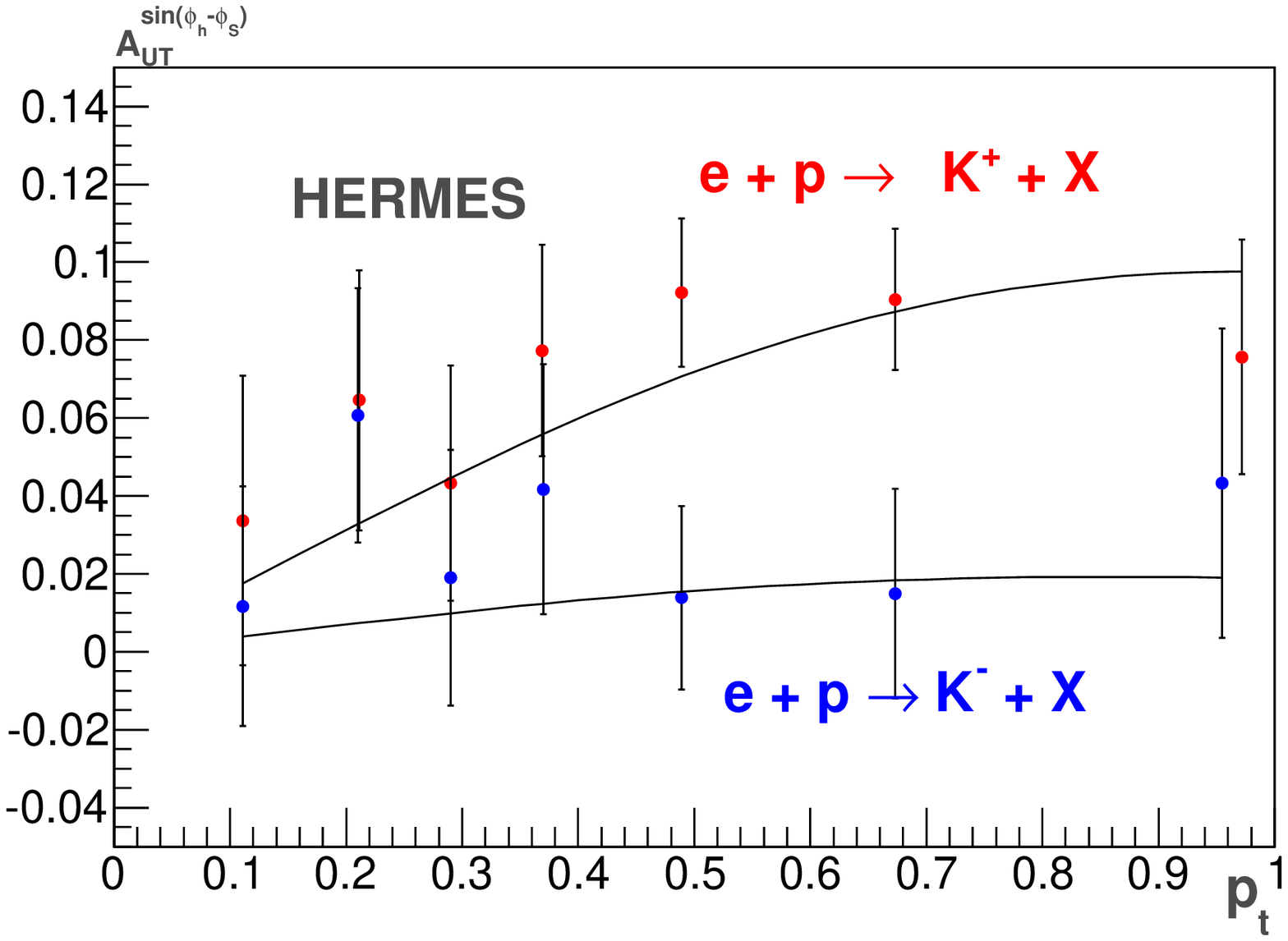}
\includegraphics[width=7.cm]{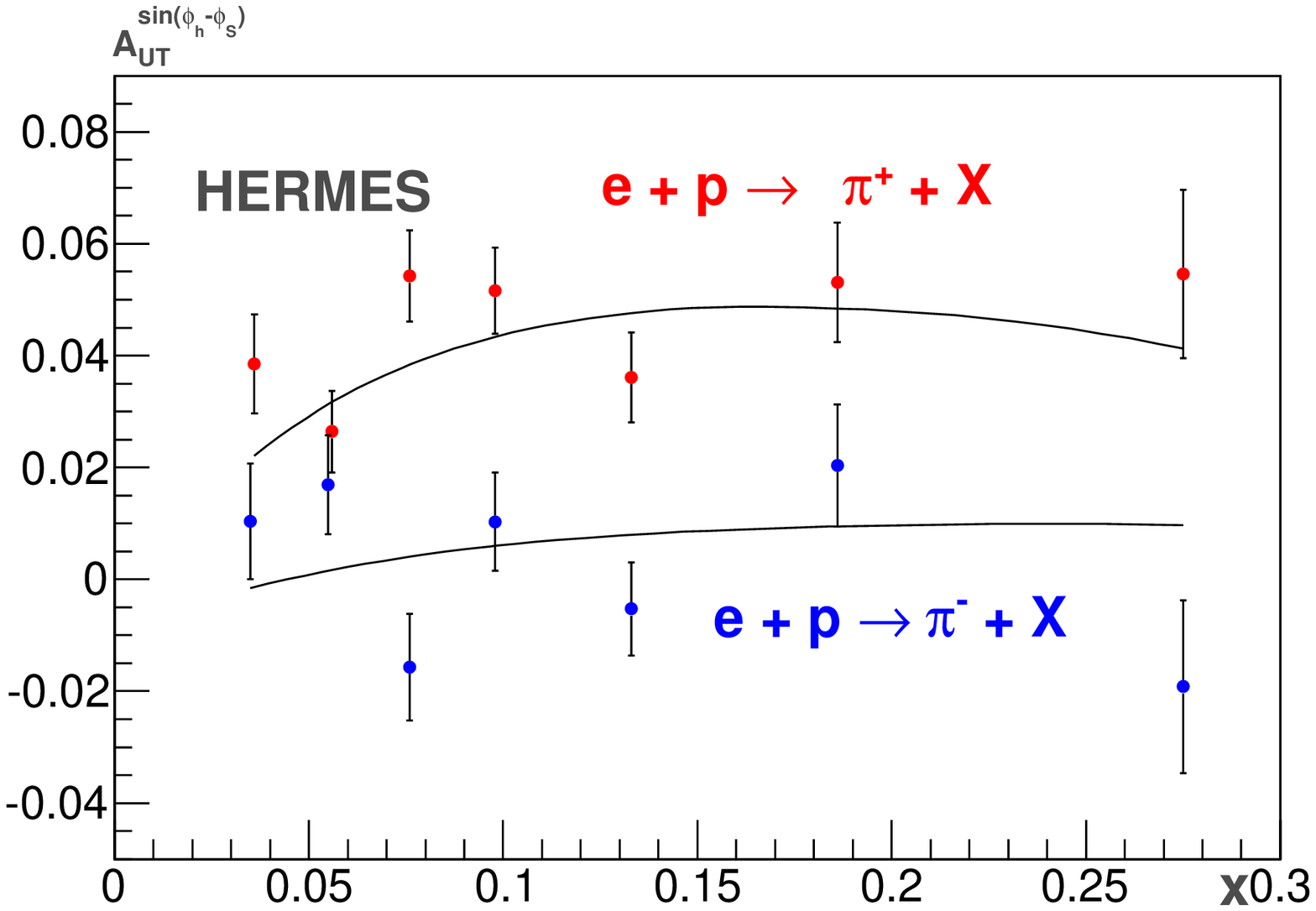}
\includegraphics[width=7.cm]{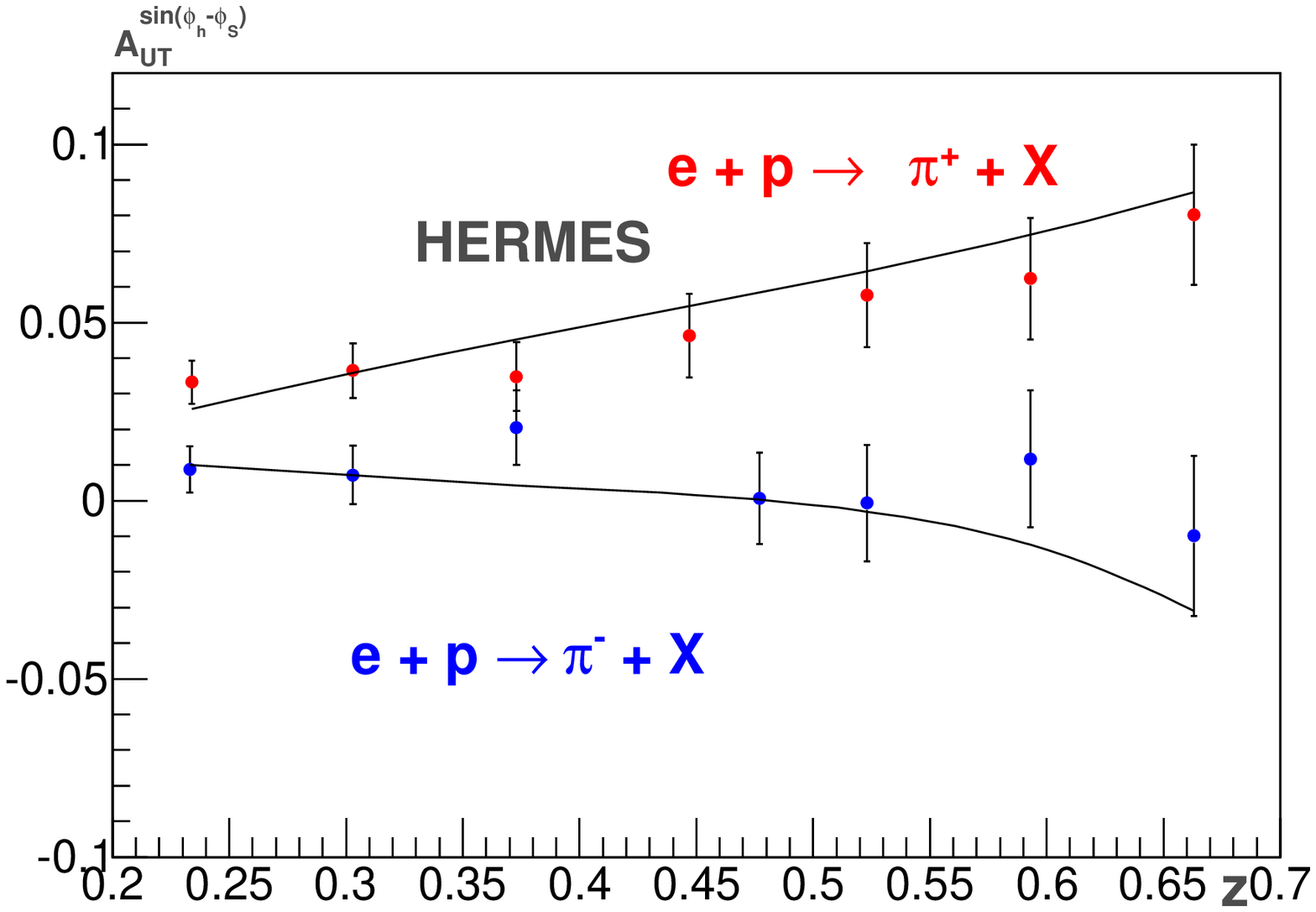}
\includegraphics[width=7.cm]{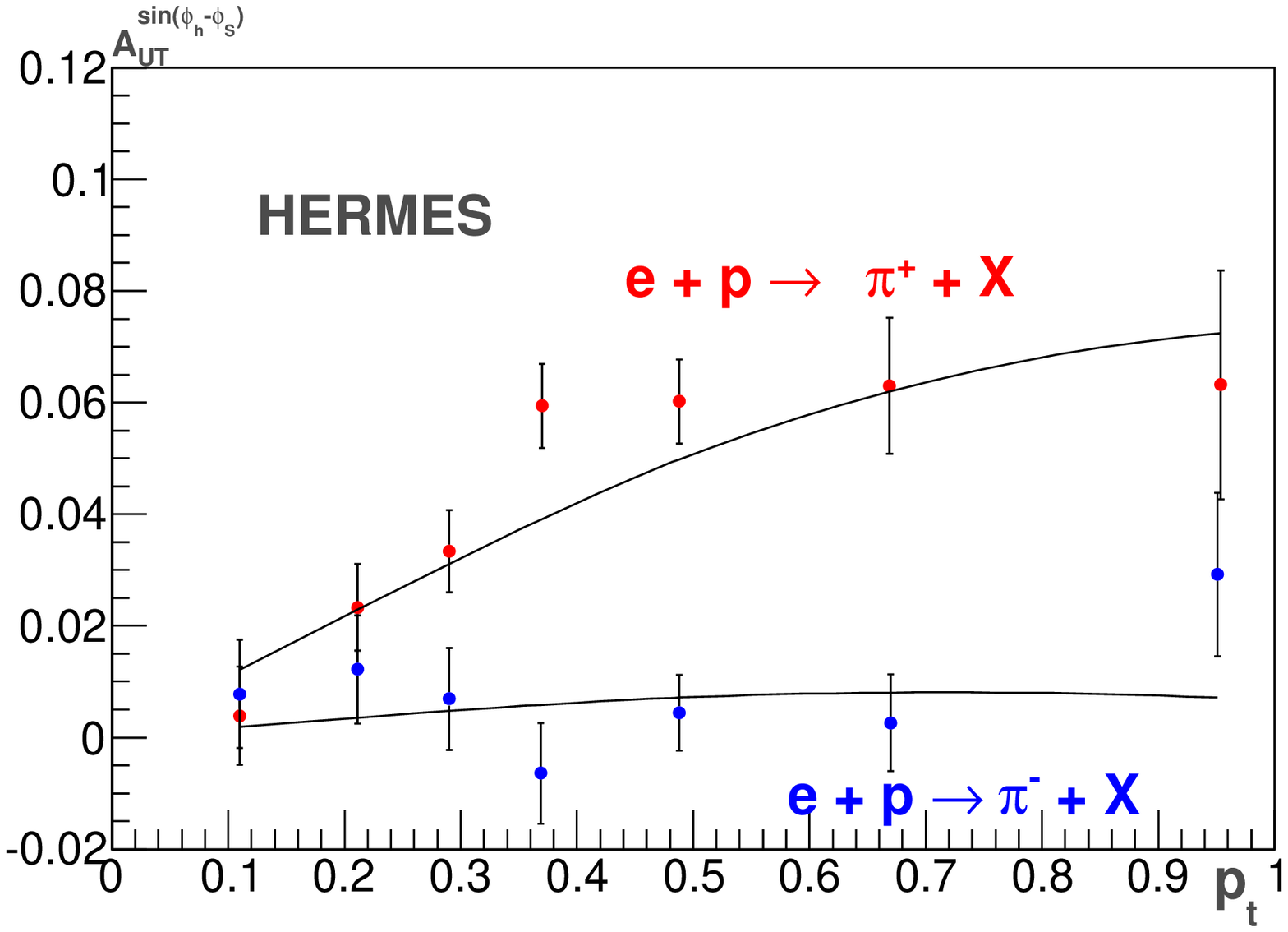}
\includegraphics[width=7.cm]{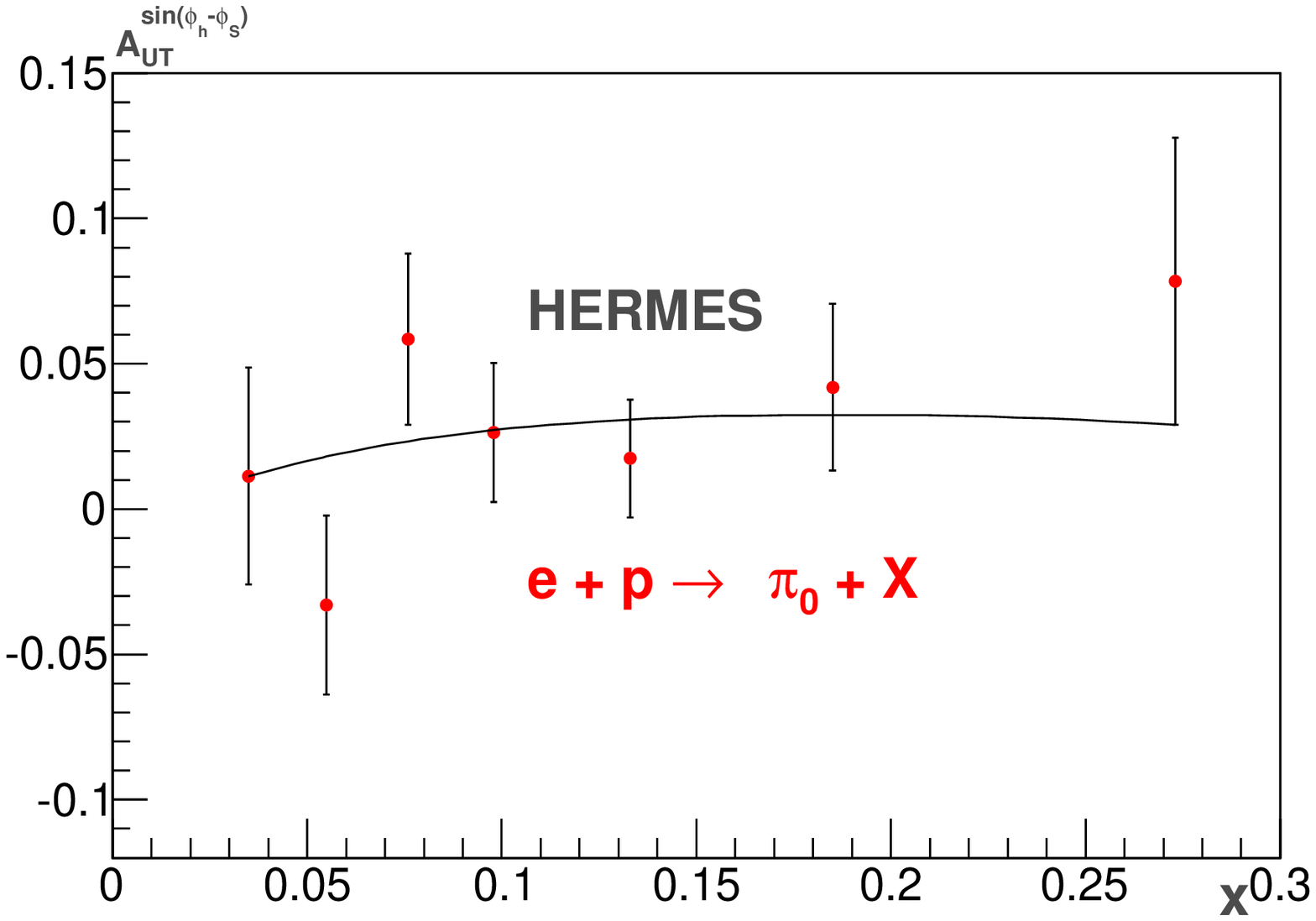}
\includegraphics[width=7.cm]{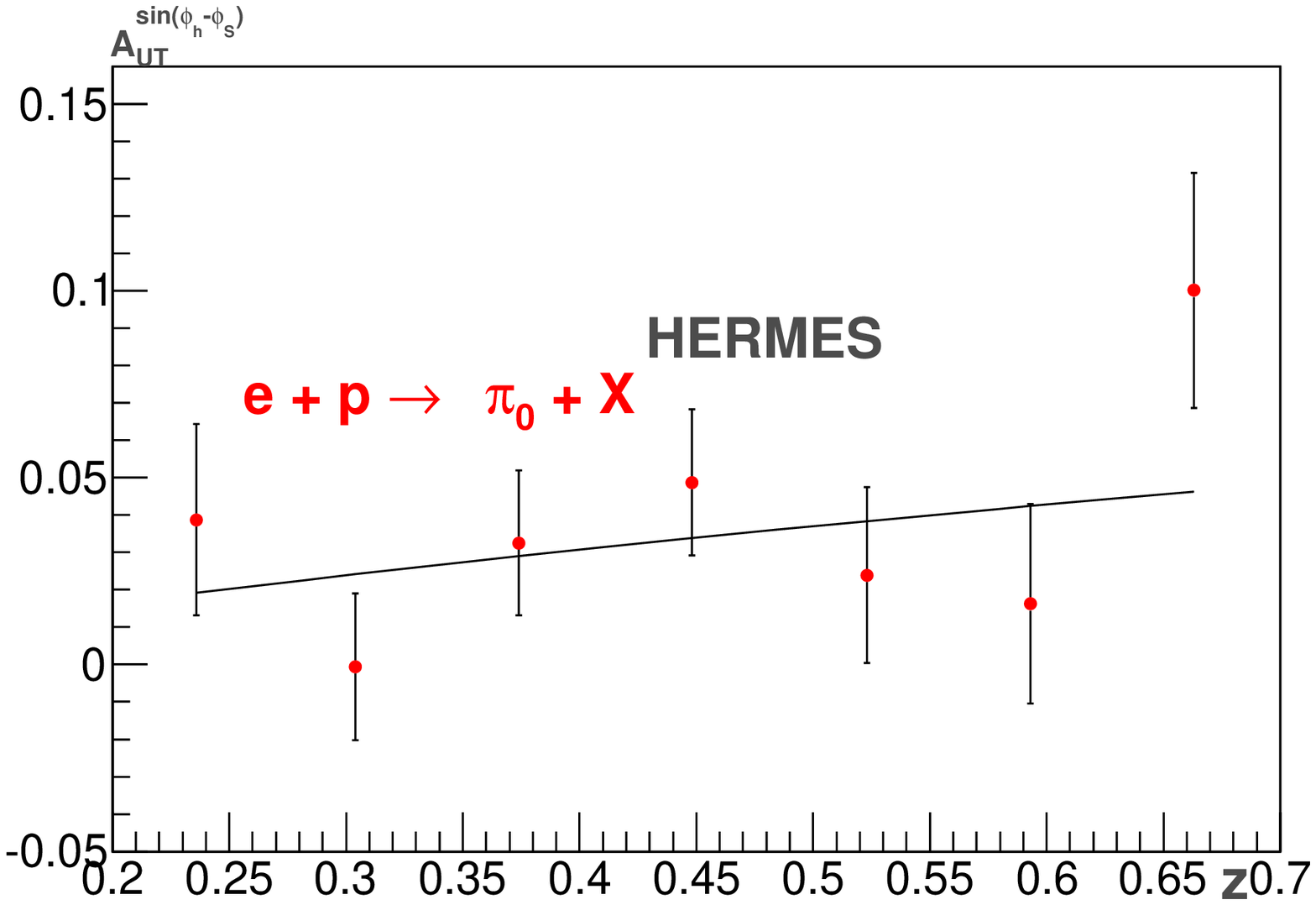}
\includegraphics[width=7.cm]{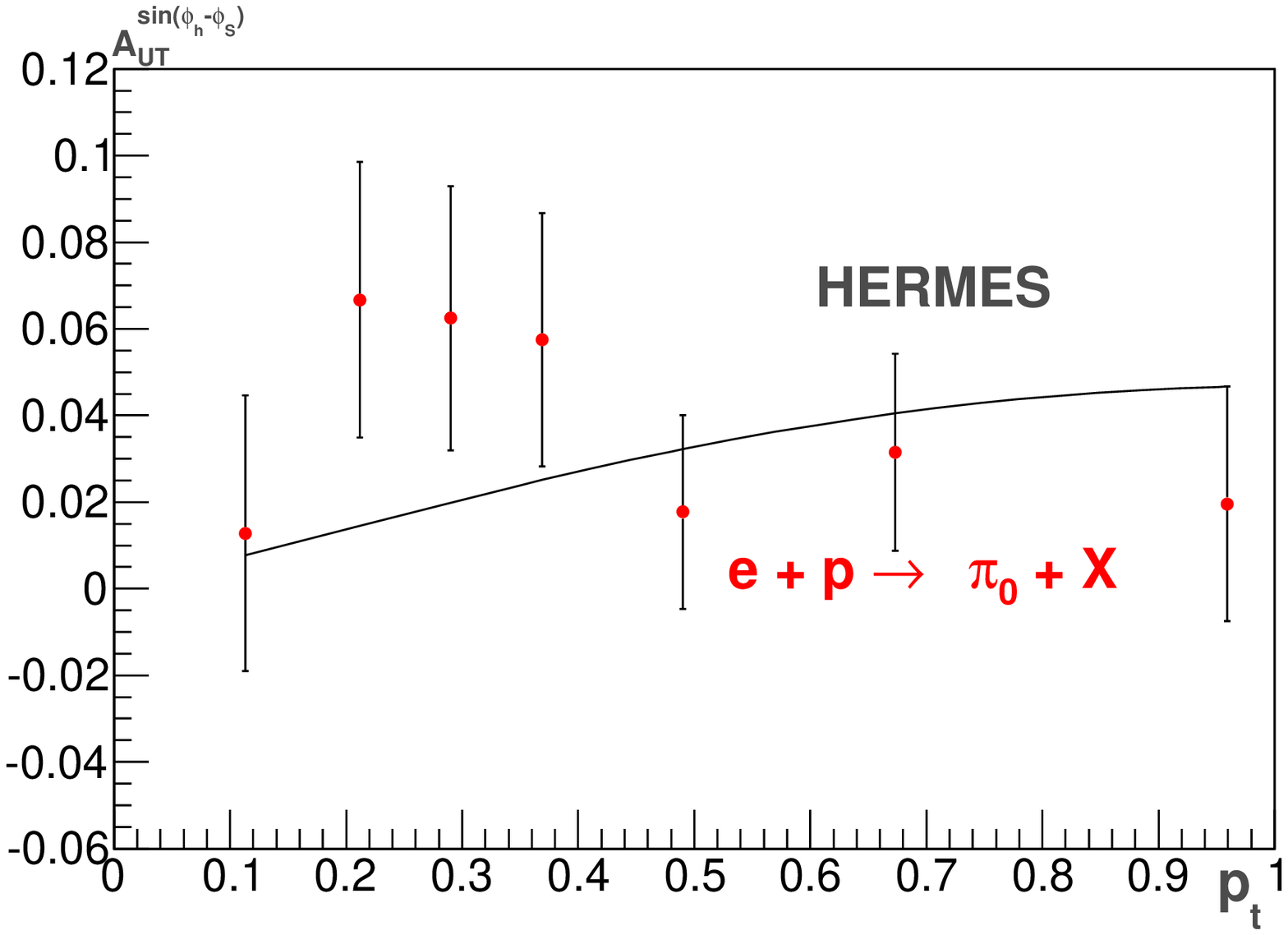}
\caption{Comparison of our fits with the experimental
data for the Sivers asymmetries as functions of $x_B$, $z_h$, and
$P_{h\perp}$: HERMES data ~\cite{Airapetian:2009ae}.}
\label{fig:hermes-sivers}
\end{figure}

\begin{figure}[tbp]
\centering
\includegraphics[width=7.1cm]{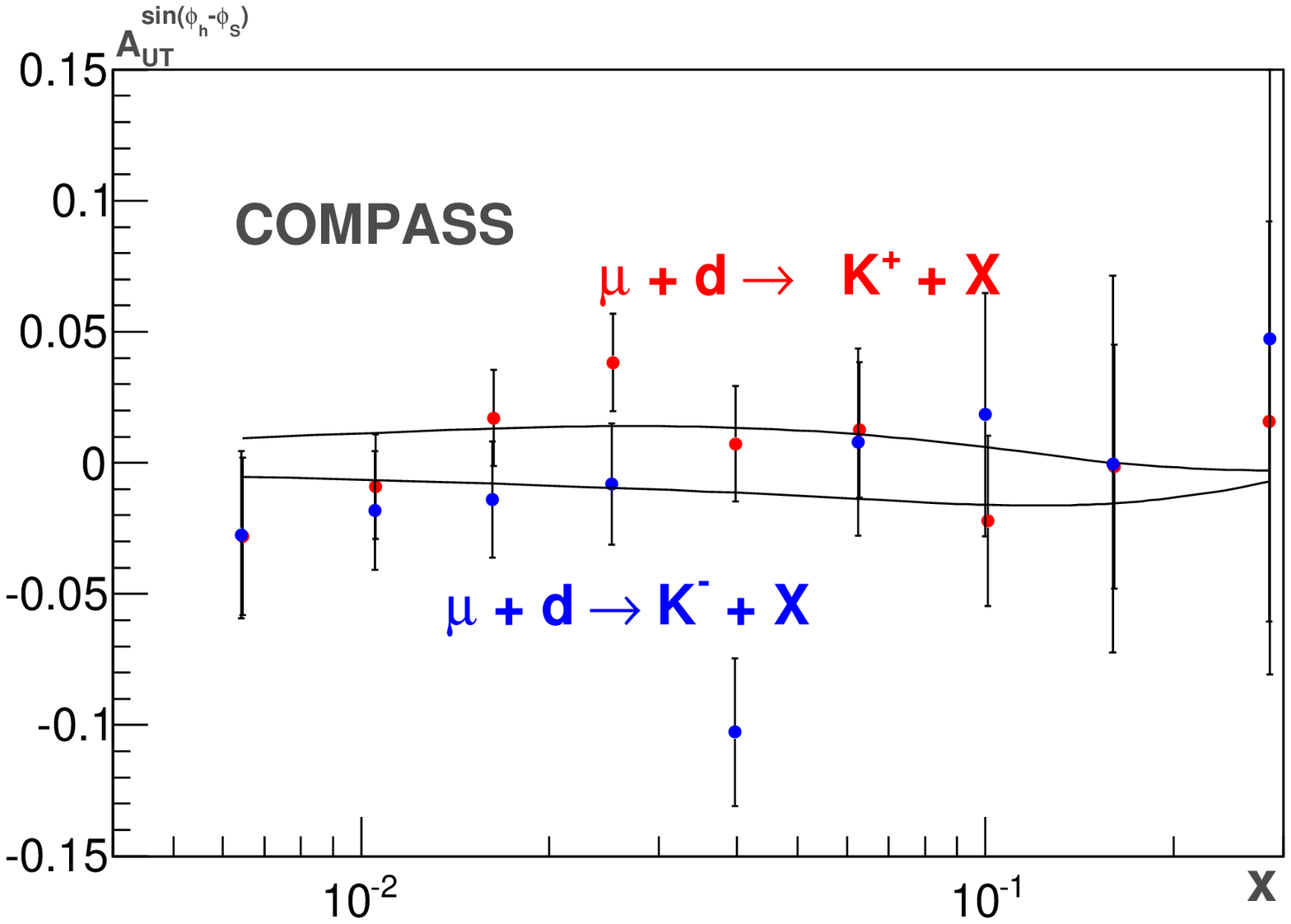}
\includegraphics[width=7.1cm]{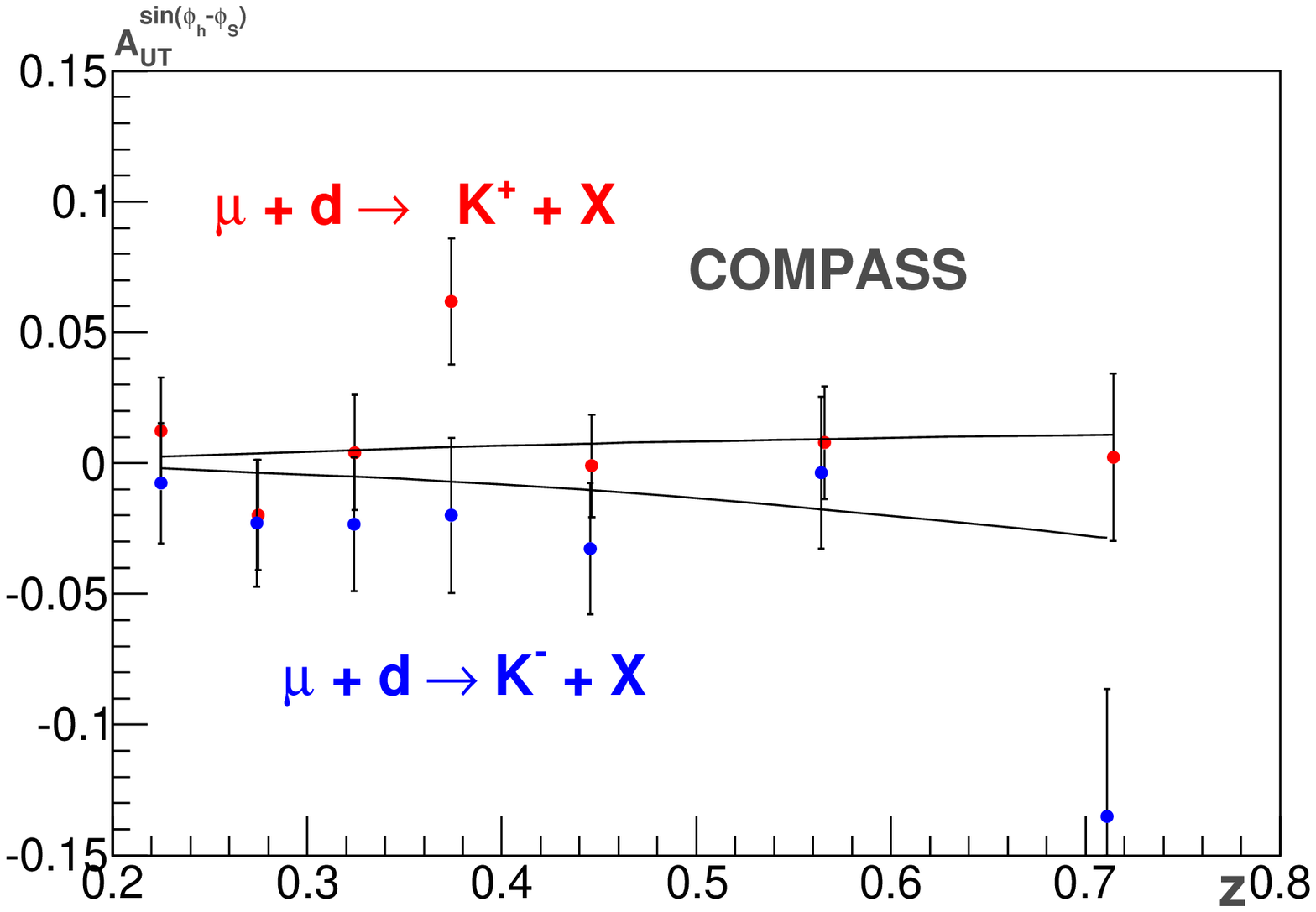}
\includegraphics[width=7.1cm]{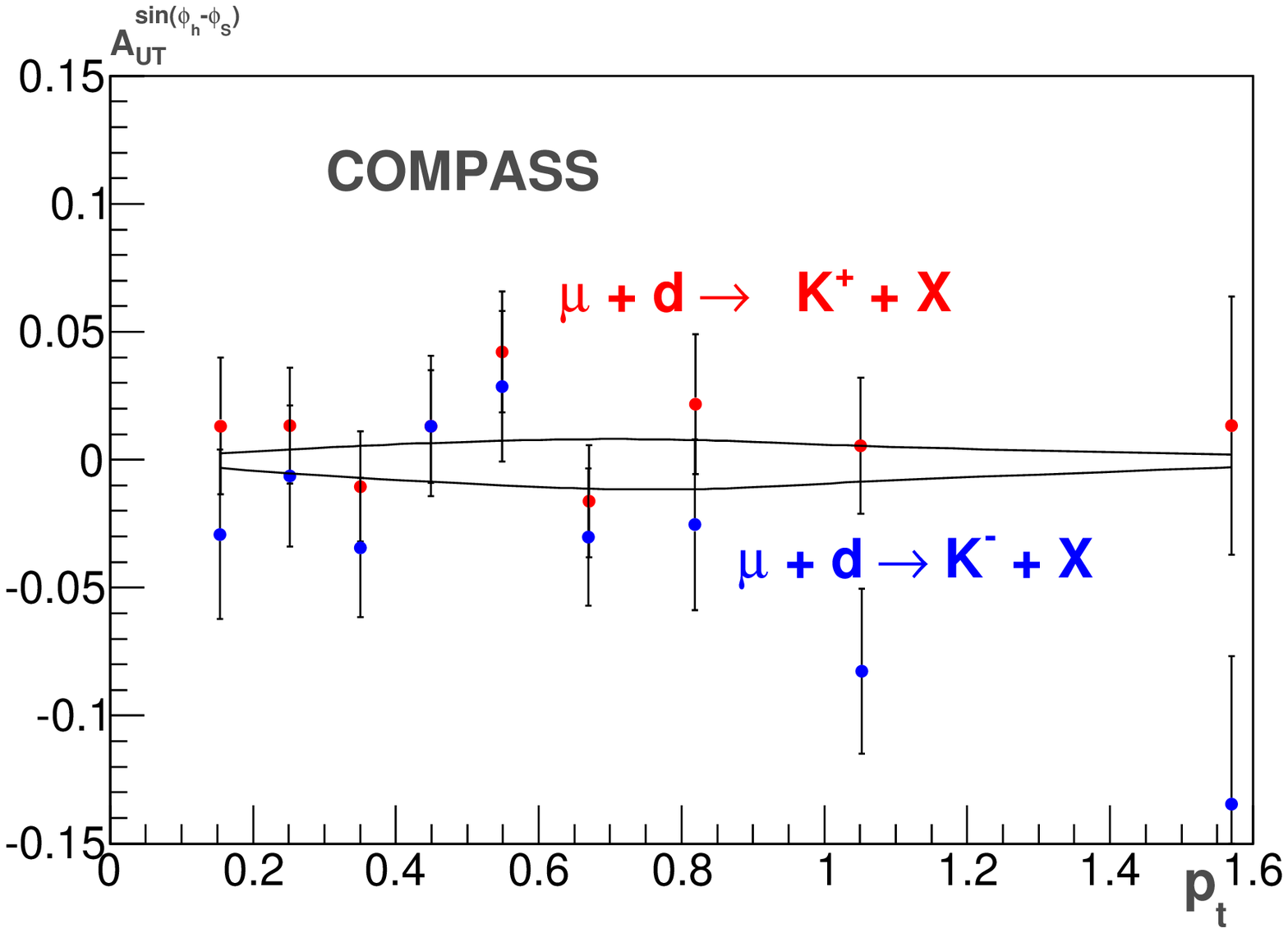}
\includegraphics[width=7.1cm]{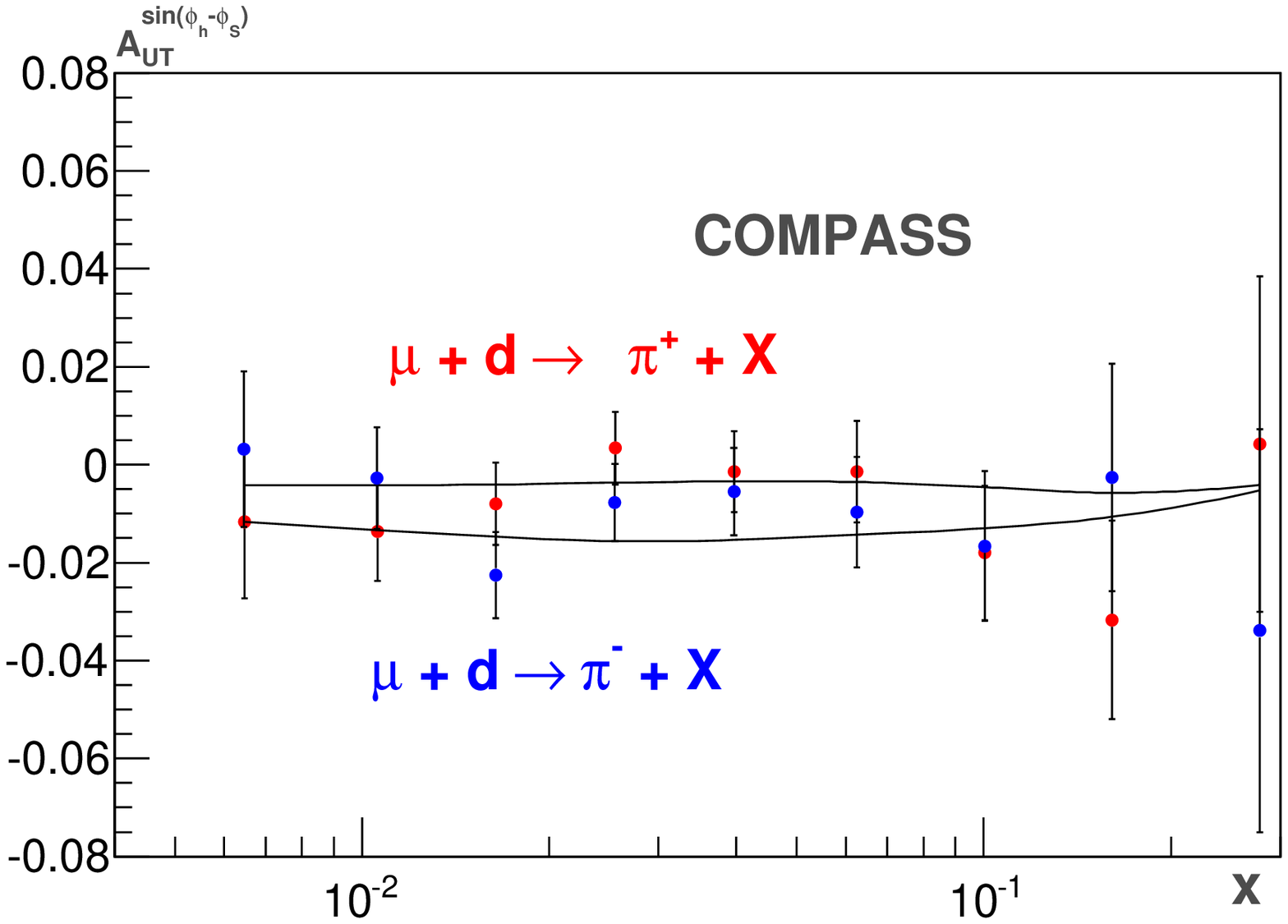}
\includegraphics[width=7.1cm]{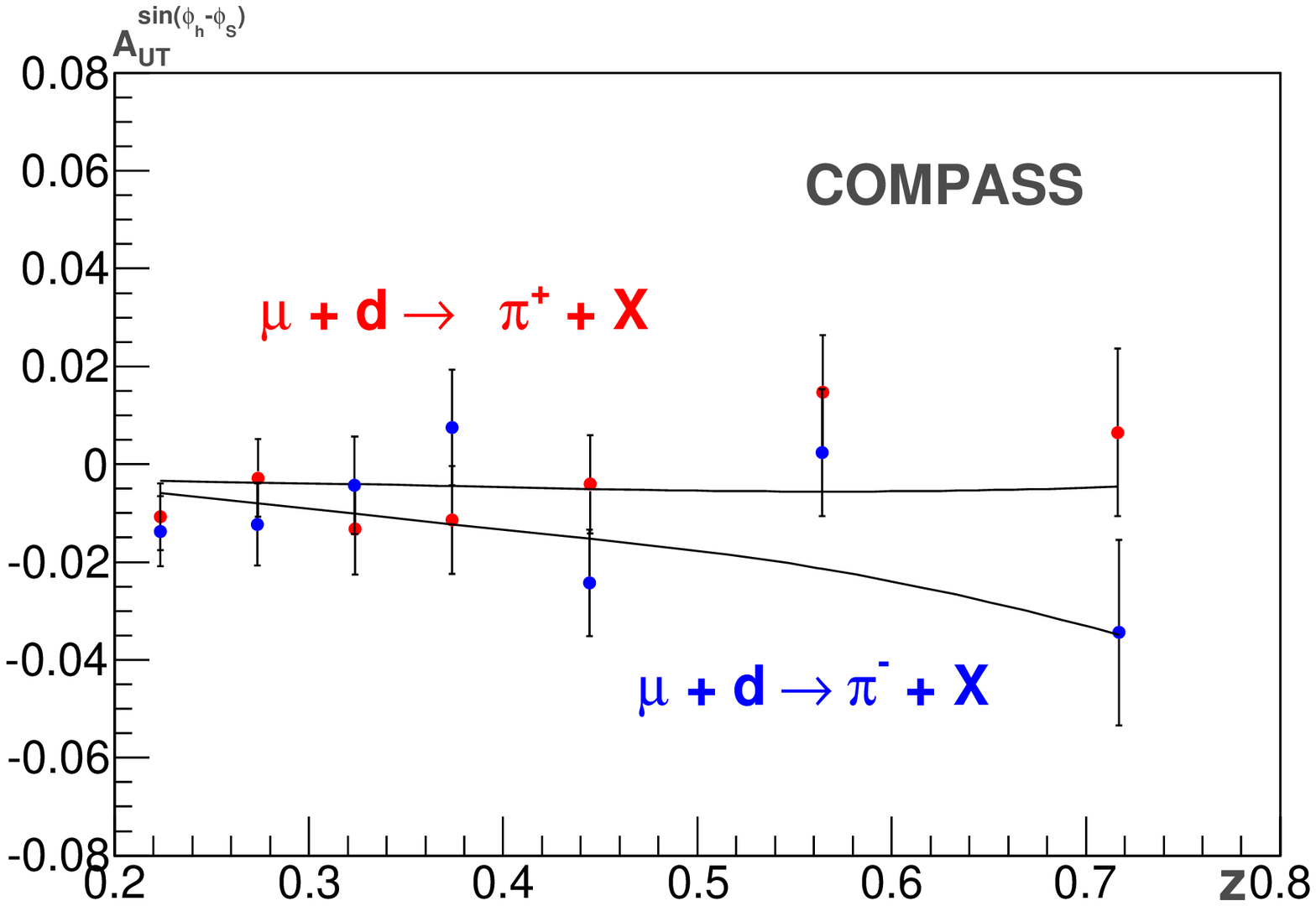}
\includegraphics[width=7.1cm]{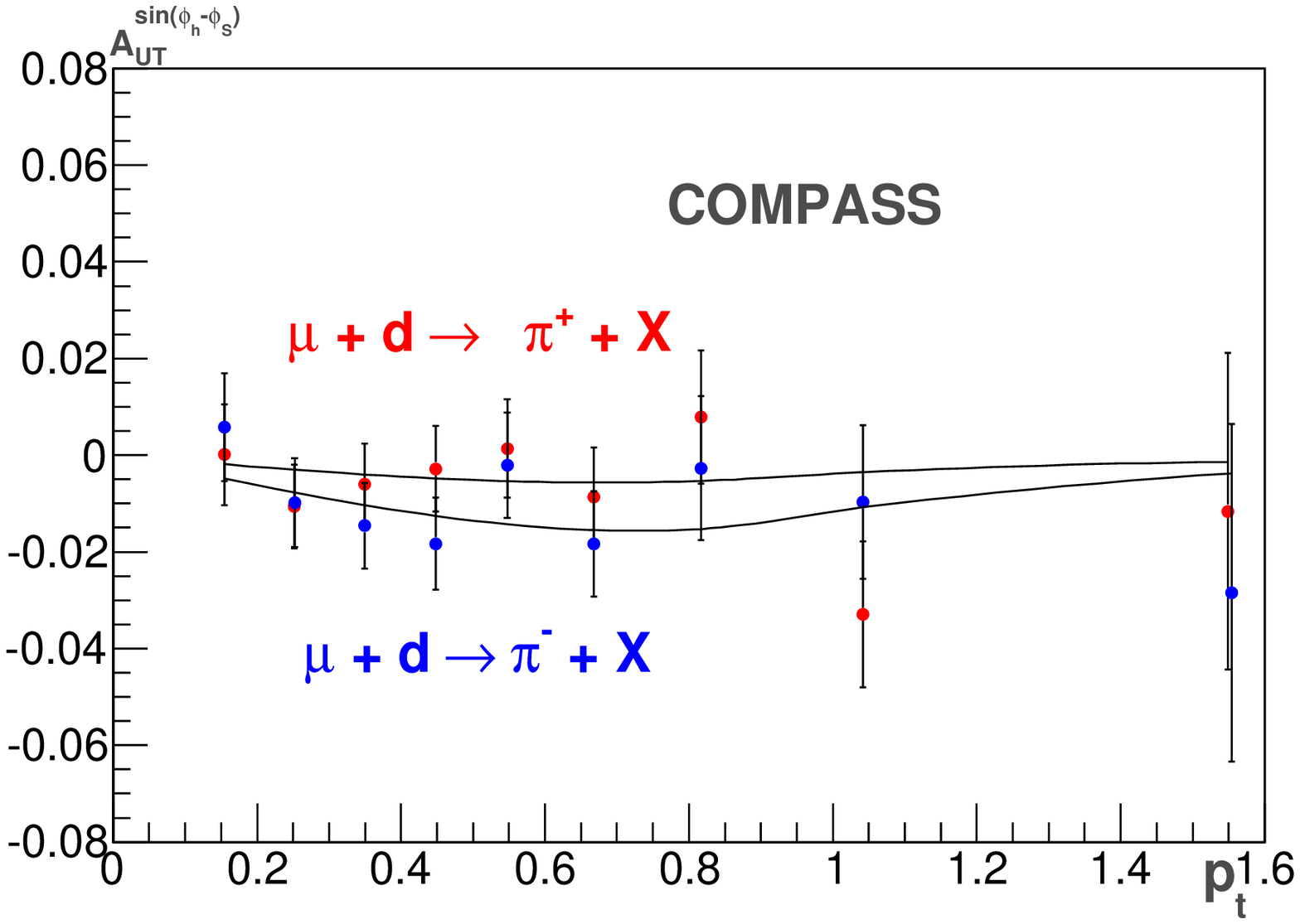}
\includegraphics[width=7.1cm]{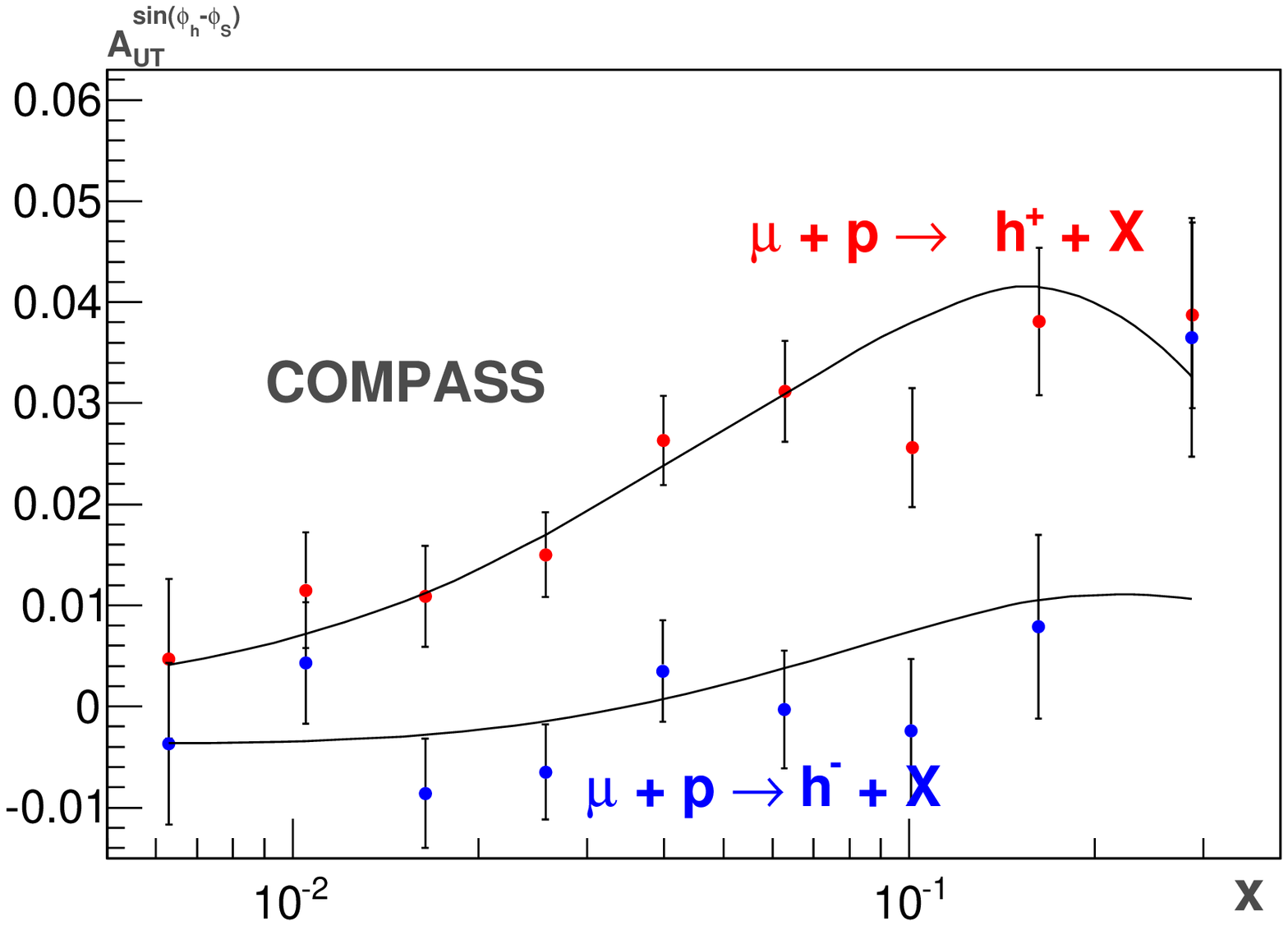}
\includegraphics[width=7.1cm]{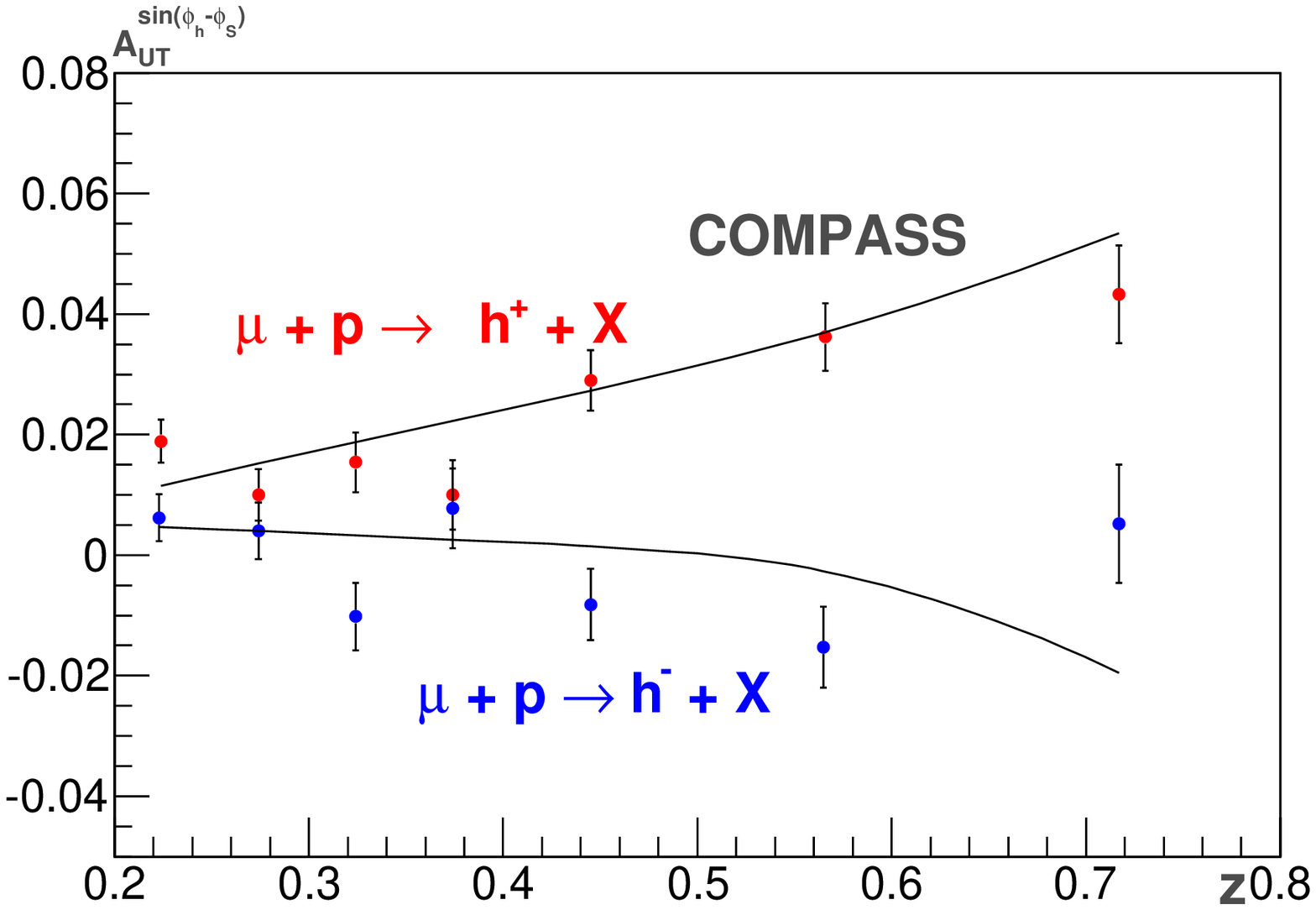}
\includegraphics[width=7.1cm]{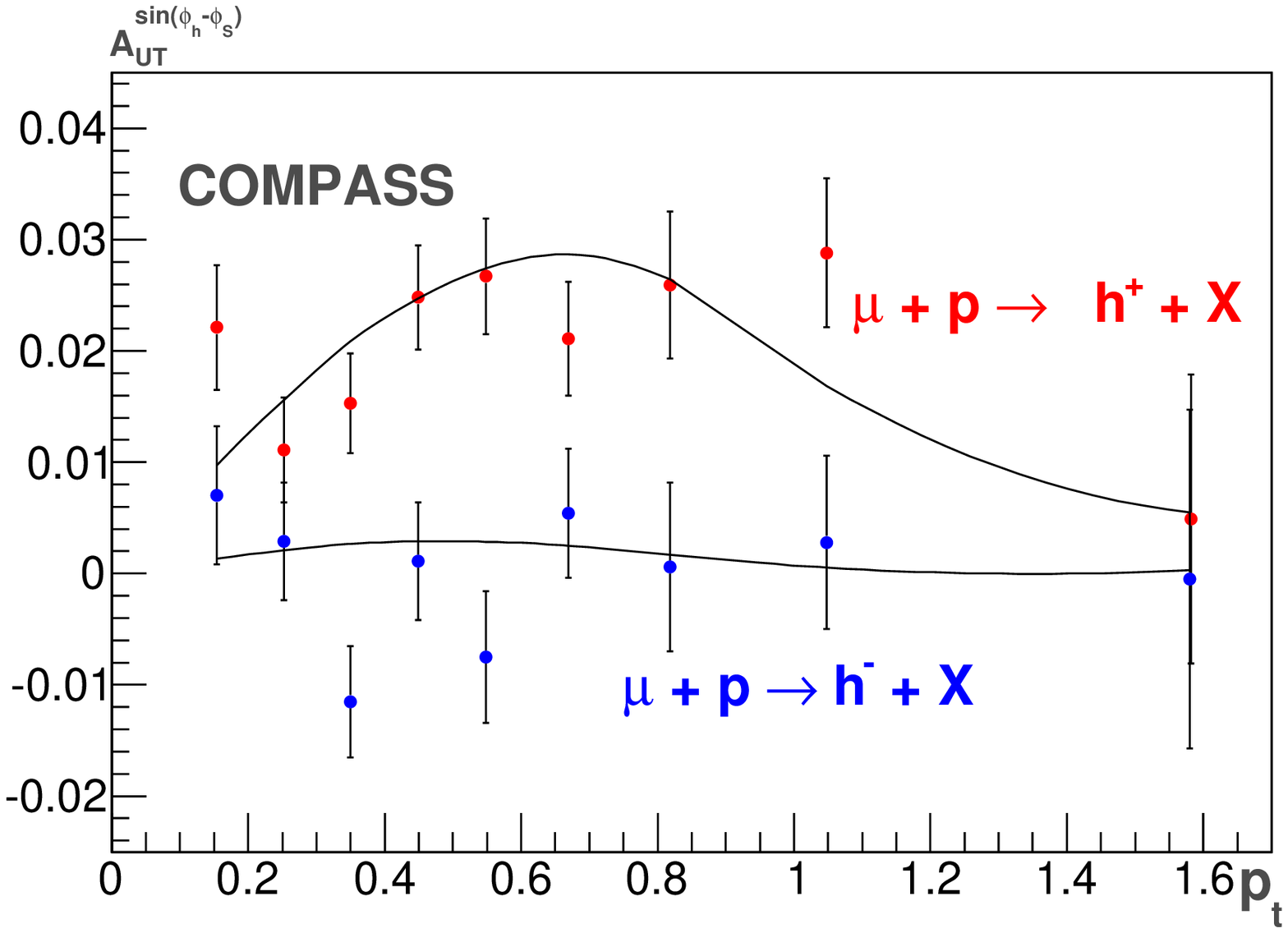}
\caption{Same as Fig.~\ref{fig:hermes-sivers} but for COMPASS ~\cite{Alekseev:2008aa,Adolph:2012sp}.}
\label{fig:compass-sivers}
\end{figure}

In total we have ten parameters in the fit: $N_u$, $N_d$ and $N_{(\bar u,\bar d,s)}$ for the normalization,
$\alpha_u$, $\alpha_d$, $\alpha_s$ and $\beta$ for $x$ and $(1-x)$ power
behavior, and $g_s$ for the transverse momentum dependence in the Sives function.
In our fit, we include the Sivers asymmetries in SIDIS,  which include
$\pi^+$, $\pi^-$, $\pi^0$, $K^+$, $K^-$ from HERMES/COMPASS,
and positive and negative charged hadrons from COMPASS.
We include all these in our fit~\footnote{The last $z$ bins from COMPASS
are too close to 1, and we do not include them in the fit.}.
The total number of data points are 255. We use
minimum $\chi^2$ fit by the MINUITE program package. The resulting fit
gives $\chi^2/d.o.f=1.08$. The parameters are found to be,
\begin{eqnarray}
&&N_u=0.13\pm 0.023,~~\alpha_u=0.81\pm  0.16, ~~\beta= 4.0\pm 1.2  \ ,\nonumber\\
&&N_d=-0.27\pm 0.12,~~\alpha_d= 1.41\pm 0.28  \ , \nonumber\\
&&N_s=0.07\pm 0.06,~~\alpha_s= 0.58\pm 0.39   \ , \nonumber\\
&&N_{\bar{u}}=-0.07\pm 0.05   \ , \nonumber\\
&&N_{\bar{d}}=-0.19\pm 0.12   \ , \nonumber\\
&&g_s=0.062\pm 0.0053 \ .
\end{eqnarray}
We plot the comparisons between our fits to the experimental data
in Figs.~\ref{fig:hermes-sivers} and \ref{fig:compass-sivers}.

\begin{figure}[tbp]
\centering
\includegraphics[width=7.1cm]{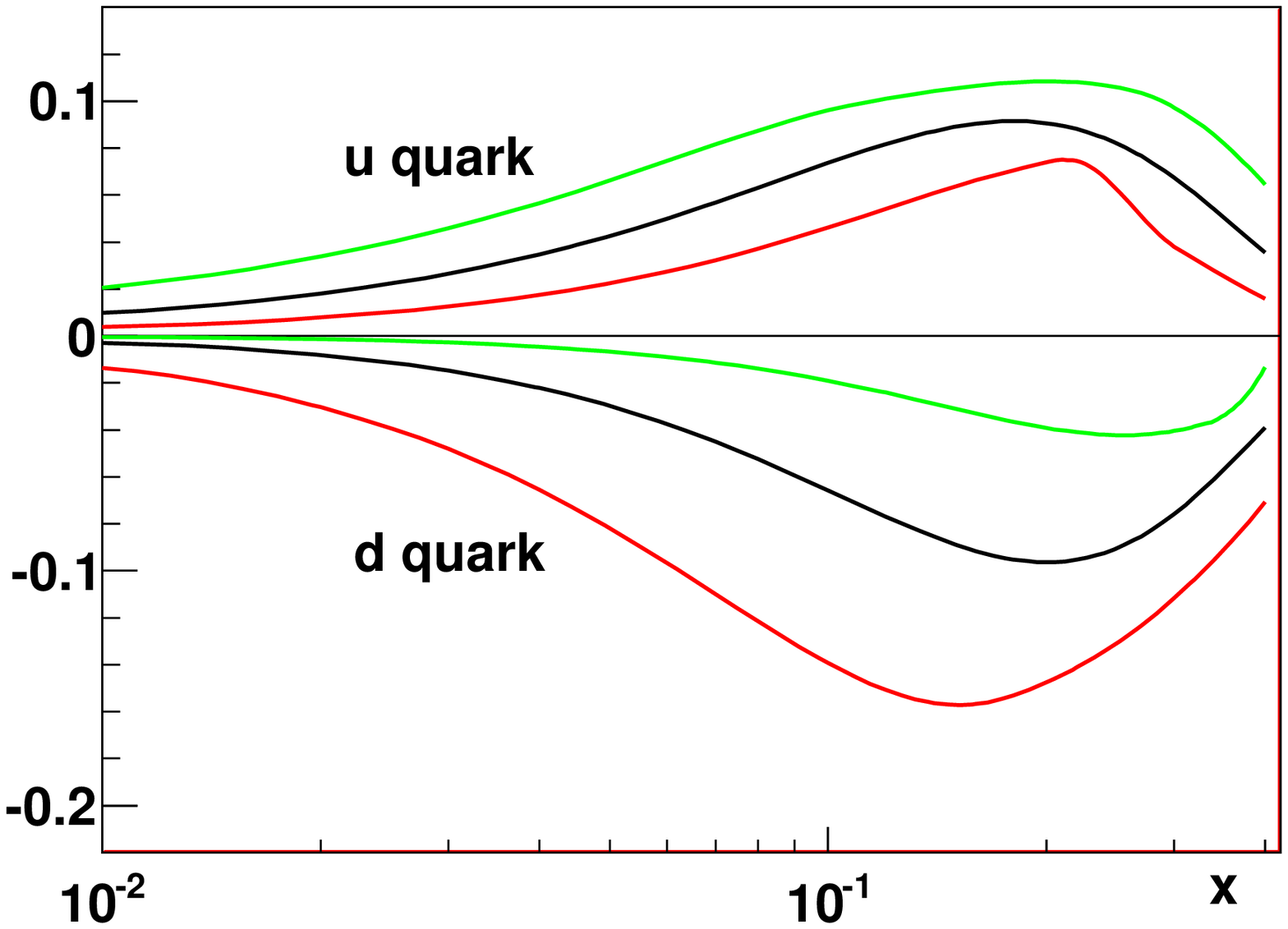}
\includegraphics[width=7.1cm]{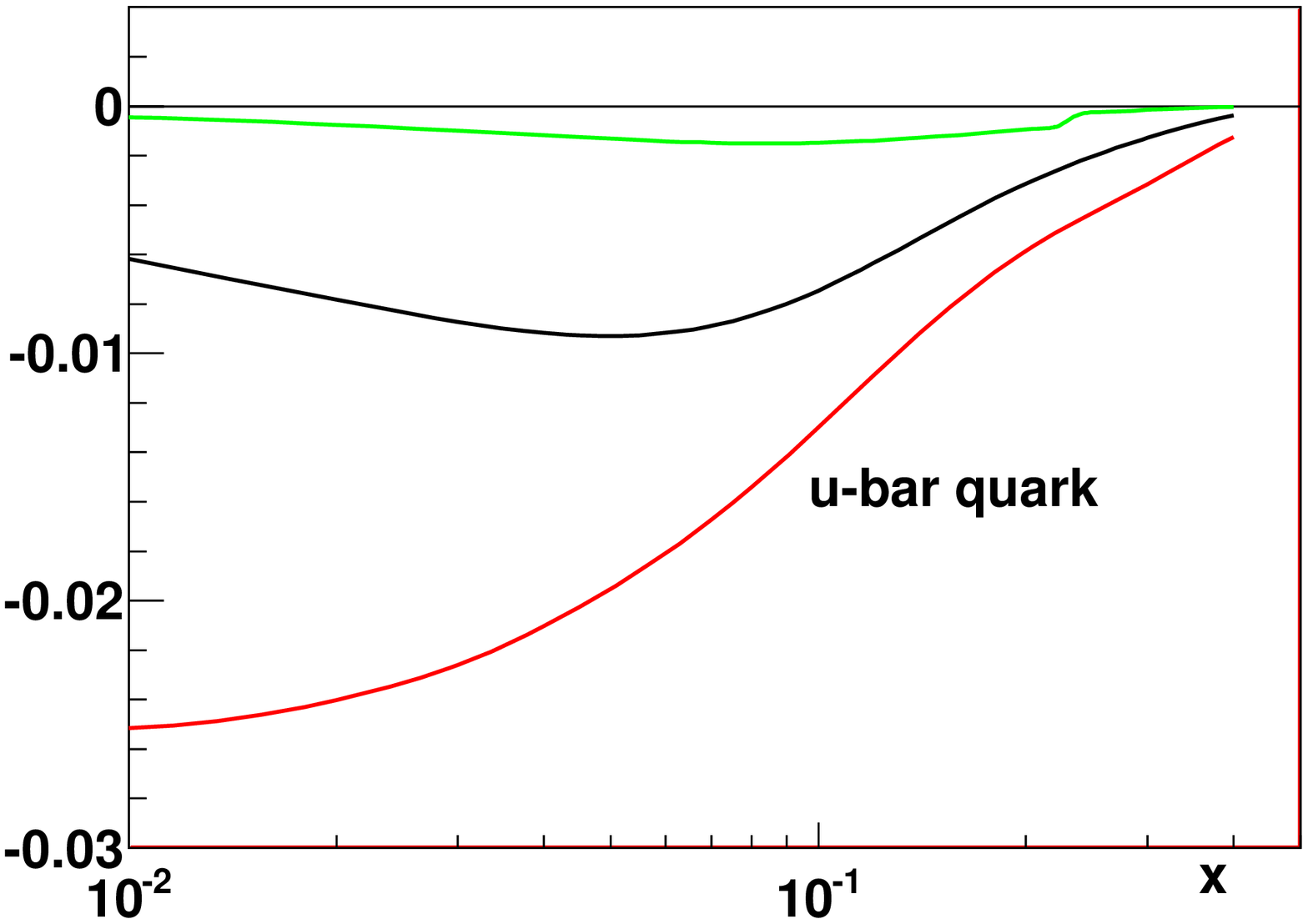}
\caption{ Moments of the quark Sivers functions $\Delta f_q=T_F(x,x)/M$ fitted to HERMES
and COMPASS data: up and down quark (left) and anti-up quark (right). Upper and
lower curves for the uncertainties.}
\label{fig:tfud}
\end{figure}

The determined Sivers functions are plotted in Fig.~\ref{fig:tfud}.
From the above fits, we clearly see that the Sivers asymmetries in SIDIS
from HERMES and COMPASS experiments have demonstrated the sizable
quark Sivers functions. However, because of the experimental error bars
are still large, the constraints on the quark Sivers functions are not strong enough
to obtain a precise picture of the up and down quark Sivers functions. But, a number
of features can be derived from the above analysis.
First, the up quark Sivers function was best determined. This is because of
the charge enhancement of the up quark as compared to the down quark.
In other words, the SIDIS with proton target is most sensitive to the up quark
contribution. Second, although the down quark Sivers was not strongly
constrained from the current experimental observations, it seems that
down quark Sivers function might be opposite to that of the up quark Sivers
function in sign, with larger uncertainties though.
Third, there is no constraints for the sea quark Sivers function at all.
We need future experiments to pin down both down quark Sivers function
and sea quark Sivers functions. As we mentioned above, SIDIS measurements
in the 12 GeV upgrade of JLab and the planed electron-ion collider experiments,
plus the Drell-Yan experiments which we will discuss in the following section,
shall help us to achieve these goals.

Quark Sivers functions have been phenomenologically
constrained from HERMES/COMPASS experiments by several groups
in the literature~\cite{Anselmino:2005an,Anselmino:2005ea,
Anselmino:2012aa,Bacchetta:2011gx,Gamberg:2013kla}. In the following, we briefly comment
on the comparisons with these studies. First, early studies
only consider the factorization of the SIDIS and the quark
Sivers functions are determined by comparing the theory calculations
with the experimental data without taking into account the $Q^2$-evolution
effects. Second, Ref.~\cite{Anselmino:2012aa} has first started an analysis with TMD evolution
following Rogers et al. approach. As we have demonstrated above, Rogers et al.
approach overestimated the TMD evolution effects. The combined analysis
of Ref.~\cite{Anselmino:2012aa} might need to be re-examined.

By comparing with previous constraints on the quark Sivers function,
we notice an interesting aspect: our results agree roughly with the fits done
without the TMD evolution effects. This comes from the fact that the HERMES and COMPASS
experiments do not differ much on $Q^2$, and the evolution effects from
our calculation is not so strong as naively expected. Certainly, more
theoretical studies are needed to check the evolution effects.

As we emphasized above, in this paper, we focus on the quark
Sivers functions in the moderate $x$ range around 0.1. this is also the
region where the sizable Sivers single spin asymmetries are observed
by HERMES/COMPASS experiments. For small-$x$ region quark Sivers
functions, there are additional theoretical uncertainties from the TMD evolution,
which has only be tested for moderate $x$ range as we showed in the last
section. Therefore, the constraints for small-$x$ region quark Sivers function
have to be taken a particular caution.

\section{Sivers SSAs in Drell-Yan and W/Z Boson Production}

In this section, we present the predictions on the Sivers single spin
asymmetries in Drell-Yan lepton pair production and W/Z boson
production in polarized $pp$ collisions. These predictions
not only serve as important base lines for the sign change tests of the
Sivers asymmetry in these experimental proposals,
but also provide guidelines for precision measurement of the
TMD quark Sivers functions. In particular, the combination
of different kinematic coverages in these experiments shall help
to identify down quark and sea quark Sivers functions beyond what
we can extract from the current HERMES/COMPASS experiments.
Therefore, in the following, we will also highlight the opportunities
in these experiments.

Since the observation of the Sivers single spin asymmetries in semi-inclusive
hadron production in deep inelastic scattering process, there have been
several proposals to measure the Sivers asymmetries in Drell-Yan lepton
pair production in $pp$ collisions. The main focus is to observe the sign
change between the Sivers asymmetries of these two processes. This
is a fundamental prediction from QCD gauge theory, which is also
crucial to understand the strong interaction dynamics, such
as the factorization and energy evolution.

Two important issues arise when we predict the SSAs for the proposed
Drell-Yan experiments. One is the flavor dependence, i.e., how precise
the quark Sivers functions can be determined from the SIDIS experiments
from HERMES/COMPASS collaborations. Another important issue is the
energy dependence of these observables. The SIDIS experiments from HERMES
and COMPASS are mainly in the relative low $Q^2\sim 3\textmd{GeV}^2$ range,
whereas the Drell-Yan processes are typically range of $Q^2$ from $ 20\textmd{GeV}^2$
to $100\textmd{GeV}^2$. We have to understand the energy evolution of the Sivers
spin asymmetries before we can make precise predictions.

The last few sections of this paper are dedicated to understand correctly
the energy evolution of the hard processes in SIDIS and Drell-Yan lepton
pair productions, including the unpolarized cross sections, and the
Sivers single spin asymmetries. Our strategy has always been to test
the evolution for the unpolarized cross sections before we can apply to
the spin-dependent observables. We have been able to show that the
energy evolution equations we used can successfully describe the
transverse momentum distributions in SIDIS from HERMES/COMPASS
experiments and Drell-Yan lepton pair production in fixed target experiments.
HERMES/COMPASS experiments are where we observed the Sivers
single spin asymmetries, whereas the Drell-Yan processes in our study
are close to the kinematics of future proposed experiments to
measure the Sivers single spin asymmetries. Most importantly, all these
experiments (including SIDIS from HERMES/COMPASS and Drell-Yan
in fixed target experiments) cover similar range of $x$. Therefore, we do
not have to worry too much on additional complication involving $x$-dependence
in the non-perturbative form factors in the TMD evolutions.

Let us briefly summarize our procedure to predict the SSAs in the Drell-Yan
processes. We will mainly use the CSS resummation formalism in terms of the
collinear parton distribution and correlation functions. The Sivers asymmetries
depend on the transverse-momentum moments of the quark Sivers functions
in the CSS resummation formula,
which we will use the parameterization we extracted from the combined fit
of the HERMES and COMPASS data from the last section. The non-perturbative
form factors will follow those in Konychev-Nadolsky result, which are fitted to
unpolarized Drell-Yan and W/Z boson production data. For the single spin
dependent cross sections, the difference in the non-perturbative form factor
$g_s$ was also fitted to the HERMES/COMPASS data. Since the $Q$-dependence
of the non-perturbative form factors are the same for the spin-average
and single-spin dependent cross sections (they obey the same evolution
equation), this difference will not depend on $Q$. The fitted $g_s$ will be
applied to the SSAs in Drell-Yan and W/Z production processes.

Before we present the detailed predictions for the Drell-Yan and W/Z processes,
we would like to show that the consistency between the evolution we studied
in Sec.IV and V and those from the CSS resummation with KN non-perturbative
form factors. This will demonstrate that we do have a consistent match between
relative low $Q$ (from which we extract the quark Sivers functions) and moderate
high $Q$ (up to W/Z boson productions).

\subsection{Matching SIDIS to Drell-Yan and W/Z Boson Production in $pp$ Collisions}

In Sec.III, we have shown that for the unpolarized Drell-Yan cross sections from the
fixed target experiments,
the TMD evolution of Sun-Yuan approach is consistent with the CSS resummation with
KN parameterization of the non-perturbative form factors, see Figs.~\ref{fig:e288} and \ref{fig:e605}.
In this subsection, we will demonstrate that for the Sivers single spin asymmetries,
both calculations will yield consistent predictions as well. This shall further strength
the matching of the TMD evolution
between relative low $Q$ SIDIS and moderate high $Q$ Drell-Yan
processes.

The Sivers single spin asymmetries in Sun-Yuan approach are calculated with
the expressions in Eqs.~(\ref{wuu-sy},\ref{wuu0-sy}) for the unpolarized
cross section, and equations of (\ref{wut-sy},\ref{wut0-sy}) for the
single spin dependent cross section. We take the quark Sivers functions
determined by the combined analysis in the last section. As mentioned before,
the ``special" universality of the quark Sivers function between Drell-Yan and
SIDIS is reflected by the sign difference in Eqs.~(\ref{wut0-sy}) and (\ref{siversq0}).

To calculate the Sivers single spin asymmetries in the CSS formalism,
we follow what has been done for the unpolarized cross section of Eq.~(\ref{blny0}),
and write down $\widetilde{W}_{UT}$ as
\begin{eqnarray}
\widetilde{W}_{UT}(Q;b)=\left(\frac{-i b_\perp^\alpha}{2}\right)e^{-{\cal S}_{pert}(Q^2,b_*)-S_{NP}^T(Q,b)}
\Sigma_{q} T_F^q(z_1,z_1;Q_0)\bar q (z_2,Q_0 ) \ ,
\end{eqnarray}
where $S_{NP}$ follows Eqs.(\ref{blny0},\ref{kn}), and $$S_{NP}^T(Q,b)=S_{NP}(Q,b)-g_sb^2$$ with
$g_s=0.062$ fit from the HERMES and COMPASS experiments.
 $T_F(z)$ is calculated from the quark Sivers function,
\begin{equation}
T_F(z,z;\mu=Q_0)=\int\frac{d^2k_\perp}{(2\pi)^2}\frac{k_\perp^2}{M}f_{1T}^{\perp(DY)}(z,k_\perp)=M\Delta f^{\rm sivers}(z) \ ,
\end{equation}
where $\Delta f^{\rm sivers}(z)$ has been determined from the combined analysis in
the last section. The above identification depends on the TMD factorization, and the
CSS resummation derived in Sec. II. Of course, the integrated parton correlation
function, such as $T_F(x,x)$ depends on the scale, which is not easy to identity in the
above equation. We argue that the TMD quark Sivers functions are determined at the low
energy $Q_0$, which indicates that the moments of the quark Sivers functions
calculated from the fit shall be in the similar range. Therefore, we shall be able to identify
the scale for $T_F$ as $Q_0$. We notice that this will introduce some theoretical uncertainties.
In particular, the scale evolution for $T_F$ is more complicated than that for
the unpolarized quark distribution. In order to estimate the uncertainties that come
from the scale setting, in the following we will make a rough estimate to see
how large the scale setting affect the final results for the single spin asymmetries.

\begin{figure}[tbp]
\centering
\includegraphics[width=9cm]{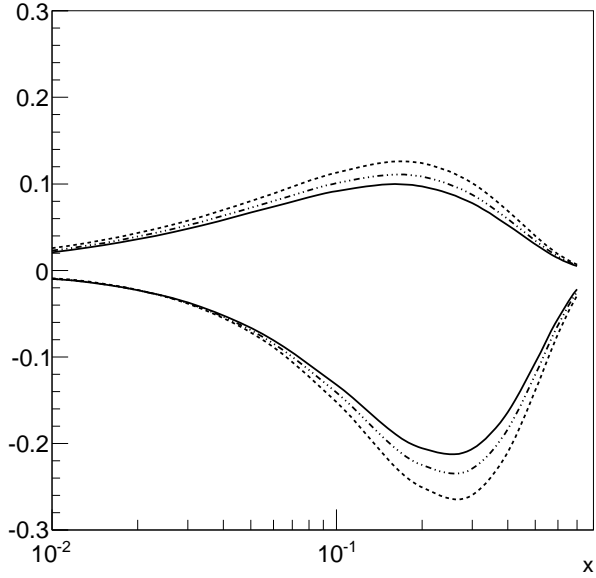}
\caption{Transverse-momentum moments of the quark Sivers functions $T_F$
divided by the unpolarized quark distributions for up and down quarks, at the scale
$\mu^2=Q_0^2=2.4\textmd{GeV}^2$, $5\textmd{GeV}^2$, and $10\textmd{GeV}^2$, respectively.
$T_F(z,\mu)$ is obtained by solving the simplified evolution equation
Eq.~(\ref{dglap0}) with input parameterization at the initial scale $Q_0$.}
\label{fig:tfratio}
\end{figure}

The evolution equation for $T_F$ has been
derived in the literature. To solve this equation completely, we need input from
$T_F(x_1,x_2)$ functions and $\tilde{T}_F$, and gluonic contributions as well.
The complete solution is not available at current stage. As a first step, we take
an approximate form of the splitting kernel, i.e., large $\xi$ limit. This should be
relevant, in particular, in current case that the quark Sivers function mainly
concentrate in the valence region. This limit has been discussed in Ref.~\cite{Braun:2009mi},
where the evolution can be simplified as
\begin{equation}
\frac{\partial}{\partial\ln\mu}T_F(z,z;\mu)=\frac{\alpha_s}{2\pi}\int\frac{dx}{x}T_F(x,x;\mu)
\left[C_F\left(\frac{1+\xi^2}{(1-\xi)}\right)_+-C_A\delta(1-\xi)\right] \ ,\label{dglap0}
\end{equation}
where $\xi=z/x$. In Ref.~\cite{Braun:2009mi}, it is further argued that
an approximate solution of the above equation would lead to the following
behavior for $T_F(z)$,
\begin{equation}
T_F(x,x,Q^2)/f_q(x,Q^2)\sim \left(\frac{\alpha_s(Q^2)}{\alpha_s(Q_0^2)}\right)^{2N_c/b_0}\ . \label{braun0}
\end{equation}
Since we have determined the $T_F$ at the scale around $\mu^2=2.4\textmd{GeV}^2$,
we will be able to solve the above evolution equation numerically, by using the
HOPPET program~\cite{Salam:2008qg}. We show
the solution in Fig.~\ref{fig:tfratio}. It is interesting to find that, indeed, in the valence
region, the above approximation works for both up and down quark Sivers functions.

\begin{figure}[tbp]
\centering
\includegraphics[width=9cm]{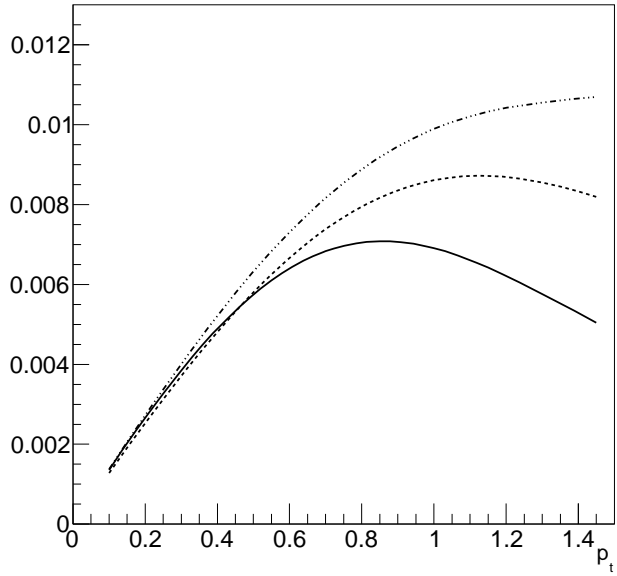}
\caption{Compare the predictions for the Sivers asymmetries in Drell-Yan lepton pair
production in polarized $pp$ collisions: center of mass energy $\sqrt{S}=20\textmd{GeV}$
at $y=0$ with $Q=5.5\textmd{GeV}$: soid curve calculated with Sun-Yuan evolution, while the
point curve and point slash curve are from the CSS resummation with $b_*$-prescription and KN paramterization
of the non-perturbative form factors with fixing parton distribution scale at 2.4 GeV$^2$ and $C_1^2/b_*^2$ respectively.}
\label{fig:match}
\end{figure}

Although the scale dependence is stronger for $T_F$ as compared to the
unpolarized quark distribution according to the evolution
equation, its effects on the $p_\perp$ distribution from the resummation
formalism may not be as strong as naively expected. This is because, in the
CSS resummation formalism Eq.~(\ref{css}), the scale for both $T_F$ and $f_q$ is set
at $\mu=1/b_*$ which does not change much with $Q$. To estimate this
effect, we calculate the transverse momentum dependence of the Sivers
single spin asymmetries in Drell-Yan lepton pair production
of $Q=5.5\textmd{GeV}$ at a typical fixed target experiment with $\sqrt{S}=20\textmd{GeV}$,
with three different assumptions: (1) CSS resummation with fixed
scale for the parton distributions and the transverse momentum moment
of the quark Sivers function $T_F$ as $\mu=Q_0$; (2) including the scale dependence in these
two distributions as $\mu=c_0/b_*$; (3) Sun-Yuan approach from a direct
integral of the Sudakov kernel from $Q_0$ to $Q$. These comparisons
are plotted in Fig.~\ref{fig:match}.

From this plot, in the low transverse momentum region, we can see that these
calculations present consistent predictions for the SSAs. At higher
transverse momentum, the TMD evolution of Sun-Yuan approach predicts smaller
SSAs than that from the CSS resummation formalism with the two
implementations (1) and (2).
This comes from the Gaussian assumption of the TMDs at lower scale $Q_0$
which leads to more suppression of the spin asymmetry at moderate transverse
momentum. However, with the CSS resummation correcting behavior for the transverse
momentum distribution, this can be improved. Another point is the evolution of $T_F$
(implementation (2)) increases the asymmetry predictions than that without $T_F$ evolution.
This is because for fixed scale, we have used $\mu^2=2.4\textmd{GeV}^2$, whereas at
moderate transverse momentum
the scale for $T_F$ in (2) is around $1/b_{max}$, and is smaller than $Q_0$. This will lead
to larger $T_F$ contribution to the Sivers single spin asymmetries. This difference
highlight the opportunities to study the QCD dynamics in this transverse
momentum region, because it is sensitive to the QCD evolution. We hope in
the future the Sivers single spin asymmetries can be measured in Drell-Yan
process in the relative high transverse momentum region.

These two plots demonstrate that we do have a consistent picture for the
Sivers single spin asymmetries in the Drell-Yan lepton pair production
in the mass region of interest in the fixed target experiments. One approach
(Sun-Yuan) uses the TMD evolution from SIDIS of HERMES/COMPASS to
Drell-Yan processes; one approach uses the resummation formulas only for
Drell-Yan processes but with the collinear correlation functions (transverse-momentum
moments) extracted from HERMES/COMPASS experiments. The agreement
between these two calculations shows that we understand the energy
evolution effects on the Sivers single spin asymmetries.

In particular, the size and slope of the SSAs calculated from the above
formulas in the low transverse momentum region agree with each other.
This region is mainly determined by the ratio between the quark Sivers
function and unpolarized quark distribution. The agreements indicate that
the identification of $T_F(z,z)$ with $M\Delta f_{\rm sivers}$ is appropriate
for the Sivers single spin asymmetry calculations. Of course, the difference
in the relative high transverse momentum region will introduce
additional theoretical uncertainties. From Fig.~\ref{fig:match}, we can
estimate that the difference is around $10\%$ for the integrated asymmetries.

In the following, we will present the predictions with the CSS formalism
and KN non-perturbative form factors. This approach does better job for
moderate transverse momentum. In addition, it is the only approach that would
predict correctly the cross section and spin asymmetries for W/Z production
at RHIC.

We would like to add one more comment. From the analysis of the last section,
we know that the uncertainties of the extracted quark Sivers functions
from the HERMES/COMPASS experiments is not negligible. Therefore, in the following
calculations, we will present the predictions with the standard one sigma
uncertainty estimate for the Sivers single spin asymmetries in the Drell-Yan processes.

\subsection{Drell-Yan Process in COMPASS at CERN}

The COMPASS collaboration has planed to run Drell-Yan experiments in
the coming years. They will use $\pi^-$ beam with energy 190GeV scattering
on the polarized hydrogen target. The kinematic coverage can be found in the
publication of Ref.~\cite{compassdy}.

\begin{figure}[tbp]
\centering
\includegraphics[width=9cm]{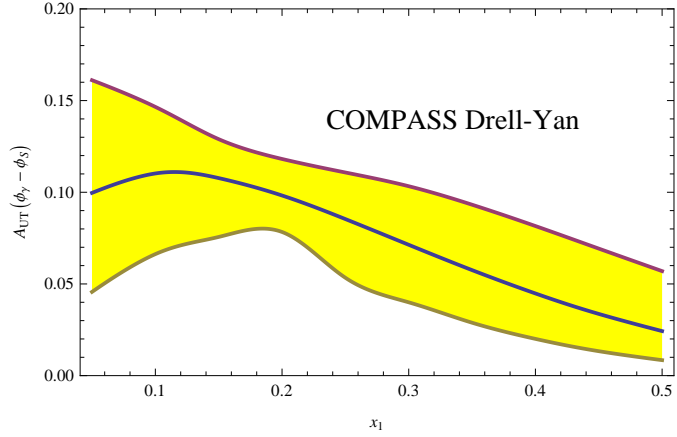}
\caption{Predictions for the Sivers single spin asymmetry for
the Drell-Yan process at COMPASS, with $\pi^-$ beam of $190\textmd{GeV}$, as function of
$x_p$. We have chosen
the average $x_\pi\approx 0.55$ and integrate transverse momentum
up to $2\textmd{GeV}$.}
\label{fig:compass-dy}
\end{figure}

In order to make predictions for this experiment, we assume
the non-perturbative form factors for
$\pi^- N$ scattering follows the same KN parameterization.  The
predicted asymmetries as functions of $x_p$ (the momentum fraction
of the proton carried by the di-lepton pair).
The average $Q^2$ from the experimental
simulation has been used in the calculations of the Sivers single spin
asymmetries.

An important feature of this experiment is that the incoming $\pi^-$
is dominated by $\bar u$ and $d$ quarks from the pion, which
leads to the dominant contribution from the up quark Sivers functions.
The up quark Sivers function is constrained by the HERMES/COMPASS
experiments. Therefore, the theory uncertainties are relative small.

There have been plan to measure the Sivers single spin asymmetries in
the low mass region between $2$ and $3$ GeV for the di-lepton production
in COMPASS experiment. However, in this mass region, we do have background
from hadronic decays. Nevertheless, it will be interesting to measure the single spin asymmetries
in these kinematics, including $J/\psi$ resonance. The latter process
shall provide some information on the gluon Sivers function in the relevant
kinematics.

\subsection{Fermilab Fixed Target Experiments}

The proposal of the polarized Drell-Yan experiments at the Fermilab contain two possible options~\cite{fermilabdy}:
polarized beam or polarized target. Both cases can be used to measure the Sivers
single spin asymmetries in the Drell-Yan lepton pair production. In the proposed
experiment, the incoming beam has energy of 120GeV.

Different from the Drell-Yan experiments at COMPASS, the Fermilab proposal
have proton-proton scattering. The flavor structure will be very different from that
in COMPASS. This is because in the proposed kinematics, the sea quark
contribution to the unpolarized cross section is not negligible. Therefore, we would
expect that the sea quark Sivers functions will play an important role as well.

\begin{figure}[tbp]
\centering
\includegraphics[width=7cm]{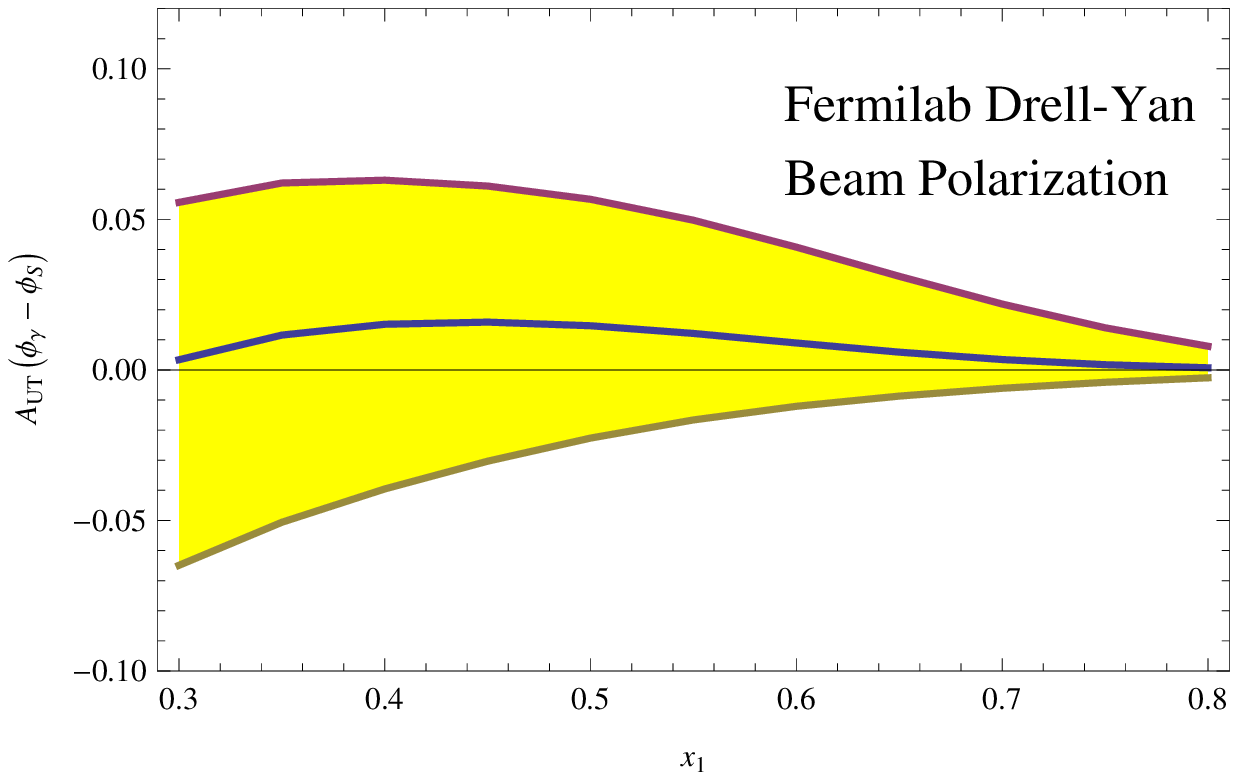}
\includegraphics[width=7cm]{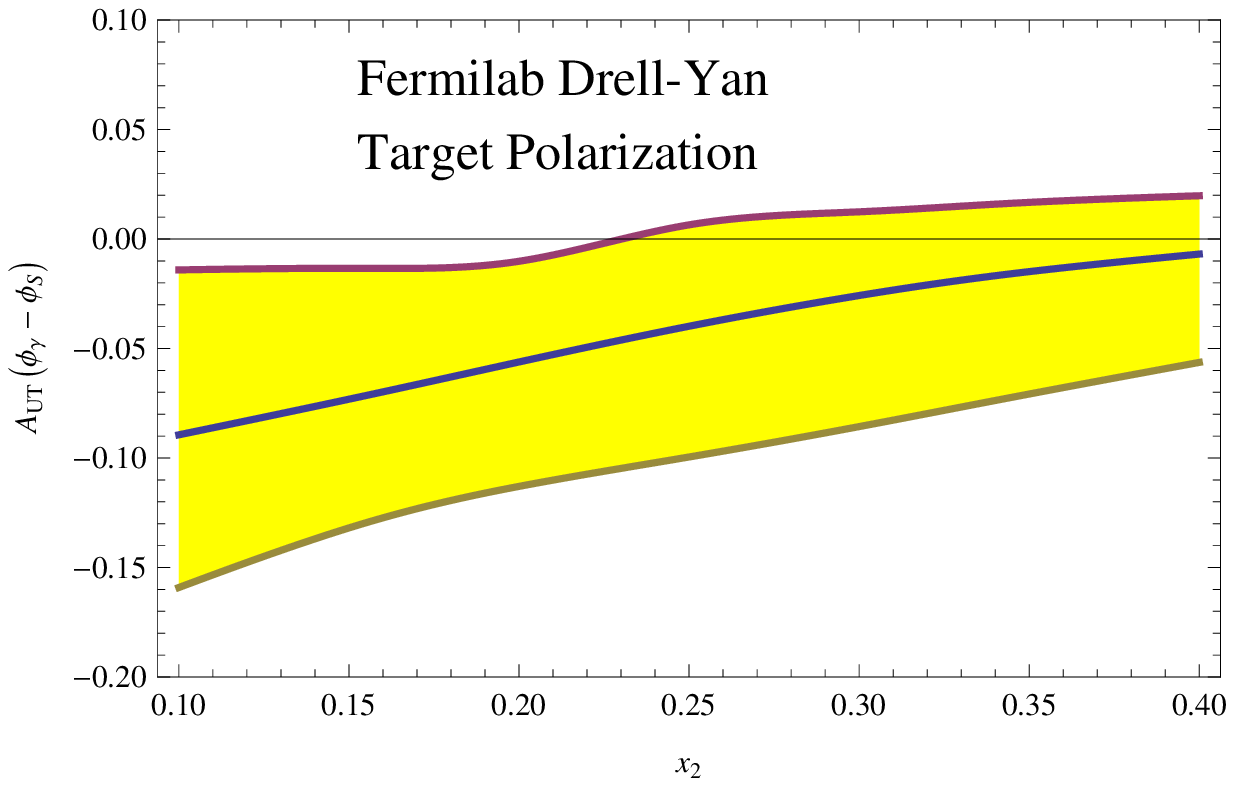}
\caption{Predictions for the Sivers single spin asymmetry for the Drell-Yan process
at Fermilab fixed target experiments, with proton beam
of $120\textmd{GeV}$,  as function of $x$ for the polarized
proton: polarized beam (left) and polarized
target (right). }
\label{fig:fermilab-dy}
\end{figure}

In Fig.~\ref{fig:fermilab-dy}, we plot our predictions
for the Sivers single spin asymmetries in the Drell-Yan
process at the fixed target experiment at Fermilab with polarized
beam (left) or polarized target (right) options. For the beam polarization
case, we show the Sivers single spin asymmetry as function of $x_1$ where
$x_1$ is the momentum fraction of the polarized proton beam carried by the
virtual photon in the final state, and we have chosen the average
$\langle x_2\rangle=0.3$ from the kinematic
simulation of the experimental
proposal. Clearly, this experiment will mostly cover the valence region of
the polarized proton, $x_1\ge 0.30$, which is beyond the current HERMES/COMPASS
measurements. Our predictions come from the extrapolation of the function
form constrained by the HERMES/COMPASS experiments. The measurements
of this asymmetry in the proposed experiment shall, for the first time,
investigate the Sivers asymmetries in this kinematic region.

On the other hand, for target polarization, because $x_1$ is still around valence
region and the quark distribution is dominated by the valence up quark, the
Sivers single spin asymmetry will be very sensitive to anti-up quark Sivers
function. If there is no sea quark Sivers function, the
asymmetry will be very small. However, with sea quark Sivers function allowed from
HERMES/COMPASS experiments, we find that the Sivers asymmetry in
Drell-Yan process at this experiment is relative sizable although the uncertainties
are large. The predictions are shown in the right panel of Fig.~\ref{fig:fermilab-dy},
where average of $\langle x_1\rangle=0.55$ has been used in the calculations.
The observation of this asymmetry will definitely signal non-zero sea quark Sivers function in
this relevant kinematics.

\subsection{Sivers Asymmetries in $W^\pm$ Production and Drell-Yan process at RHIC}

\begin{figure}[tbp]
\centering
\includegraphics[width=10cm]{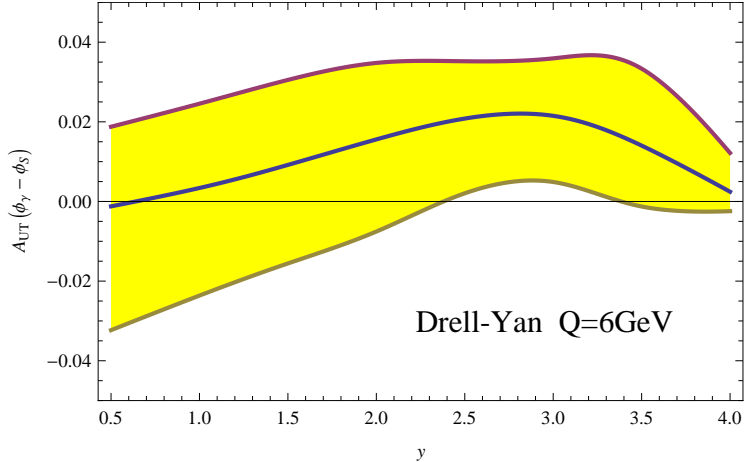}
\caption{Predictions of the Sivers single spin asymmetries for Drell-Yan process
as function of rapidity at RHIC with $\sqrt{S}=500\textmd{GeV}$.}
\label{fig:rhic-dy}
\end{figure}

Drell-Yan lepton pair production at RHIC of Brookhaven National Laboratory has
been proposed for quite some time. We have presented the predictions for this
experiment in Ref.~\cite{Sun:2013dya}. With the CSS resummation and KN
parameterization for the non-perturbative form factors, we also estimated
the asymmetries in Drell-Yan process at RHIC, and we obtain similar results
as we showed in~\cite{Sun:2013dya}. In Fig.~\ref{fig:rhic-dy}, we show the
predictions for the Drell-Yan process in $\sqrt{S}=500\textmd{GeV}$.

Besides the Drell-Yan process, RHIC experiments can, in principle,  measure
the Sivers single spin asymmetries in $W^\pm$ boson production in polarized
proton-proton collision st $\sqrt{S}=500\textmd{GeV}$. Early calculations have emphasized
the unique opportunity to test the sign change~\cite{Kang:2009bp,Metz:2010xs}. In the following,
we present the asymmetries calculated with the quark Sivers functions determined
from HERMES/COMPASS experiments with the evolution effects taken into
account.

\begin{figure}[tbp]
\centering
\includegraphics[width=6.9cm]{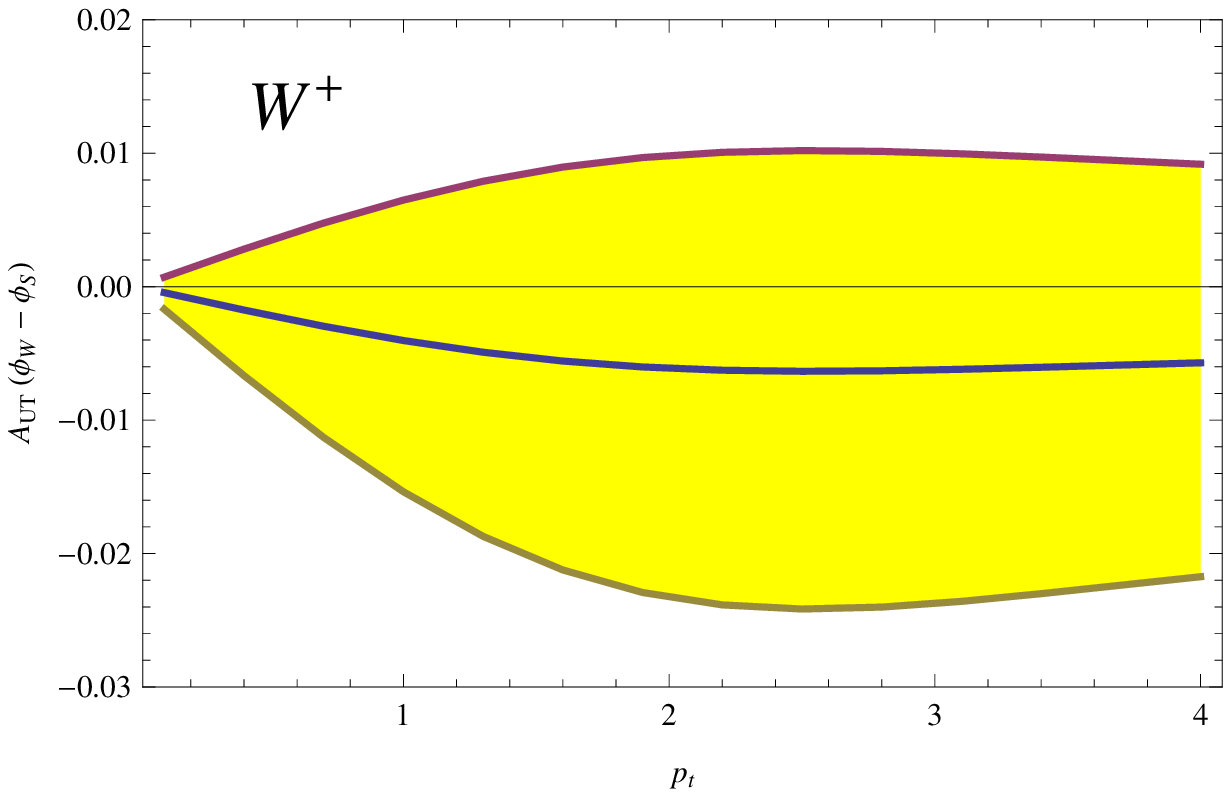}
\includegraphics[width=7cm]{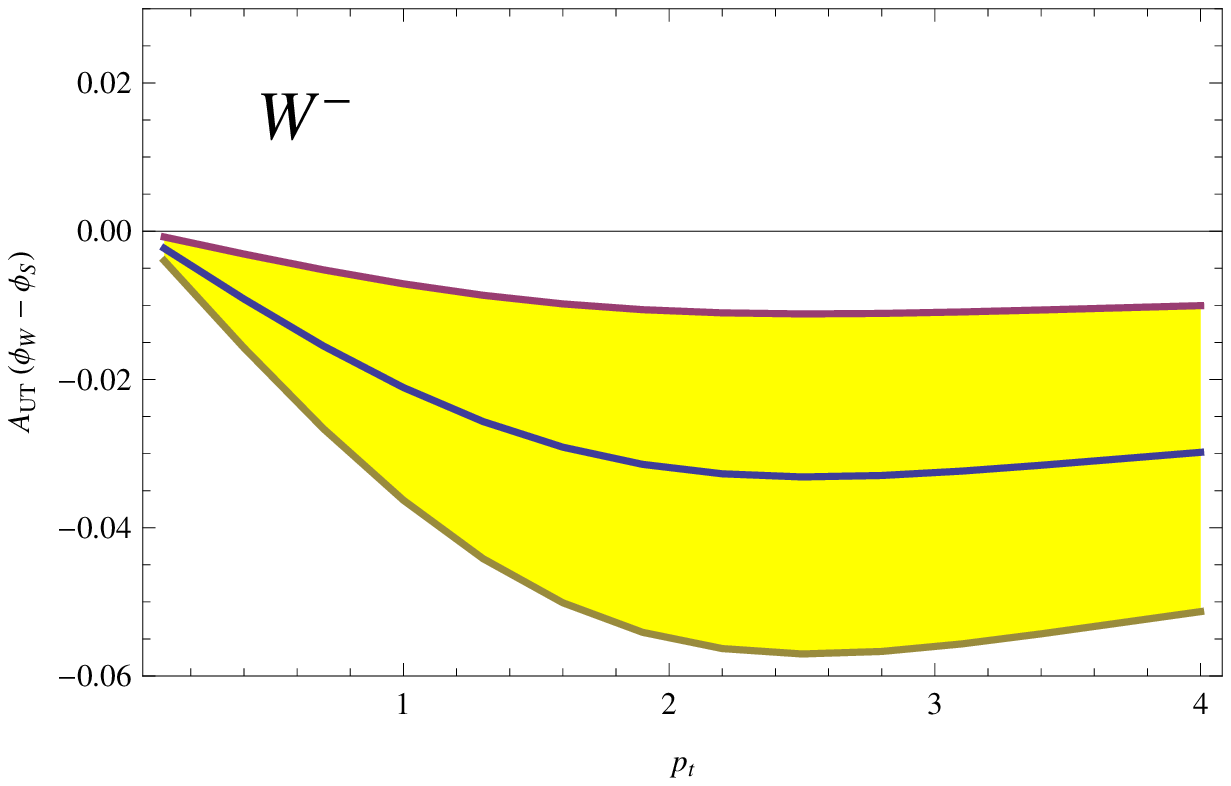}
\caption{Predictions for the Sivers single spin asymmetries for $W^+$ (left) and $W^-$ (right) productions
at RHIC with $\sqrt{S}=500\textmd{GeV}$ at mid-rapidity as functions of the transverse
momenta of the vector bosons.}
\label{fig:rhic-w}
\end{figure}

For $W^+$ production, it is dominated by the Sivers up quark and anti-down quark
from the polarized proton. As shown in Fig.~\ref{fig:rhic-w}, two important features can be found from our
calculations: first, the prediction is much reduced as compared to the previous calculations;
second, anti-down quark Sivers function also contributes significantly.
However, because the uncertainties associated with the sea quark contribution
is not negligible, even the sign of the Sivers asymmetries is not well constrained.

On the other hand, $W^-$ production is dominated by down quark and anti-up quark
Sivers functions. Again, both quantities are not well constrained from HERMES/COMPASS
experiments. Therefore,  the observation of this asymmetry will tell us
on the sea quark polarization.

We would like to emphasize that for $W/Z$ boson production, the TMD evolution directly
from low scale can not describe the $p_\perp$ spectrum. For this, we have to use CSS
resummation. The main reason is that the perturbative gluon radiation dominates
the low transverse momentum $W/Z$ production, whereas the low scale TMD is not
easy to generate these contributions. It may be improved by taking into account perturbative
tail in the TMD quark distribution at low $Q$ scale.

\section{summary and conclusion}

In this paper, we have investigated the TMD evolution effects
on semi-inclusive DIS and low transverse momentum Drell-Yan lepton pair production
in $pp$ collisions, consistently describing both unpolarized and single transverse spin
Sivers asymmetries. In particular, we have built up a framework to match SIDIS
and Drell-Yan, which can cover the TMD physics with $Q^2$ from 2 $\textmd{GeV}^2$ to $10^4\textmd{GeV}^2$
(for W/Z boson production). By doing so, we constrained the transverse-momentum moments
of the quark Sivers functions from the combined analysis of the HERMES/COMPASS
data on the Sivers single spin asymmetries in semi-inclusive hadron production in
DIS process with the TMD evolution for the spin-average and single-spin dependent
differential cross sections. The TMD evolution was carefully examined against the
transverse momentum distribution of unpolarized SIDIS and Drell-Yan processes
in the $Q^2$ range from 2 to 100 $\textmd{GeV}^2$. Our approach agrees well with
the existing experimental data.

Most importantly, we have demonstrated the matching between our evolution
calculation with the well-established CSS resummation formalism with $b_*$-prescription
and the KN parameterization of the non-perturbative form factors, for the spin-averaged
cross sections, as well as the Sivers single spin asymmetries, in Drell-Yan lepton
pair production in $pp$ collisions. This shows that our calculations are consistent
within SIDIS from HERMES/COMPASS and Drell-Yan lepton pair production.
Future experiments will provide important cross checks of our results.

A number of improvements shall follow. First, for the Drell-Yan lepton pair
production, we shall include the DGLAP evolution in the CSS
resummation, although the effects is not so large for low to moderate
transverse momentum region. As we have shown in Fig.~\ref{fig:match}, in the relative high transverse momentum
region, the DGLAP evolution is evident, and shall be able to distinguish different
dynamics in the TMD evolution. Hope we will have precise experimental data
in the future, which will provide unique opportunity to study the associated
dynamics.

Second, our approach builds connection between relative low $Q$ hard
processes to those with high $Q^2$ (up to $W/Z$ production): low $Q$ with
direct integral of the evolution kernel; high $Q$ with CSS resummation
with $b_*$ prescription. It will be nice to have a single framework to calculate
both unpolarized and spin-dependent cross sections. To do that, we have
to modify the current assumption on the non-perturbative form factors used
in the literature. This will need additional research effort. We will come back to
this issue later.

Third, in this paper, we only focused on the moderate $x$ range, where
the $x$-dependence in the non-perturbative form factors in the CSS resummation
is not so evident. In the future, we shall extend the studies to the small-$x$
region as well. For this, the HERA experiments have published data on SIDIS
processes, and shall be taken into account for a global analysis.

In addition, the TMD evolutions we derived for the spin-averaged and spin-dependent
cross sections can serve as guidelines for future theoretical developments.
An immediate extension is to analyze the Collins asymmetries in SIDIS and
di-hadron production in $e^+e^-$ annihilations. We will present a detailed
calculations in a separate publication.

\section{acknowledgements }

We thank Z.~Kang and B.~Xiao for early collaboration related to Ref.~\cite{Kang:2011mr}.
We thank A.~Bressan, A.~Martin, G.~Schnell for communications concerning HERMES and COMPASS
experimental data. We also thank Center of High Energy Physics, Peking University,
for the warm host of our visits, during which this paper is finished. This work was partially supported by the U.
S. Department of Energy via grant DE-AC02-05CH11231.

\appendix

\section{Useful Fourier Transform Formulas}

Let us start with the simple Fourier transform of $1/q_\perp^2$, in the $n=2-2\epsilon$ dimension,
\begin{eqnarray}
\frac{\alpha_s}{2\pi^2}\int\frac{d^n q_\perp}{(2\pi)^n}\frac{1}{q_\perp^2}e^{iq_\perp\cdot b_\perp}
&=&\frac{\alpha_s}{8\pi^3}\left(\frac{4}{4\pi b_\perp^2\mu^2}\right)^{-\epsilon}\Gamma(-\epsilon)\nonumber\\
&=&\frac{\alpha_s}{8\pi^3}\left[-\frac{1}{\epsilon}+\ln\frac{c_0^2}{b_\perp^2\mu^2}\right ]_{\rm \overline{MS}} \ ,
\end{eqnarray}
where $c_0=2e^{-\gamma_E}$ with $\gamma_E$ Euler constant, and the last equation
is done with ${\rm \overline{MS}}$ subtraction. The leading double logarithmic term
leads to the double pole contribution,
\begin{eqnarray}
&&\frac{\alpha_s}{2\pi^2}\int\frac{d^n q_\perp}{(2\pi)^n}\frac{1}{q_\perp^2}\ln\frac{Q^2}{q_\perp^2}
e^{iq_\perp\cdot b_\perp} =\frac{\alpha_s}{8\pi^3}\left(\frac{4}{4\pi b_\perp^2\mu^2}\right)^{-\epsilon}\lim_{\alpha\to 0} \partial_\alpha\left[\left(\frac{Q^2b_\perp^2}{4}\right)^\alpha\frac{\Gamma(-\epsilon-\alpha)}{\Gamma(1+\alpha)}\right]\nonumber\\
&&~~~=\frac{\alpha_s}{8\pi^3}\left[\frac{1}{\epsilon^2}-\frac{1}{\epsilon}\ln\frac{Q^2}{\mu^2}
+\frac{1}{2}\left(\ln\frac{Q^2}{\mu^2}\right)^2-\frac{1}{2}\left(\ln\frac{Q^2b_\perp^2}{c_0^2}\right)^2-\frac{\pi^2}{12}\right ]_{\rm \overline{MS}} \ .
\end{eqnarray}
For the Sivers spin asymmetry, we have $q_\perp^\beta$ in the Fourier transform,
which can be related to the above integral,
\begin{eqnarray}
\int\frac{d^n q_\perp}{(2\pi)^n}\frac{q_\perp^\beta}{(q_\perp^2)^2}e^{iq_\perp\cdot b_\perp} &=&
\left(\frac{ib_\perp^\beta}{2}\right)\int\frac{d^n q_\perp}{(2\pi)^n}\frac{1}{q_\perp^2}e^{iq_\perp\cdot b_\perp} \ .
\end{eqnarray}
However, for the leading double logarithmic term, we have an additional term,
\begin{eqnarray}
\int\frac{d^n q_\perp}{(2\pi)^n}\frac{q_\perp^\beta}{(q_\perp^2)^2}\ln\frac{Q^2}{q_\perp^2}e^{iq_\perp\cdot b_\perp} &=&
\left(\frac{ib_\perp^\beta}{2}\right)\left[
\int\frac{d^n q_\perp}{(2\pi)^n}\frac{1}{q_\perp^2}\ln\frac{Q^2}{q_\perp^2}e^{iq_\perp\cdot b_\perp}
-\int\frac{d^n q_\perp}{(2\pi)^n}\frac{1}{q_\perp^2}e^{iq_\perp\cdot b_\perp} \right]
\ . \label{ftlog}
\end{eqnarray}
This additional term comes from the fact that the Sivers function is 2-dimension vector depending
on the transverse momentum.



\begin{thebibliography}{99}

\bibitem{Boer:2011fh}
  D.~Boer 
  {\it et al.},
  arXiv:1108.1713 [nucl-th];
  A.~Accardi
  {\it et al.},
  arXiv:1212.1701 [nucl-ex].


\bibitem{Collins:1981uk}
J.~C.~Collins and D.~E.~Soper,
Nucl.\ Phys.\ B {\bf 193}, 381 (1981) [Erratum-ibid.\ B {\bf 213},
545 (1983)];
Nucl.\ Phys.\ B {\bf 197}, 446 (1982).


\bibitem{Collins:1984kg}
J.~C.~Collins, D.~E.~Soper and G.~Sterman,
Nucl.\ Phys.\ B {\bf 250}, 199 (1985).


\bibitem{Sudakov:1954sw}
  V.~V.~Sudakov,
  Sov.\ Phys.\ JETP {\bf 3}, 65 (1956)
  [Zh.\ Eksp.\ Teor.\ Fiz.\  {\bf 30}, 87 (1956)].


\bibitem{Dokshitzer:1978dr}
  Y.~L.~Dokshitzer, D.~Diakonov and S.~I.~Troian,
  Phys.\ Lett.\ B {\bf 78}, 290 (1978);
  Phys.\ Lett.\ B {\bf 79}, 269 (1978);
  Phys.\ Rept.\  {\bf 58}, 269 (1980).

\bibitem{Parisi:1979se}
  G.~Parisi and R.~Petronzio,
  Nucl.\ Phys.\ B {\bf 154}, 427 (1979).

\bibitem{Landry:2002ix}
  F.~Landry, R.~Brock, P.~M.~Nadolsky and C.~P.~Yuan,
  Phys.\ Rev.\ D {\bf 67}, 073016 (2003);
  Phys.\ Rev.\ D {\bf 63}, 013004 (2001).

\bibitem{Konychev:2005iy}
  A.~V.~Konychev and P.~M.~Nadolsky,
  Phys.\ Lett.\ B {\bf 633}, 710 (2006).


\bibitem{Qiu:2000ga}
  J.~-w.~Qiu and X.~-f.~Zhang,
  Phys.\ Rev.\ Lett.\  {\bf 86}, 2724 (2001)
  [hep-ph/0012058];
  Phys.\ Rev.\ D {\bf 63}, 114011 (2001)
  [hep-ph/0012348].

\bibitem{Kulesza:2002rh}
  A.~Kulesza, G.~F.~Sterman and W.~Vogelsang,
  Phys.\ Rev.\ D {\bf 66}, 014011 (2002)
  [hep-ph/0202251];
  Phys.\ Rev.\ D {\bf 69}, 014012 (2004)
  [hep-ph/0309264].

\bibitem{Catani:2000vq}
  S.~Catani, D.~de Florian and M.~Grazzini,
  Nucl.\ Phys.\ B {\bf 596}, 299 (2001)
  [hep-ph/0008184].
\bibitem{Catani:2003zt}
  S.~Catani, D.~de Florian, M.~Grazzini and P.~Nason,
  JHEP {\bf 0307}, 028 (2003)
  [hep-ph/0306211].

\bibitem{Bozzi:2003jy}
  G.~Bozzi, S.~Catani, D.~de Florian and M.~Grazzini,
  Phys.\ Lett.\ B {\bf 564}, 65 (2003)
  [hep-ph/0302104].
  Nucl.\ Phys.\ B {\bf 737}, 73 (2006)
  [hep-ph/0508068].
  Nucl.\ Phys.\ B {\bf 791}, 1 (2008)
  [arXiv:0705.3887 [hep-ph]].
  Nucl.\ Phys.\ B {\bf 815}, 174 (2009)
  [arXiv:0812.2862 [hep-ph]].
  Phys.\ Lett.\ B {\bf 696}, 207 (2011)
  [arXiv:1007.2351 [hep-ph]].

\bibitem{Ji:2004wu}
  X.~Ji, J.~P.~Ma and F.~Yuan,
  Phys.\ Rev.\ D {\bf 71}, 034005 (2005);
Phys.\ Lett.\ B {\bf 597}, 299 (2004).

\bibitem{Collins:2004nx}
  J.~C.~Collins and A.~Metz,
  Phys.\ Rev.\ Lett.\  {\bf 93}, 252001 (2004).


\bibitem{Collins}
J.C.Collins, {\it Foundations of Perturbative QCD}, Cambridge University Press, Cambridge, 2011.

\bibitem{Aybat:2011zv}
  S.~M.~Aybat and T.~C.~Rogers,
  Phys.\ Rev.\ D {\bf 83}, 114042 (2011).

\bibitem{Aybat:2011ge}
  S.~M.~Aybat, J.~C.~Collins, J.~-W.~Qiu and T.~C.~Rogers,
  Phys.\ Rev.\ D {\bf 85}, 034043 (2012).


\bibitem{Boer:2001he}
  D.~Boer,
  Nucl.\ Phys.\ B {\bf 603}, 195 (2001);
  Nucl.\ Phys.\ B {\bf 806}, 23 (2009);
  arXiv:1304.5387 [hep-ph].


\bibitem{Idilbi:2004vb}
  A.~Idilbi, X.~Ji, J.~P.~Ma and F.~Yuan,
  Phys.\ Rev.\  D {\bf 70}, 074021 (2004).

\bibitem{Kang:2011mr}
  Z.~-B.~Kang, B.~-W.~Xiao and F.~Yuan,
  Phys.\ Rev.\ Lett.\  {\bf 107}, 152002 (2011).



\bibitem{Mantry:2009qz}
  S.~Mantry and F.~Petriello,
  Phys.\ Rev.\ D {\bf 81}, 093007 (2010)
  [arXiv:0911.4135 [hep-ph]];
  Phys.\ Rev.\ D {\bf 83}, 053007 (2011)
  [arXiv:1007.3773 [hep-ph]];
  Phys.\ Rev.\ D {\bf 84}, 014030 (2011)
  [arXiv:1011.0757 [hep-ph]].

\bibitem{Becher:2010tm}
  T.~Becher and M.~Neubert,
  Eur.\ Phys.\ J.\ C {\bf 71}, 1665 (2011)
  [arXiv:1007.4005 [hep-ph]];
  T.~Becher, M.~Neubert and D.~Wilhelm,
  JHEP {\bf 1202}, 124 (2012)
  [arXiv:1109.6027 [hep-ph]];
  JHEP {\bf 1305}, 110 (2013)
  [arXiv:1212.2621 [hep-ph]].

\bibitem{GarciaEchevarria:2011rb}
  M.~G.~Echevarria, A.~Idilbi and I.~Scimemi,
  JHEP {\bf 1207}, 002 (2012)
  [arXiv:1111.4996 [hep-ph]];
  arXiv:1211.1947 [hep-ph];
  M.~G.~Echevarria, A.~Idilbi, A.~Schafer and I.~Scimemi,
  arXiv:1208.1281 [hep-ph].

\bibitem{Chiu:2012ir}
  J.~-Y.~Chiu, A.~Jain, D.~Neill and I.~Z.~Rothstein,
  JHEP {\bf 1205}, 084 (2012)
  [arXiv:1202.0814 [hep-ph]].

\bibitem{Collins:2012uy}
  J.~C.~Collins and T.~C.~Rogers,
  arXiv:1210.2100 [hep-ph].



\bibitem{Brodsky:2002cx}
S.~J.~Brodsky, D.~S.~Hwang and I.~Schmidt,
Phys.\ Lett.\ B {\bf 530}, 99 (2002);
Nucl.\ Phys.\ B {\bf 642}, 344 (2002).


\bibitem{Collins:2002kn}
J.~C.~Collins,
Phys.\ Lett.\ B {\bf 536}, 43 (2002).




\bibitem{Airapetian:2009ae}
  A.~Airapetian {\it et al.}  [HERMES Collaboration],
  Phys.\ Rev.\ Lett.\  {\bf 103}, 152002 (2009).

\bibitem{Airapetian:2010ds}
  A.~Airapetian {\it et al.}  [HERMES Collaboration],
  Phys.\ Lett.\ B {\bf 693}, 11 (2010).

\bibitem{Alekseev:2008aa}
  M.~Alekseev {\it et al.}  [COMPASS Collaboration],
  Phys.\ Lett.\ B {\bf 673}, 127 (2009).


\bibitem{Adolph:2012sp}
  C.~Adolph {\it et al.}  [COMPASS Collaboration],
  Phys.\ Lett.\ B {\bf 717}, 383 (2012).
\bibitem{Qian:2011py}
  X.~Qian {\it et al.}  [Jefferson Lab Hall A Collaboration],
  Phys.\ Rev.\ Lett.\  {\bf 107}, 072003 (2011)
  [arXiv:1106.0363 [nucl-ex]].

\bibitem{compassdy}
COMPASS proposal at CERN, \url{http://wwwcompass.cern.ch/compass/proposal/compass-II_proposal/compass-II_proposal.pdf}


\bibitem{fermilabdy}
{\it Polarized Drell-Yan Measurements with the Fermilab Main Injector}, W. Lorenzon, P.E. Reimer, {\it et al.},
\url{http://www.fnal.gov/directorate/program_planning/June2012Public/P-1027_Pol-Drell-Yan-proposal.pdf}


\bibitem{futurerhic} 
  E.~C.~Aschenauer, A.~Bazilevsky, K.~Boyle, K.~O.~Eyser, R.~Fatemi, C.~Gagliardi, M.~Grosse-Perdekamp and J.~Lajoie {\it et al.},
  arXiv:1304.0079 [nucl-ex].

\bibitem{Vogelsang:2005cs}
  W.~Vogelsang and F.~Yuan,
  Phys.\ Rev.\ D {\bf 72}, 054028 (2005)
  [hep-ph/0507266].

\bibitem{Collins:2005rq}
  J.~C.~Collins, A.~V.~Efremov, K.~Goeke, M.~Grosse Perdekamp, S.~Menzel, B.~Meredith, A.~Metz and P.~Schweitzer,
  Phys.\ Rev.\ D {\bf 73}, 094023 (2006)
  [hep-ph/0511272].


\bibitem{Anselmino:2005ea}
  M.~Anselmino, {\it et al.},
  Phys.\ Rev.\ D {\bf 72}, 094007 (2005)
  [Erratum-ibid.\ D {\bf 72}, 099903 (2005)].


\bibitem{Anselmino:2005an}
  M.~Anselmino, M.~Boglione, J.~C.~Collins, U.~D'Alesio, A.~V.~Efremov, K.~Goeke, A.~Kotzinian and S.~Menzel {\it et al.},
  hep-ph/0511017.

\bibitem{Anselmino:2009st}
  M.~Anselmino, M.~Boglione, U.~D'Alesio, S.~Melis, F.~Murgia and A.~Prokudin,
  Phys.\ Rev.\ D {\bf 79}, 054010 (2009)
  [arXiv:0901.3078 [hep-ph]].

\bibitem{Kang:2009sm}
  Z.~-B.~Kang and J.~-W.~Qiu,
  Phys.\ Rev.\ D {\bf 81}, 054020 (2010)
  [arXiv:0912.1319 [hep-ph]].

\bibitem{Bacchetta:2011gx}
  A.~Bacchetta and M.~Radici,
  Phys.\ Rev.\ Lett.\  {\bf 107}, 212001 (2011)
  [arXiv:1107.5755 [hep-ph]].

\bibitem{Gamberg:2013kla}
  L.~Gamberg, Z.~-B.~Kang and A.~Prokudin,
  Phys.\ Rev.\ Lett.\  {\bf 110}, 232301 (2013)
  [arXiv:1302.3218 [hep-ph]].

\bibitem{Aybat:2011ta}
  S.~M.~Aybat, A.~Prokudin and T.~C.~Rogers,
  Phys.\ Rev.\ Lett.\  {\bf 108}, 242003 (2012).



\bibitem{Anselmino:2012aa}
  M.~Anselmino, M.~Boglione and S.~Melis,
  Phys.\ Rev.\ D {\bf 86}, 014028 (2012).





\bibitem{Sun:2013dya}
  P.~Sun and F.~Yuan,
  arXiv:1304.5037 [hep-ph].


\bibitem{Meng:1995yn}
  R.~Meng, F.~I.~Olness and D.~E.~Soper,
  Phys.\ Rev.\ D {\bf 54}, 1919 (1996)
  [hep-ph/9511311].

\bibitem{Nadolsky:1999kb}
  P.~M.~Nadolsky, D.~R.~Stump and C.~P.~Yuan,
  Phys.\ Rev.\ D {\bf 61}, 014003 (2000)
  [Erratum-ibid.\ D {\bf 64}, 059903 (2001)]
  [hep-ph/9906280];
  Phys.\ Rev.\ D {\bf 64}, 114011 (2001)
  [hep-ph/0012261].

\bibitem{Efremov:1981sh}
  A.~V.~Efremov and O.~V.~Teryaev,
  Sov.\ J.\ Nucl.\ Phys.\  {\bf 36}, 140 (1982)
  [Yad.\ Fiz.\  {\bf 36}, 242 (1982)];
  A.~V.~Efremov and O.~V.~Teryaev,
  Phys.\ Lett.\ B {\bf 150}, 383 (1985).

\bibitem{Qiu:pp}
J.W.~Qiu and G.~Sterman,
Phys.\ Rev.\ Lett.\  {\bf 67}, 2264 (1991);
  Nucl.\ Phys.\ B {\bf 378}, 52 (1992);
Phys.\ Rev.\ D {\bf 59}, 014004 (1998).


\bibitem{Ji:2006ub}
  X.~Ji, J.~W.~Qiu, W.~Vogelsang and F.~Yuan,
Phys.\ Rev.\ Lett.\ {\bf 97}, 082002 (2006);
  Phys.\ Rev.\ D {\bf 73}, 094017 (2006);
  Phys.\ Lett.\ B {\bf 638}, 178 (2006);
Y.~Koike, W.~Vogelsang and F.~Yuan,
Phys.\ Lett.\  B {\bf 659}, 878 (2008).

\bibitem{bbdm} A.~Bacchetta, D.~Boer, M.~Diehl and P.~J.~Mulders,
  JHEP {\bf 0808}, 023 (2008).

\bibitem{Bacchetta:2004jz}
  A.~Bacchetta, U.~D'Alesio, M.~Diehl and C.~A.~Miller,
  Phys.\ Rev.\ D {\bf 70}, 117504 (2004)
  [hep-ph/0410050].

\bibitem{Vogelsang:2009pj}
  W.~Vogelsang and F.~Yuan,
  Phys.\ Rev.\  D {\bf 79}, 094010 (2009).

\bibitem{Kang:2008ey}
  Z.~B.~Kang and J.~W.~Qiu,
Phys.\ Rev.\  D {\bf 79}, 016003 (2009);
 [arXiv:0811.3101 [hep-ph]].
\bibitem{Zhou:2008mz}
  J.~Zhou, F.~Yuan and Z.~T.~Liang,
  Phys.\ Rev.\  D {\bf 79}, 114022 (2009);
  [arXiv:0812.4484 [hep-ph]].
\bibitem{Braun:2009mi}
  V.~M.~Braun, A.~N.~Manashov and B.~Pirnay,
  Phys.\ Rev.\  D {\bf 80}, 114002 (2009).
  [arXiv:0909.3410 [hep-ph]].

\bibitem{Schafer:2012ra}
  A.~Schafer and J.~Zhou,
  Phys.\ Rev.\ D {\bf 85}, 117501 (2012)
  [arXiv:1203.5293 [hep-ph]].
\bibitem{Kang:2012em}
  Z.~-B.~Kang and J.~-W.~Qiu,
  Phys.\ Lett.\ B {\bf 713}, 273 (2012)
  [arXiv:1205.1019 [hep-ph]].
\bibitem{Ma:2012xn}
  J.~P.~Ma and Q.~Wang,
  Phys.\ Lett.\ B {\bf 715}, 157 (2012)
  [arXiv:1205.0611 [hep-ph]].


\bibitem{Bacchetta:2006tn}
  A.~Bacchetta, M.~Diehl, K.~Goeke, A.~Metz, P.~J.~Mulders and M.~Schlegel,
  JHEP {\bf 0702}, 093 (2007).


\bibitem{Balazs:1997xd}
  C.~Balazs and C.~P.~Yuan,
  Phys.\ Rev.\ D {\bf 56}, 5558 (1997)
  P.~M.~Nadolsky, D.~R.~Stump and C.~P.~Yuan,
  Phys.\ Rev.\ D {\bf 61}, 014003 (2000)
  [Erratum-ibid.\ D {\bf 64}, 059903 (2001)]

\bibitem{Schweitzer:2010tt}
  P.~Schweitzer, T.~Teckentrup and A.~Metz,
  Phys.\ Rev.\ D {\bf 81}, 094019 (2010).

\bibitem{Lai:2010vv}
  H.~-L.~Lai, M.~Guzzi, J.~Huston, Z.~Li, P.~M.~Nadolsky, J.~Pumplin and C.~-P.~Yuan,
  Phys.\ Rev.\ D {\bf 82}, 074024 (2010)
  [arXiv:1007.2241 [hep-ph]].

\bibitem{deFlorian:2007aj}
  D.~de Florian, R.~Sassot and M.~Stratmann,
  Phys.\ Rev.\ D {\bf 75}, 114010 (2007).

\bibitem{Schweitzer:2012hh}
  P.~Schweitzer, M.~Strikman and C.~Weiss,
  JHEP {\bf 1301}, 163 (2013)
  [arXiv:1210.1267 [hep-ph]].

\bibitem{Airapetian:2012ki}
  A.~Airapetian {\it et al.}  [HERMES Collaboration],
  Phys.\ Rev.\ D {\bf 87}, 074029 (2013).



\bibitem{Adolph:2013stb}
  C.~Adolph {\it et al.}  [COMPASS Collaboration],
  arXiv:1305.7317 [hep-ex].



\bibitem{Ito:1980ev}
  A.~S.~Ito, {\it et al.},
  Phys.\ Rev.\ D {\bf 23}, 604 (1981).

\bibitem{Moreno:1990sf}
  G.~Moreno, C.~N.~Brown, W.~E.~Cooper, D.~Finley, Y.~B.~Hsiung, A.~M.~Jonckheere, H.~Jostlein and D.~M.~Kaplan {\it et al.},
  Phys.\ Rev.\ D {\bf 43}, 2815 (1991).

\bibitem{Salam:2008qg}
  G.~P.~Salam and J.~Rojo,
  Comput.\ Phys.\ Commun.\  {\bf 180}, 120 (2009)
  [arXiv:0804.3755 [hep-ph]].

\bibitem{Kang:2009bp}
  Z.~-B.~Kang and J.~-W.~Qiu,
  Phys.\ Rev.\ Lett.\  {\bf 103}, 172001 (2009)
  [arXiv:0903.3629 [hep-ph]].
\bibitem{Metz:2010xs}
  A.~Metz and J.~Zhou,
  Phys.\ Lett.\ B {\bf 700}, 11 (2011)
  [arXiv:1006.3097 [hep-ph]].


\end{thebibliography}
\end{document}